\documentclass[letterpaper,aps,pre,showpacs,preprintnumbers,superscriptaddress]{revtex4}

\usepackage{graphicx}
\usepackage{epsfig}
\usepackage{srcltx}
\usepackage{citesort}

\usepackage{amsmath}
\usepackage{amssymb}

\newcommand\T{\rule{0pt}{3.0ex}}

\newcommand{\bi}[1]{\mathbf{#1}}
\newcommand{\eref}[1]{\eqref{#1}}
\newcommand{\rmi}{{\rm i}}
\newcommand{\rmd}{{\rm d}}
\newcommand{\rme}{{\rm e}}
\newcommand{\Or}{{\rm O}}
\newcommand{\tref}{\ref}

\begin{document}
\title{Dynamical Renormalization Group Study for a Class of Non-local Interface Equations}

\author{Matteo Nicoli}
\affiliation{Physique de la Mati\`ere Condens\'ee, \'Ecole Polytechnique, CNRS, 91128 Palaiseau, France}

\author{Rodolfo Cuerno}
\affiliation{Departamento de Matem\'aticas and Grupo Interdisciplinar de Sistemas Complejos (GISC),
Universidad Carlos III de Madrid, Avenida de la Universidad 30, 28911 Legan\'es, Spain}
\author{Mario Castro}
\affiliation{GISC and Grupo de Din\'amica No Lineal (DNL), Escuela T\'ecnica Superior de Ingenier\'ia (ICAI),
Universidad Pontificia Comillas, 28015 Madrid, Spain}

\begin{abstract}
We provide a detailed Dynamic Renormalization Group study for a class of stochastic equations
that describe non-conserved interface growth mediated by non-local interactions. We
consider explicitly both the morphologically stable case, and the less studied case in which
pattern formation occurs, for which flat surfaces are linearly unstable to periodic perturbations.
We show that the latter leads to non-trivial scaling behavior in an appropriate parameter range
when combined with the Kardar-Parisi-Zhang (KPZ) non-linearity, that nevertheless does not correspond
to the KPZ universality class. This novel asymptotic behavior is characterized by two scaling laws that fix the critical
exponents to dimension-independent values, that agree with previous reports from numerical
simulations and experimental systems. We show that the precise form of the linear stabilizing terms
does not modify the hydrodynamic behavior of these equations. One of the scaling laws, usually
associated with Galilean invariance, is shown to derive from a vertex cancellation
that occurs (at least to one loop order) for any choice of linear terms in the equation of motion and is independent on the
morphological stability of the surface, hence generalizing this well-known property of the KPZ equation.
Moreover, the argument carries over to other systems like the Lai-Das Sarma-Villain equation, in which vertex
cancellation is known {\em not to} imply an associated symmetry of the equation.
\end{abstract}
\maketitle

\section{Introduction}

Many surface growth systems can be described by stochastic differential equations that capture the evolution
of the surface height $h(\bi{r},t)$ at time $t$ and point $\bi{r}$ on a $d$-dimensional substrate \cite{Barabasi:1995,Krug:1997}.
In many cases these equations are derived from first principles \cite{cuerno:2007,Misbah:2010}, thus linking the small scale physics of the growth process with a coarse grained description of the surface height. Sometimes, this procedure ends up phrasing a physical problem in which interactions are short-range and local, into a stochastic equation in which {\em effective} interactions turn out to be long-range and/or non-local due to kinetic constraints. An example is provided by diffusion limited growth \cite{Cuerno:2001,Nicoli:2008} (or erosion \cite{Krug:1991}) of an aggregate, in which
the local growth velocity depends on the global morphology of the evolving cluster, due to shadowing effects produced by the most prominent surface peaks over less exposed areas. Thus, a local transport mechanism such as diffusion induces long-range correlations among positions at the aggregate surface. Dynamically, this may result into a morphological instability in the case of a growing surface (protuberances {\em grow} faster, amplifying deviations from a flat interface), or morphological stability in the case of an eroding one (protuberances {\em decay} faster so that deviations from a flat interface are attenuated).

In principle, the non-local morphologically unstable case is more interesting than the non-local morphologically stable case, since the latter is known to lead to trivial scaling behavior, in the sense that the corresponding scaling exponents can be obtained from simple dimensional analysis as applied to the corresponding linearized equation \cite{Krug:1991,Biler:1999}.
On the other hand, morphological instabilities are key to pattern forming systems \cite{Cross:2009}, {\em non-local} linear instabilities
having been the subject of extensive studies in the past. Celebrated examples include the Mullins-Sekerka (MS) instability in solidification \cite{Pelce:2004}, the Saffman-Taylor (ST) hydrodynamic instability in fluid flow \cite{Saffman:1958,Bensimon:1986},
or the Daerrius-Landau (DL) instability \cite{Darrieus:1938,Landau:1944,Clanet:1998,Bychkov:2000} occurring in the
propagation of a premixed laminar flame, 
to cite a few.

The stochastic equations that describe morphologically unstable surfaces typically combine linear terms that account for the selection of a typical wavelength associated with that one among the height field modes, $h_{{\bi k}}(t)={\mathcal F}[h({\bi r},t)]_{{\bi k}}$, that has the largest growth rate \cite{Cross:2009}, with additional non-linear terms that stabilize the exponential growth associated with the morphological instability, and prevent blow up the surface height field. Here, ${\mathcal F}$ stands for space Fourier transform and ${\bi k}$ is wave vector. In the context of surface growth, the Kardar-Parisi-Zhang (KPZ)~\cite{Kardar:1986} non-linearity $(\nabla h)^2$ is expected whenever the interface evolves in absence of conservation laws \cite{Krug:1997,Castro:2007}. Thus, a natural example for a model of unstable surface growth subject to external fluctuations is given e.g.\ by
\begin{equation}
\partial _th= \nu \nabla^2h - {\cal K} \nabla^4 h +\frac{\lambda}{2} (\nabla h)^2+\eta,
\label{nks_eq}
\end{equation}
where $\eta$ is an uncorrelated Gaussian noise with zero mean and constant variance $\Pi_0$, i.e.,
\begin{equation}
\langle\eta({\bi{x},t})\eta({\bi{x}',t'}) \rangle=
2\Pi_0 \, \delta({\bi{x}-\bi{x}'})\, \delta({t-t'}).
\label{noise_0}
\end{equation}
In equation \eref{nks_eq}, $\nu <0$, ${\cal K} >0$, and $\lambda$ are parameters. A {\em negative} value for $\nu$ implements the morphological instability, that is countered both by the biharmonic and by the nonlinear terms. Equation (\ref{nks_eq}) is the noisy Kuramoto-Sivashinsky (nKS) equation \cite{Misbah:2010}, whose deterministic limit is a paradigmatic
model for chaotic spatially extended systems, arising in a variety of physical contexts, like thin solid films, interfaces
between viscous fluids, waves in plasmas and chemical reactions, reaction-diffusion systems, and others \cite{Cross:2009}.
Actually, the large scale behavior of the nKS systems is in the KPZ universality class both in one and in two spatial
dimensions \cite{Cuerno:1995,Cuerno:1995b,Ueno:2005,Nicoli:2010,Pradas:2011}. This is a remarkable manifestation of renormalization, by which the ``surface tension'' parameter $\nu$ flows from a bare {\em negative} value to a positive renormalized value at large scales, as initially conjectured by Yakhot thirty years ago \cite{Yakhot:1981}.

Indeed, the KPZ (or noisy Burgers \cite{McComb:1991}) equation reads
\begin{equation}
\partial _th=\nu \nabla^2h+\frac{\lambda}{2} (\nabla h)^2+\eta,
\label{kpz_eq}
\end{equation}
where $\nu >0$. The KPZ nonlinear term is invariant under translations in the growth direction ($h\rightarrow h+c$, with $c$ an arbitrary constant), in the substrate direction ($x\rightarrow x+a$) and, besides, it is symmetric under rotations and inversion about the growth direction \cite{Barabasi:1995}. Its ubiquity has led to a huge amount of work in the past, the KPZ equation having become one of the central problems in non-equilibrium interface dynamics~\cite{Barabasi:1995,Krug:1997,Halpin-Healy:1995}. Actually, based on Dynamical Renormalization Group (DRG) arguments \cite{Kardar:1986,Medina:1989}, KPZ scaling is generically expected at the asymptotic state for non-conserved growth systems. However, such a theoretical expectation has been too scarcely confirmed by experiments~\cite{Cuerno:2004}. One possible explanation is that short-time morphological instabilities (of the type occurring e.g.\ in the noisy Kuramoto-Sivashinsky system) may hinder the observation of the asymptotic behavior~\cite{Cuerno:2001}. For instance, for appropriate choices of parameter values in the nKS equation, KPZ scaling may become unobservable even in numerical simulations \cite{Nicoli:2010}. Moreover, some of the hypothesis made in the derivation of the KPZ equation may not be realistic for many experimental situations. In particular, complex processes may take place in the system, such as diffusion, advection in a fluid, elastic stress in the aggregate, etc.\ that break down the assumption of {\em local} interactions.

Following the above ideas, a number of non-local versions of the KPZ equation have been proposed ---see e.g.~\cite{Tang:2001,Mann:2001,Katzav:2003} and further references in \cite{Cuerno:2004}--- in order to account for scaling exponent values seen in experimental conditions that differ from the KPZ values. However, some of these are {\em ad-hoc} generalizations that lack a clear physical justification. Moreover, most of them consider the morphologically stable condition [e.g, $\nu>0$ in Eqs.\ (\ref{nks_eq}), (\ref{kpz_eq})] which, for uncorrelated time-dependent noise, leads to scaling behavior associated with linear equations as mentioned above.

\subsection{A family of non-local interface equations}

There are other ways to generalize interface equations, that do have a consistent physical interpretation. For instance, one can generate additional terms on the rhs of, say, Eqs.\ (\ref{nks_eq}), (\ref{kpz_eq}), if one enforces full compatibility of the equation with the symmetries of the system~\cite{Castro:2007}. Otherwise, starting from a first principles formulation of a specific growth system in terms of, e.g., a moving boundary problem, one can derive explicitly the evolution equation for $h({\bi r},t)$ through projection of the whole bulk dynamics onto the interface, under suitable approximations. For non-conserved diffusion-limited growth, this has been accomplished for instance in \cite{Cuerno:2002,Nicoli:2008}. For epitaxial growth systems, see a review in, e.g., \cite{Misbah:2010}. Working along these lines, we have recently identified a whole family of non-local equations that is formulated as a single stochastic differential equation~\cite{Nicoli:2009b}. This equation has many relevant, well-known physical systems as particular cases. In Fourier space, the
equation takes the simple form
\begin{equation}
\partial_t h_\bi{k}(t) = (-\nu k^{\mu} - \mathcal{K} k^m- \mathcal{N} k^n)
    h_\bi{k}(t) + \frac{\lambda}{2} \mathcal{F}[(\nabla h)^2]_\bi{k} +
    \eta_\bi{k}(t),
\label{eq_gral}
\end{equation}
where $k$ denotes the magnitude of the $d$ dimensional wave-number $\bi{k}$ and $\eta_{\bi{k}}(t) = \mathcal{F}(\eta)$
with $\eta$ as in equation \eref{noise_0}. Here, $\mu$, $m$, $n$, $\mathcal{K}$, and $\mathcal{N}$ are {\em positive} parameters such that $0 <\mu \leq 2$, $m \geq 2$, and $n > m$, see below. We will consider both the morphologically unstable case in which $\nu<0$, as in e.g.\ the noisy Kuramoto-Sivashinsky system, and the morphologically stable case $\nu>0$. The term with coefficient ${\cal N}$ represents higher order stabilizing mechanisms that are required to make sense of parameter choices in which fluctuations do not dissipate at small scales, i.e.\ $-\nu k^{\mu} - \mathcal{K} k^m\to+\infty$ for $k\to+\infty$, as in the $\mu \to m=2$ nKS-type limit. Still, one of the results of the present work is to explicitly prove the irrelevance of the term with coefficient ${\cal N}$ for the asymptotic scaling. Note that, whenever $\mu$ is {\em not} an even integer, the corresponding linear term in equation (\ref{eq_gral}) is a fractional Laplacian acting on the height field,
\begin{equation}
(-\nabla^2)^{\mu/2} h(\mathbf{r}) = c_{d,\mu} \, {\rm PV} \int_{\mathbb{R}^d}
\frac{h(\mathbf{r})-h(\mathbf{r}')}{|\mathbf{r}-\mathbf{r}'|^{d+\mu}}
\, {\rm d}\mathbf{r}' , \;\;\; 0 < \mu \leq 2 ,
\label{lap}
\end{equation}
where ${\rm PV}$ denotes Cauchy principal value and $c_{d,\mu}$ are appropriate numerical constants \cite{Samko:2002,Silvestre:2007}. The Fourier representation of (\ref{lap}) is given by
\begin{equation}
\mathcal{F}[(-\nabla^2)^{\mu/2} h] = k^{\mu} h_{\mathbf{k}}.
\label{lap_F}
\end{equation}
Indeed, only if $\mu$ is an even integer can (\ref{lap}) be inverse Fourier transformed to get a local operator in real space, as in the noisy Kuramoto-Sivashinsky case. Otherwise, the fractional Laplacian corresponds to convolution of height differences with an algebraically decaying kernel. In the context of equilibrium systems with long-range interactions, equation \eref{lap} corresponds to the so-called ``weak'' form of the latter, see \cite{Mukamel:2008} for a review. For the morphologically stable case, relaxation operators with the same shape as \eref{lap} have also been advocated to account for long-range interactions in processes defined both on Euclidean and on fractal lattices, see e.g.\ \cite{Krug:1997,Katzav:2003,Kim:2010} and references therein.

For the morphologically unstable case, the linear dispersion relation that appears in equation (\ref{eq_gral}), $\sigma_{\bi{k}} = -\nu k^{\mu} - \mathcal{K} k^m$, is common to many celebrated pattern forming systems. Thus, for $\mu=1$ and $m=3$ we get the Mullins-Sekerka or Saffman-Taylor dispersion relation, while the Darrieus-Landau and Kuramoto-Sivashinsky cases are obtained for $\mu=1$, $m=2$, or for $\mu=2$, $m=4$, respectively. Thus, (\ref{eq_gral}) allows us to study a large class of models in which a non-local morphological instability is coupled with the KPZ non-linearity, which includes a number of equations that are important on their own rights, but also local models as nKS equation that have been exhaustively studied in the literature.

Apart from the noisy Kuramoto-Sivashinsky equation, we have the stochastic generalization of the Michelson-Sivashinsky (sMS) equation \cite{Sivashinsky:1977,Michelson:1977} for $\mu=1$, $m=2$, that provides a standard model to describe propagation of premixed laminar flames \cite{Bychkov:2000}. Moreover, the sMS equation is the small slope approximation of a nonlinear equation found for growing interfaces controlled by {\em ballistic} transport \cite{Bales:1990}, and has been also derived from first principles for reactive infiltration in porous media \cite{Kechagia:2001}. Likewise, an interface equation that we term MSKPZ, in which the Mullins-Sekerka dispersion relation ($\mu=1$, $m=3$) is combined with the KPZ nonlinearity, describes thin film growth by chemical vapor deposition (CVD) or by electrochemical deposition (in galvanostatic conditions) \cite{Nicoli:2008} when the attachment kinetics of aggregating particles is fast compared to mean velocity of the growing interface, and has been recently employed to quantitatively describe CVD experiments \cite{Castro:2011}. Moreover, the linear dispersion relation appearing in (\ref{eq_gral})
has been found to provide a phenomenological description for experimental systems in which transport is of a diffusive nature, using
$\mu$ as a fitting parameter \cite{Nicoli:2009}.

In \cite{Nicoli:2009b}, a numerical study has been reported for the family of equations (\ref{eq_gral}) in one and two dimensions, together with a preliminary DRG study for fixed $m=2$ as a representative case.
Particular attention was paid to the morphologically unstable case $\nu<0$ for which, strikingly, non-trivial (non-linear) scaling occurs. Thus, under such conditions simulations suggest that, for each dimension $d$, the universality class changes at $\mu = z_{\rm KPZ}(d)$.
Unstable equations characterized by $\mu$ such that $z_{\rm KPZ}(d) \leq \mu \leq 2$ are all in the KPZ universality class (as for the noisy Kuramoto-Sivashinsky equation), while the scaling behavior for small values $0 < \mu < z_{\rm KPZ}(d)$ is characterized by dimension independent exponents that are {\em different} from those predicted by dimensional analysis. Here, $z_{\rm KPZ}(d)$ denotes the dynamic exponent of the KPZ class for the given value of $d$. For the definition of the critical exponents in these systems see section \ref{scaling_anal}.

In the present work, we perform a detailed DRG study of (\ref{eq_gral}) for arbitrary substrate dimensions and considering
both the morphologically stable and unstable conditions. Although the perturbative DRG has well known limitations \cite{McComb:1991,Halpin-Healy:1995}, it has recently shown a remarkable explanatory power for various types of systems that include conserved or non-conserved versions of the KPZ nonlinearity, under morphologically stable
(see \cite{Haselwandter:2007-prl,Haselwandter:2010} and references therein) or unstable \cite{Keller:2011} conditions. We expect relevant information to be also obtained from the DRG flow on the asymptotic properties of equation (\ref{eq_gral}). Indeed, our present work extends previous analytical results in a number of important aspects: {\em (i)} By generalizing the linear dispersion relation and the nonlinearity appearing in the stochastic equation to more general forms, we prove that a vertex cancellation occurs to one loop order in our perturbative expansion, inducing a Galilean scaling relation at the non-linear fixed points found in the numerical simulations of \cite{Nicoli:2009b}.
The ensuing equation generalizes both the KPZ and the Lai-Das Sarma-Villain \cite{Lai:1991,Villain:1991} equations. Thus, explicit symmetry under a Galilean transformation, that occurs for the former equation but not for the latter, is seen not to be a necessary condition for the mentioned scaling relation to apply. Moreover, this scaling relation also
holds at the asymptotic state numerically characterized for sufficiently small $\mu$ values in the morphologically unstable
case of equation \eref{eq_gral}. {\em (ii)} We consider in detail the RG flow for arbitrary $d$, obtaining fixed points and flow properties that remained unexplored in \cite{Nicoli:2009b}. In particular, we assess in detail the extent to which the renormalization flow of (\ref{eq_gral}) inherits the structure of the perturbative KPZ flow. {\em (iii)} Working on the reference $m=2$ case for $\mu=1$, we prove explicitly that the additional interactions $-{\cal N} k^{n} h_{\mathbf{k}}$ do not modify the asymptotic behavior for $n=3, 4$; this justifies numerical results in \cite{Nicoli:2009b} that exploited this fact and had remained analytically unproven thus far. As a bonus, this result also justifies our parameter restriction to the case $\mu \leq 2$ in the definition of equation (\ref{eq_gral}), since taking $\mu>2$ in the equation generates $k^2 h_{\bi{k}}$ terms under renormalization, that become more relevant than the bare $k^{\mu} h_{\bi{k}}$ ones.

The paper is organized as follows: 
As a preliminary, we start in section \ref{scaling_anal} with a scaling analysis of the family of equations \eref{eq_gral} through power counting and through a more refined Flory-type approach \cite{Hentschel:1991}, and with considerations on their naive upper critical dimensions.
Given the insufficiency of mere scaling analysis, in section \ref{RGsketch} we outline the perturbative DRG procedure as applied to equation \eref{eq_gral}. We provide the proof on the mentioned vertex cancellation leading to the Galilean scaling relation. In section \ref{DRGflow} we discuss the most convenient parameter space in which to study the DRG flow. This is addressed in detail in section \ref{results}, where we provide the main results (fixed points, stability and critical exponents) for dimensions $d=1$, $d=2$, and $d>2$, assessing how dimensionality changes the overall properties of the equation. Finally, in section \ref{discussion}, we discuss the significance of our results both for the equations studied and from a more general perspective for surface kinetic roughening, and give our conclusions and an outlook. In order to make our presentation self-contained while trying not to obscure it, a number of technical details are provided in appendices.

\section{Scaling analysis and upper critical dimension}
\label{scaling_anal}

In order to gain some intuition on the scaling behavior that can be expected for equation \eref{eq_gral}, we can perform a scaling analysis. Straightforward power counting first leads to the conclusion that, under a rescaling of the height field and space and time coordinates as $\tilde{h}(\bi{r},t)=b^{-\alpha} h(b\bi{r},b^z)$ with $b>1$, an equation with the same shape as \eref{eq_gral} holds for $\tilde{h}$, with rescaled parameters
\begin{equation}
\tilde{\nu}=b^{z-\mu} \nu, \quad \tilde{\cal K} = b^{z-m} {\cal K}, \quad \tilde{\lambda}= b^{\alpha+z-2} \lambda, \quad \tilde{\Pi}_0 = b^{z-2\alpha-d} \Pi_0.
\label{scaling}
\end{equation}
Here, $\alpha$ and $z$ are, respectively, the roughness and dynamic exponents that characterize the scaling behavior of the system. The behavior of the ratio $\tilde{\cal K}/\tilde{\nu}$ immediately implies that ${\cal K}$ is less relevant than $\nu$ under rescaling, since $\mu < m$. Likewise, under morphologically stable conditions, we expect $\alpha < 1$, leading to the dominance of the non-local term over the KPZ term for $\mu \leq 1$. Conversely, for morphologically unstable conditions we can assume $\alpha > 1$, so that we would expect $\nu$ to be irrelevant as compared with $\lambda$ for $\mu \geq 1$.

Power counting is indeed limited in its predictions regarding precise values of scaling exponents and
expected behavior outside the exponent regions considered in the previous paragraph. One can improve over
these results by performing a scaling argument \`a la Flory \cite{Hentschel:1991} in which one estimates the
magnitude of individual terms present in \eref{eq_gral}. We assume that at long times $t\gg t_l$, and
averaged over length scales $l$, the height-height correlation function \cite{Barabasi:1995} scales as
$C(l,t)\sim h_l^2 \sim l^{2\alpha}$, and that at long times the relaxation time of these fluctuations is
$t_l \sim l^z$. Note that $t_l$ and $h_l$ depend on the length scale $l$. The various terms
in equation \eref{eq_gral} can be estimated as
\begin{equation}
\left\langle \left|\partial_t h \right|\right\rangle_l\sim \frac{h_l}{t_l},
\qquad
\nu\left\langle \left| \mathcal{F}^{-1}[k^\mu h_{\bi k}] \right|\right\rangle_l \sim
	\nu\frac{h_l}{l^\mu}, \qquad
\lambda\left\langle \left|(\nabla h)^2 \right|\right\rangle_l \sim \lambda
	\frac{h_l^2}{l^2}.
\label{flory}
\end{equation}
For white noise we can estimate its mean-square fluctuations on length
scales $l$ and times scales $t_l$ as $\eta_l\sim (\Pi_0/S_l t_l)^{1/2}$,
where $S_l$ is the average surface area of the interface on scale $l$, and thus $S_l\sim (h_l^2+l^2)^{d/2}$
\cite{Hentschel:1991}. For morphologically stable conditions, the $l^2$ term in $S_l$ dominates the height fluctuations
(namely, $\alpha < 1$); instead, for morphologically unstable surfaces the dominant term is $h_l^2$.
Let us start by considering the former case, and assume that $\nu$ dominates over $\lambda$.
Then, by equating the non-local term with the inertial term (the one proportional to the time derivative
of $h$), we obtain a simple condition for the characteristic time of the height fluctuations
$t_l\sim l^\mu/\nu$ and, consequently, the relation $z=\mu$. Under these conditions
we estimate $\eta_l\sim (\Pi_0/ l^d t_l)^{1/2}$, and equating in turn the inertial term with
the noise fluctuations we obtain $h_l\sim l^{(\mu-d)/2}$. This expression gives us the value of the roughness exponent
$\alpha=(\mu-d)/2$, which combined with $z=\mu$ leads to the so-called
hyperscaling relation $2\alpha+d=z$, see section \ref{DRGflow}. This set
of exponents has indeed been observed in our numerical simulations of the morphologically stable
condition \cite{Nicoli:2009b}. In the morphologically unstable case, we expect the noise term to scale as $\eta_l\sim (\Pi_0/ h_l^d t_l)^{1/2}$. If we try to equate this with the inertial term, we expect a violation of hyperscaling. Led by the previous power counting analysis, we now equate the inertial and KPZ terms, leading to $h_l \sim l^2 t_l^{-1} \sim l^{2-z} \sim l^{\alpha}$, i.e.\ we obtain the so-called Galilean exponent relation $\alpha+z=2$.

In principle we have not assumed any restrictions on the values of $\mu$ when deriving the previous scaling relations and exponent values. As mentioned in the previous section, numerical simulations of the morphologically unstable
condition \cite{Nicoli:2009b} indicate that the scaling behavior depends non-trivially on the precise value of $\mu$.
Moreover, the present argument does not allow us to assess possible changes for different substrate dimension $d$. For these
reasons, we need to resort to more refined scaling analysis, among which the DRG is a natural choice.

Before leaving this section, we can also make estimates on the naive critical dimension of equation \eref{eq_gral}
by making use of Eqs.\ \eref{scaling}. As in the standard KPZ case \cite{Kardar:2007}, the idea is
to employ the exponents associated with the relevant linear theory in order to derive conditions under which
$\tilde{\lambda}$ scales to zero, or else grows unboundedly for increasing $b$. Typically such conditions depend on the
value of $d$, hence the appearance of a critical dimension. In our case, given that the linearized equation \eref{eq_gral}
blows up exponentially in time for the morphologically unstable condition, such an estimate can only be meaningful,
if at all, in the morphologically stable condition. Thus, employing the exponent values $z=\mu$ and $\alpha=(z-d)/2$,
we get $\tilde{\lambda} = b^{n_{\lambda}} \lambda$ with $n_{\lambda}=(3\mu-4-d)/2$. Hence, the nonlinearity becomes irrelevant for $d>d_c(\mu)=3\mu-4$, which is the case for $d\geq2$ and any value of $\mu$ (note $d_c=0$ for $\mu =4/3$). Alternatively, this implies that in $d<2$ dimensions there exists a critical value $\mu_c(d) = (4+d)/3$ such that for $0 < \mu < \mu_c(d)$ the critical exponents are again given by the linear equation (mean-field). For $\mu_c(d) < \mu < 2$ the scaling behavior should be controlled by the nonlinearity; we reconsider this result in Section \ref{results} below. In general, for $\mu \geq 2$ we expect also KPZ-type behavior, but that remains beyond power counting analysis, as it requires renormalization. Note the strong similarities between these results and those expected for equilibrium systems with so-called weak long range interactions \cite{Mukamel:2008}, 
in particular about the possibility that non-localities can modify the values of the critical exponents \cite{Fisher:1972},
and the occurrence of mean-field type behavior as a function of $\mu$.

\section{Sketch of the renormalization procedure}\label{RGsketch}

Previous studies on the morphologically stable version of \eref{eq_gral} (namely, $\nu>0$) for $\mu=1$
show that the equation indeed produces scale invariant interfaces whose critical exponents follow from simple dimensional analysis like the one performed in the previous section \cite{Krug:1991}. Actually, under conditions of morphological stability, one can even obtain analytically the exponents for arbitrary $\mu$ values in the deterministic limit \cite{Biler:1999}, that are expected to apply also for the stochastic case \cite{Mann:2001}. However,
the behavior for the morphologically unstable cases has been studied only recently \cite{Nicoli:2009b}. In that work we integrated equation \eref{eq_gral} numerically using a pseudospectral scheme \cite{Giada:2002,Gallego:2007} for $d=1,2$, and different values of $\mu$. The exponents measured numerically were confirmed through a DRG analysis using the Forster-Nelson-Stephen (FNS)~\cite{Forster:1977} scheme. In this section we sketch the general ideas of the latter approach, and show that the KPZ nonlinearity does not renormalize for any linear dispersion relation that is  a linear combination of arbitrary powers of the wavenumber $k$, a result that actually is independent on the morphological stability/instability of the system i.e., on the sign of $\nu$, and generalizes known results for the KPZ and Lai-Das Sarma-Villain equations.

For the sake of generality, it is convenient to consider the evolution equation for a generic linear dispersion relation $\sigma_{\bi{k}}$. Thus,
\begin{equation}
\partial_t h_\bi{k}(t) = \sigma_{\bi{k}}
    h_\bi{k}(t) + \frac{\lambda}{2} \mathcal{F}[(\nabla h)^2]_\bi{k} +
    \eta_\bi{k}(t),
\label{gen_eq}
\end{equation}
where the noise term is as in equation \eref{eq_gral}.
The DRG procedure starts by writing
equation \eref{gen_eq} after time Fourier transform,
\begin{eqnarray}
\left[-\sigma_\bi{k}-\rmi \omega\right]\, h_{\bi{k},\omega}=
\eta_{\bi{k},\omega}-\frac{\lambda}{2}  \int_{|\bi{q}|\leq\Lambda}
\frac{\rmd \bi{q}}{(2\pi)^d}\int_{-\infty}^{+\infty}\, \frac{\rmd \Omega}{2\pi}\,
\bi{q}\cdot(\bi{k}-\bi{q})\,
 h_{\bi{q},\Omega}\, h_{\bi{k}-\bi{q},\omega-\Omega},
\label{iee:start}
\end{eqnarray}
where $\omega$ is the time frequency, $\Lambda=\pi/\Delta x$ is the wave-number cut-off in the system,
$\Delta x$ being the lattice spacing in real space.
In Fourier space, the noise term in \eref{iee:start}
still has zero mean $\langle\eta_{\bi{k},\omega}\rangle=0$ and
is delta correlated, but its variance is rescaled by a constant as
\begin{equation}
\langle\eta_{\bi{k},\omega}\, \eta_{\bi{k}',\omega'} \rangle=
2\Pi_0 (2\pi)^{d+1}\, \delta_{\bi{k}+\bi{k}'}\, \delta_{\omega+\omega'}.
\end{equation}
Following the standard FNS procedure
\cite{McComb:1991,Halpin-Healy:1995,Forster:1977},
the height and the noise fields are split into two types of components,
fast modes $h^<_{\bi{k},\omega}$, $\eta^<_{\bi{k},\omega}$ for wave-numbers
$0 < k < \Lambda/b$, and slow modes $h^>_{\bi{k},\omega}$, $\eta^>_{\bi{k},\omega}$
for $\Lambda/b \le k \le \Lambda$, where $b=\rme^{\delta l}$ is a rescaling parameter
(assumed to be larger that one, i.e.\ $\delta l>0$).
Then one eliminates a ``small number'' (namely, $\delta l$ is infinitesimal) of the fast
modes by solving the growth equation perturbatively for $h^>_{\bi{k},\omega}$, substituting
the solution into the equation for the slow modes and assuming statistical independence between
high and low-frequency components. This procedure leads to an effective equation in which the
fast modes are thus integrated out,
\begin{equation}
\left[-\sigma_\bi{k}-\Sigma(\bi{k},0)-\rm{i}\omega\right]  \, h^<_{\bi{k},\omega}
=\eta^<_{\bi{k},\omega}-
\frac{\lambda}{2}
\int_<
\frac{\rm{d}\bi{q}}{(2\pi)^d}
\int
\frac{\rm{d}\Omega}{2\pi}\,
\bi{q}\cdot(\bi{k}-\bi{q}) \,
h^<_{\bi{q},\Omega}\, h^<_{\bi{k}-\bi{q},\omega-\Omega} + {\rm O}(\lambda^3),
\label{iee:renorm_eq}
\end{equation}
where the effect of this coarse-graining step (the elimination of fast modes) is obtained by solving perturbatively the
following integral [for the result in a particular representative case of equation \eref{eq_gral}, see appendix \ref{ax:nonlocal}]
\begin{equation}
\Sigma(\bi{k},\omega)=
2\lambda^2 \Pi_0 \hskip -0.1cm
\int_>  \frac{\rm{d}\bi{q}}{(2\pi)^d}
\int\frac{\rm{d}\Omega}{2\pi}\, [\bi{q}\cdot (\bi{k}-\bi{q})] \, [-\bi{q}\cdot\bi{k}] \,
G_0(\bi{q},\Omega) G_0(-\bi{q},-\Omega)
G_0(\bi{k}-\bi{q},\omega-\Omega).
\label{sigma}
\end{equation}
In Eqs.\ (\ref{iee:renorm_eq}), (\ref{sigma}) we have denoted integrals over the fast (slow) modes with $\int_>$ ($\int_<$),
and we have omitted the integration limits in the frequency domain [note that $\Omega\in(-\infty,\infty)$].
In equation \eref{sigma}, $G_0(\bi{k},\omega)=\left[-\sigma_\bi{k}-\rm{i}\omega\right]^{-1}$
is the bare propagator, and the integral in $\bi{q}$ is calculated only for the (fast) modes within the
{\em shell} \mbox{$q\in [\Lambda/b,\Lambda]$}. In equation \eref{iee:renorm_eq} the coarse-grained propagator
appears,
\begin{equation}
\label{G-renorm}
G^<_0(\bi{k},\omega) \equiv \left[-\sigma_\bi{k}-\Sigma(\bi{k},0)-\rmi\omega\right]^{-1} .
\end{equation}
From this expression, the effect of coarse-graining on the propagator is seen to merely amount to a modification
of the system parameters. For example, $\Sigma(\bi{k},0)=\Sigma_\nu k^\mu+\Sigma_\mathcal{K} k^2$
for equation \eref{eq_gral} for $m=2$, so that the renormalized propagator has the same functional shape as the bare propagator, but with modified coefficients $\nu^<=\nu+\Sigma_\nu$ and $\mathcal{K}^<=\mathcal{K}+\Sigma_\mathcal{K}$.
Actually, in equation \eref{iee:renorm_eq} we have omitted higher order terms that can in principle be associated with
a change in the value of $\lambda \to \lambda^<$ as induced by the present coarse-graining step, see section \ref{sec_vertex}.

On the other hand, the coarse-graining of the noise variance is performed
by starting from equation
\begin{equation}
\left\langle h^<_{\bi{k},\omega} h^<_{-\bi{k},-\omega}\right\rangle =
	2\Pi^<_0 \, G(\bi{k},\omega)G(-\bi{k},-\omega),
\end{equation}
in which the exact propagator appears. This equation leads, after a little algebra, to
\begin{equation}
\langle\eta^<_{\bi{k},\omega}\, \eta^<_{\bi{k}',\omega'} \rangle=
2\left[\Pi_0+\Phi(\bi{k},0)\right]
(2\pi)^{d+1}\, \delta_{\bi{k}+\bi{k}'}\, \delta_{\omega+\omega'} + {\rm O}(\lambda^3),
\end{equation}
where the coarse-grained noise variance is given by
\begin{equation}
\Phi(\bi{k},\omega)=\lambda^2 \Pi_0^2\int_>  \frac{\rm{d}\bi{q}}{(2\pi)^d}
\int\frac{\rm{d}\Omega}{2\pi}\, [\bi{q}\cdot(\bi{k}-\bi{q})]^2 |G_0(\bi{q},\Omega)|^2
|G_0(\bi{k}-\bi{q},\omega-\Omega)|^2.
\label{noise}
\end{equation}
Hence, the noise variance in the equation for the slow modes is $\Pi_0^<=\Pi_0+\Phi$.
For the result in a particular representative case of equation \eref{eq_gral}, see appendix \ref{ax:nonlocal}.

\subsection{Vertex cancellation for generalized equations}
\label{sec_vertex}
In this section we show that, to within our one-loop approximation, the coarse-graining procedure
does {\em not} lead to a modification in the coefficient of the nonlinearity for equation \eref{gen_eq}
with quite general $\sigma_\bi{k}$, namely, $\lambda^< = \lambda$. Specifically, we consider any
linear dispersion relation that can be written as
\begin{equation}
\sigma_{\bi{k}}=-\sum_{i=1}^n c_i |\bi{k}|^{r_i},
\label{gen_disp_rel}
\end{equation}
where the set $\{r_i\}$ is formed by a collection of positive real exponents
that increase with the index, $r_1<r_2<\dots<r_n$, and coefficients $c_i$ are generic,
but with the restriction that $c_n$ is positive in order to obtain a well behaved
dispersion relation at high wave numbers, i.e.\ $\sigma_{\bi{k}} \to
-\infty$ for $k\to \infty$. Given this choice, we consider an even more general equation
of which \eref{gen_eq} is a particular case, namely,
\begin{equation}
\partial_t h_\bi{k}(t) = \sigma_{\bi{k}}
    h_\bi{k}(t) - \frac{\lambda}{2} k^{\sigma} \mathcal{F}[(\nabla h)^2]_\bi{k} +
    \eta_\bi{k}(t),
\label{super_gen_eq}
\end{equation}
where $ \sigma \geq 0$, see \cite{Lauritsen:1995} for the special case in which
$\sigma_{\bi{k}}=-\nu k^{\mu +\sigma}$.
Note that equation \eref{super_gen_eq} reduces to \eref{gen_eq} for $\sigma=0$.
In the same way as the latter includes the KPZ equation as a particular case, equation \eref{super_gen_eq}
generalizes also the celebrated Lai-Das Sarma-Villain equation \cite{Lai:1991,Villain:1991},
that is obtained for $\mu=\sigma=2$. In short, the latter is a conserved version of the KPZ equation with
non-conserved noise, and has played an important role in the study of kinetic roughening for thin films grown
by MBE \cite{Krug:1997}.

As is well known and is reviewed in detail elsewhere \cite{McComb:1991,Halpin-Healy:1995}, the
iterative solution of equations like (\ref{iee:start}) or \eref{super_gen_eq} along the lines sketched above can be conveniently cast into a diagrammatic formulation (for notation and diagrams, see e.g.\ \cite{Medina:1989}). Within such scheme, it is seen that the coarse-graining process is complete at one loop order once the coarse-grained coupling $\lambda^<$ of the non-linear term is evaluated from
\begin{align}
-\frac{\lambda}{2}^{<} k_1^{\sigma} & \int_>\frac{\rmd\bi{k}_2}{(2\pi)^d} \int
	\frac{\rmd\Omega_2}{2\pi} \, \left(\frac{\bi{k}_1}{2}+\bi{k}_2\right)
	\cdot \left(\frac{\bi{k}_1}{2}-\bi{k}_2\right)
  	h^<_{\bi{k}_1 / 2 + \bi{k}_2,\omega_1 /2 + \Omega_2} \,
	h^<_{\bi{k}_1 / 2 - \bi{k}_2,\omega_1 /2 - \Omega_2}  \nonumber \\[7pt]
 &\hskip -1 cm  \equiv
-\frac{\lambda}{2} k_1^{\sigma} \int_>\frac{\rmd\bi{k}_2}{(2\pi)^d} \int
	\frac{\rmd\Omega_2}{2\pi} \,
	\left(1+\sum_{j=1}^3 \Gamma_j \right)
	\left(\frac{\bi{k}_1}{2}+\bi{k}_2\right)
	\cdot \left(\frac{\bi{k}_1}{2}-\bi{k}_2\right)
	h^<_{\bi{k}_1 / 2 + \bi{k}_2,\omega_1 /2 + \Omega_2} \,
	h^<_{\bi{k}_1 / 2 - \bi{k}_2,\omega_1 /2 - \Omega_2}\hskip -0.1 cm
	+ {\rm O}(\lambda^5) ,
\label{lambda_corr}
\end{align}
where the original variables $\bi{k}$ and $\bi{q}$ in (\ref{iee:start}) are replaced by the symmetric variables
$\bi{k}_1 = \bi{k}$ and $\bi{k}_2 = \bi{q} - \bi{k} / 2$, and $\Gamma_j$ are appropriate integrals that are
evaluated next.
The first one arises from the contraction of fast noise components with wave vectors
$\pm \left(\frac{\bi{k}_1}{2}+\bi{k}_2-\bi{q}\right)$, and reads
\begin{align}
\Gamma_1(\bi{k}_1,\bi{k}_2,\omega_1,\omega_2) & =
\frac{2 \lambda^2 \Pi_0}{\left(\frac{k_1^2}{4}-k_2^2\right)} {\displaystyle \int_>\frac{\rmd\bi{q}}{(2\pi)^d}
 \int \frac{\rmd\Omega}{2\pi}} \,
\left(q \left|\bi{k}_1 - \bi{q} \right|\right)^\sigma
\,
\bi{q}\cdot (\bi{k}_1-\bi{q}) \left(\frac{\bi{k}_1}{2}+\bi{k}_2\right)
\cdot \left(\bi{q}-\frac{\bi{k}_1}{2}-\bi{k}_2\right) \nonumber\\[7pt]
& \hskip 0.5 cm \times
\left(\frac{\bi{k}_1}{2}-\bi{k}_2\right)
   \cdot \left(\frac{\bi{k}_1}{2}+\bi{k}_2-\bi{q}\right)\,
 G_0\left(\hat{q}\right) G_0\left(\hat{k}_1-\hat{q}\right)
G_0\left(\hat{q}-\frac{\hat{k}_1}{2}-\hat{k}_2\right)
G_0\left(\frac{\hat{k}_1}{2}+\hat{k}_2-\hat{q}\right) ,
\label{Gammaa1}
\end{align}
where we have used a shorthand notation, e.g.,
$G\left(\hat{k}_1 - \hat{q}\right)\equiv G\left(\bi{k}_1-\bi{q},\Omega_1 - \Omega \right)$.
Using the change of variables $\bi{j}=\bi{q}-\bi{k}_1/2$, we get
\begin{align}
\left(q \left|\bi{k}_1 - \bi{q} \right|\right)^\sigma
	\bi{q}\cdot (\bi{k}_1-\bi{q}) \left(\frac{\bi{k}_1}{2}+\bi{k}_2\right)
\cdot \left(\bi{q}-\frac{\bi{k}_1}{2}-\bi{k}_2\right) & \left(\frac{\bi{k}_1}{2}-\bi{k}_2\right)
   \cdot \left(\frac{\bi{k}_1}{2}+\bi{k}_2-\bi{q}\right) \nonumber \\[5pt]
& = j^{2\sigma+ 4}\left( \frac{1}{4}\cos^2 \left(\theta_1\right) \,
		k_1^2 - k_2^2 \cos^2\left(\theta_2\right) \right)+\Or(k^2_i k_j) ,
\end{align}
where $\bi{k}_i \cdot \bi{j} \equiv k j \cos \theta_i$. The contribution due to the bare propagator is a little involved,
but we need to retain only the zero-th order contribution in $k_i$ of the perturbative expansion,
which strongly simplifies the calculation. Thus,
\begin{align}
 & \lim_{\omega_1,\omega_2 \to 0}  G_0\left(\hat{j}+\frac{\hat{k}_1}{2}\right)
	G_0\left(\frac{\hat{k}_1}{2}-\hat{j}\right)
	G_0\left(\hat{j}-\hat{k}_2\right)
	G_0\left(\hat{k}_2-\hat{j}\right) \nonumber\\[5pt]
& \hskip 1cm =\left[\left(\sum_{i=1}^n c_i j^{r_i}\right)^2+\Or(k_1)+\Omega^2\right]^{-1}\,
	\left[\left(\sum_{i=1}^n c_i j^{r_i}\right)^2+\Or(k_2)+\Omega^2\right]^{-1}
=  \left(c_n j^{r_n}\right)^{-4} \,
	\left(\hat{\Delta}_j^2+\hat{z}^2\right)^{-2},
\end{align}
where
\begin{equation}
\hat{z}=\frac{\Omega}{c_n j^{r_n}},
\qquad
\hat{\Delta}_j=1+\sum_{i=1}^{n-1} \frac{c_i}{c_n} j^{r_i-r_n}=
-\frac{\sigma_j}{c_n j^{r_n}}.
\end{equation}
The function $\hat{\Delta}_j$ evaluated at $j=\Lambda$ takes a positive
value $\hat{\Delta}_\Lambda$
due to the signs of  $c_n$ and $\sigma_\Lambda$  ($c_n>0$
and $\sigma_\Lambda<0$, see above and section \ref{DRGflow}). Hence,
after integration in $\hat{z}$,
\begin{equation}
\int_{-\infty}^{+\infty} \rmd\hat{z} \left(\hat{\Delta}_j^2+\hat{z}^2 \right)^{-2} = \frac{\pi}{2\hat{\Delta}_j^3},
\end{equation}
this contribution reduces to
\begin{align}
\Gamma_1(\bi{k}_1,\bi{k}_2,0,0)&=
\frac{2\lambda^2 \Pi_0}{\left(\frac{k_1^2}{4}-k_2^2\right)} \int_> \frac{\rmd\bi{j}}{(2\pi)^d}
	\frac{j^{2\sigma+ 4}}{4 \hat{\Delta}_j^3 (c_n j^{r_n})^3} \left( \frac{1}{4}\cos^2 \left(\theta_1\right)
	\, k_1^2 - k_2^2 \cos^2\left(\theta_2\right) \right) \nonumber \\[7pt]
&=\frac{\lambda^2 \Pi_0 S_{d-1}}{2\left(\frac{k_1^2}{4}-k_2^2\right)} \int_> \frac{\rmd j}{(2\pi)^d}
	\frac{j^{d+2\sigma+ 3}}{\hat{\Delta}_j^3 (c_n j^{r_n})^3}
	\int_0^\pi \rmd\theta \, \sin^{d-2}\left( \theta \right)
	\left( \frac{1}{4}\cos^2 \left(\theta\right) \, k_1^2 - k_2^2 \cos^2\left(\theta\right) \right)
	 \nonumber \\[7pt]
 &=\frac{\lambda^2 \Pi_0}{2d} K_d \int_> \rmd j \, \frac{j^{d+2\sigma+3}}{(c_n j^{r_n})^3} \, \hat{\Delta}_j^{-3} ,
\label{Gammaa2}
\end{align}
where $K_d=2/[(4\pi)^{d/2} \Gamma(d/2)]$ is the surface area of the $d$-dimensional unit sphere
$S_d$ divided by $(2\pi)^d$,
as defined in \eref{k_d}. We stop the calculation at this point because the remaining contributions to the
vertex renormalization in (\ref{lambda_corr}) will be seen to cancel \eref{Gammaa2} exactly.
Indeed, the function $\Gamma_2$ is given by the sum of the terms
obtained from the noise contraction of $\bi{k}_1-\bi{q}$ with $\bi{q}-\bi{k}_1$, i.e.,
\begin{align}
\Gamma_2(\bi{k}_1,\bi{k}_2,\omega_1,\omega_2) & =
\frac{2 \lambda^2 \Pi_0}{\left(\frac{k_1^2}{4}-k_2^2\right)} {\displaystyle \int_>\frac{\rmd\bi{q}}{(2\pi)^d}
 \int  \frac{\rmd\Omega}{2\pi}} \,
\left(q \left|\bi{q} - \dfrac{\bi{k}_1}{2} - \bi{k}_2 \right|\right)^\sigma
\,
\bi{q}\cdot (\bi{k}_1-\bi{q}) \left(\frac{\bi{k}_1}{2}+\bi{k}_2\right)
\cdot \left(\bi{q}-\frac{\bi{k}_1}{2}-\bi{k}_2\right) \nonumber\\[7pt]
& \hskip 0.5 cm \times
    \left(\frac{\bi{k}_1}{2}-\bi{k}_2\right)
   \cdot \left(\bi{q}-\bi{k}_1\right) \,
 G_0\left(\hat{q}\right) G_0\left(\hat{k}_1-\hat{q}\right)
G_0\left(\hat{q}-\hat{k}_1\right)
G_0\left(\hat{q}-\frac{\hat{k}_1}{2}-\hat{k}_2\right) ,
\label{Gammaba1}
\end{align}
where
\begin{align}
\left(q \left|\bi{q} - \dfrac{\bi{k}_1}{2} - \bi{k}_2 \right|\right)^\sigma\,
	\bi{q}\cdot (\bi{k}_1-\bi{q})  \left(\frac{\bi{k}_1}{2}+\bi{k}_2\right)
\cdot \left(\bi{q}-\frac{\bi{k}_1}{2}-\bi{k}_2\right) & \left(\frac{\bi{k}_1}{2}-\bi{k}_2\right)
   \cdot \left(\bi{q}-\bi{k}_1\right) \nonumber \\[5pt]
 & \hskip -1 cm =-j^{2\sigma + 4}\left( \frac{1}{4}\cos^2 \left(\theta_1\right) \, k_1^2 -
 	k_2^2 \cos^2\left(\theta_2\right) \right)+\Or(k^2_i k_j) .
\end{align}
and
\begin{equation}
\lim_{\omega_1,\omega_2 \to 0}  G_0\left(\hat{j}+\frac{\hat{k}_1}{2}\right)
	G_0\left(\frac{\hat{k}_1}{2}-\hat{j}\right)
	G_0\left(\hat{j}-\frac{\hat{k}_1}{2}\right)
	G_0\left(\hat{j}-\hat{k}_2\right)
=  \left(c_n j^{r_n}\right)^{-4} \,
\left[(\hat{\Delta}_j^2+\hat{z}^2)(\hat{\Delta}_j - \rmi\hat{z})^2\right]^{-1}.
\end{equation}
Following similar steps to the case of $\Gamma_1$, after integration in
$\hat{z}$,
\begin{equation}
\int_{-\infty}^{+\infty} \rmd\hat{z} \left[(\hat{\Delta}_j^2+\hat{z}^2)(\hat{\Delta}_j \pm \rmi\hat{z})^2\right]^{-1}
	= \frac{\pi}{4\hat{\Delta}_j^3},
\end{equation}
we get
\begin{equation}
\Gamma_2(\bi{k}_1,\bi{k}_2,0,0)=
-\frac{\lambda^2 \Pi_0}{4d} K_d \int_> \rmd j\,  \frac{j^{d+2\sigma+3}}{(c_n j^{r_n})^3} \, \hat{\Delta}_j^{-3}.
\end{equation}
Finally, the contribution $\Gamma_3$ in (\ref{lambda_corr}) arises from a contraction of noise components with wave-vectors $\bi{q}$ and $-\bi{q}$. This contribution reads
\begin{align}
\Gamma_3(\bi{k}_1,\bi{k}_2,\omega_1,\omega_2) & =
\frac{2 \lambda^2 \Pi_0}{\left(\frac{k_1^2}{4}-k_2^2\right)} {\displaystyle \int_>\frac{\rmd\bi{q}}{(2\pi)^d}
 \int \frac{\rmd\Omega}{2\pi}} \,
 \left(\left|\bi{k}_1 - \bi{q} \right|  \left|\dfrac{\bi{k}_1}{2} + \bi{k}_2 - \bi{q} \right|\right)^\sigma
 \,
\bi{q}\cdot (\bi{k}_1-\bi{q}) \left(-\bi{q}\right)
\cdot \left(\frac{\bi{k}_1}{2}+\bi{k}_2\right)
     \nonumber\\[7pt]
& \hskip 0.5 cm \times
   \left(\frac{\bi{k}_1}{2}-\bi{k}_2\right) \cdot \left(\frac{\bi{k}_1}{2}+\bi{k}_2-\bi{q}\right)
\,  G_0\left(\hat{q}\right) G_0\left(-\hat{q}\right)
G_0\left(\hat{k}_1-\hat{q}\right)
G_0\left(\frac{\hat{k}_1}{2}+\hat{k}_2-\hat{q}\right) ,
\label{Gammac1}
\end{align}
where
\begin{align}
\left(\left|\bi{k}_1 - \bi{q} \right|  \left|\dfrac{\bi{k}_1}{2} + \bi{k}_2 - \bi{q} \right|\right)^\sigma
\,
\bi{q}\cdot (\bi{k}_1-\bi{q}) \left(-\bi{q}\right)
\cdot \left(\frac{\bi{k}_1}{2}+\bi{k}_2\right) &
    \left(\frac{\bi{k}_1}{2}-\bi{k}_2\right)
   \cdot \left(\frac{\bi{k}_1}{2}+\bi{k}_2-\bi{q}\right) \nonumber \\[5pt]
& = - j^{2\sigma + 4}\left( \frac{1}{4}\cos^2 \left(\theta_1\right) \, k_1^2 -
	k_2^2 \cos^2\left(\theta_2\right) \right)+\Or(k^2_i k_j) ,
\end{align}
and
\begin{equation}
\lim_{\omega_1,\omega_2 \to 0}  G_0\left(\hat{j}+\frac{\hat{k}_1}{2}\right)
	G_0\left(-\hat{j}-\frac{\hat{k}_1}{2}\right)
	G_0\left(\hat{j}-\frac{\hat{k}_1}{2}\right)
	G_0\left(\hat{j}-\hat{k}_2\right)
=  \left(c_n j^{r_n}\right)^{-4} \,
\left[(\hat{\Delta}_j^2+\hat{z}^2)(\hat{\Delta}_j + \rmi\hat{z})^2\right]^{-1}.
\end{equation}
Thus, in parallel with the previous cases, in the $\omega_{1,2} \to 0$ limit we obtain $\Gamma_3 = \Gamma_2 = - \Gamma_1/2$, hence from equation \eref{lambda_corr} finally $\lambda^< = \lambda$,
so that the non-linear term does not renormalize at one-loop order. Notice, this result
does not make any requirement on the morphological stability of the system described by equation
\eref{super_gen_eq}, and has the analogous vertex cancellations for both the KPZ and the
Lai-Das Sarma-Villain (LDV) equations as immediate corollaries. As is well known, for the former this
property leads to the scaling relation $\alpha+z = 2$ at the nonlinear fixed point, and
is usually associated with the explicit invariance of the equation under
a Galilean transformation \cite{Barabasi:1995,Halpin-Healy:1995,Krug:1997}. However, for the
LDV equation an analogous symmetry is known {\em not to} exist \cite{Janssen:1997}.
This fact makes it appear as a coincidence that the KPZ equation does combine such type of
(Galilean) symmetry {\em with} vertex cancellation. Admittedly, vertex cancellation also
occurs at two loop order in the KPZ equation \cite{Frey:1994,Sun:1994}, but not in the LDV case
\cite{Janssen:1997}, so that the interplay between Galilean symmetry and the so-called
Galilean scaling relation is subtle. Still, more recent works also question a causal relation between
the two. Thus, it has been argued that for the Navier-Stokes (NS) and related equations, such as
Burgers', Galilean invariance does not enforce vertex non-renormalization \cite{McComb:2005}.
These arguments have been further substantiated more recently \cite{Berera:2007,Berera:2009}
within a field theoretical approach to the stochastic NS system. Likewise, recently
models or discretization schemes of the KPZ equation have been shown to explicitly break Galilean
invariance while fulfilling the $\alpha+z=2$ scaling relation, see \cite{Wio:2010-a,Wio:2011} and references therein.

\section{Dynamic renormalization group flow and fixed points}
\label{DRGflow}

Coming back to equation \eref{eq_gral}, in order to account for its scaling properties, we
restrict our DRG analysis to the case $m=2$ and $\mathcal{N}=0$, that is,
\begin{equation}
\partial_t h_\bi{k}(t) = (-\nu k^{\mu} - \mathcal{K} k^2)
    h_\bi{k}(t) + \frac{\lambda}{2} \mathcal{F}[(\nabla h)^2]_\bi{k} +
    \eta_\bi{k}(t) .
\label{eq_drg}
\end{equation}
For clarity of the exposition, the details of the corresponding DRG calculations are left to appendix \ref{ax:nonlocal}.
Actually, the scaling properties of the equation do not change when we take into account a dispersion relation
with additional stabilizing relaxation terms, less relevant than $k^2$, namely, $n>2$ in equation \eref{eq_gral}.
A proof of this is given in appendix \ref{ax:complex} for the specific $\mu=1$ case, considering $\mathcal{N}\neq 0$, with $n=3$, and $4$.

After coarse graining of propagator and noise variance (as in appendix \ref{ax:nonlocal}), and the vertex (as in the previous section), the final step in the DRG method is the application of a rescaling that restores the initial value of the wave-vector cut-off. Moreover, this allows us to write parameter renormalization in a differential form, taking $l$ as the independent variable. We can actually employ the formulae \eref{scaling}, where the role of unscaled parameters is now taken by our coarse-grained estimates $\nu^<$, etc.\ and scaled quantities $\tilde{\nu}$, etc.\ become finally the renormalized parameters. This makes explicit the non-trivial effect
of the coarsening step as, e.g., $\nu^<$ differs from the bare parameter $\nu$ in the original equation.
After taking the $\delta l\to 0$ limit, the ensuing infinitesimal parameter variation constitutes the one loop RG flow for arbitrary substrate dimension $d$ (see appendix \ref{ax:nonlocal}),
\begin{align}
& \frac{\rmd \nu}{\rmd l} = \nu \left[z-\mu\right] , \qquad
	\frac{\rmd \lambda}{\rmd l} = \lambda \left[\alpha+z-2\right] , \label{ecs:nu,lambda} \\[7pt]
&\frac{\rmd \mathcal{K}}{\rmd l} = \mathcal{K} \left[z-2 -
    \frac{\lambda^2 \Pi_0 K_d}{4d} \, \frac{
    (d-2)\mathcal{K}+(d-\mu)\nu}{\mathcal{K}\left(\mathcal{K}+\nu\right)^3}\right] , \label{ec:K} \\[7pt]
& \frac{\rmd \Pi_0}{\rmd l} = \Pi_0 \left[z-2\alpha-d +
    \frac{\lambda^2 \Pi_0 K_d}{4\left(\mathcal{K}+\nu\right)^3}\right], \label{ec:D}
\end{align}
Without loss of generality, in equations \eref{ecs:nu,lambda}-\eref{ec:D}
we have fixed the lattice cut-off, $\Lambda=1$, bearing in mind that
in order to avoid the finite pole of the bare propagator,
the coarse graining step has been performed in an infinitesimal shell
$k \in [\Lambda (1-\delta l), \Lambda]$ within the band of linearly
stable (large) wave-vector values \cite{Cuerno:1995b}. The dispersion relation for $k=\Lambda$ (the largest wavenumber)
can be written as
\begin{equation}
\sigma_\Lambda = -\mathcal{K}\Lambda^2\left(1+\frac{\nu}{\mathcal{K}}
\Lambda^{\mu-2}\right) = -\mathcal{K}\Lambda^2\Delta_\Lambda,
\label{Delta}
\end{equation}
where $\Delta_k = 1+\nu\mathcal{K}^{-1} |k|^{\mu-2}$
is employed in the evaluation of the integrals in equations \eref{sigma}
and \eref{noise}. The sign of $\sigma_\Lambda$ gives us a criterion to identify a
physically consistent DRG flow (see appendix \ref{ax:nonlocal} for details). Note that $\sigma_\Lambda$ has to
be negative in order to have a well behaved dispersion relation and that, for the same reason,
$\mathcal{K}$ has to be positive. Thus, equation \eref{Delta} implies that $\Delta_\Lambda$ has
to be a positive quantity. All the fixed points obtained within the DRG approach have to
satisfy this requirement.

Equations \eref{ecs:nu,lambda}-\eref{ec:D} generalize the KPZ flow in a natural way for non-zero
$\nu$, and inherit the known analytical limitations of the latter \cite{Krug:1997,Barabasi:1995,Halpin-Healy:1995}.
Still, they carry valuable information. Introducing the following coupling parameters~\cite{Haselwandter:2007-prl}
\begin{equation}
\label{iee:couplings}
g=\frac{\lambda^2\Pi_0 K_d}{4(\mathcal{K}+\nu)^3}, \qquad
f=\frac{\mathcal{K}}{\mathcal{K}+\nu},
\end{equation}
we can calculate their flow using equations \eref{ecs:nu,lambda}-\eref{ec:D}. Thus,
\begin{align}
& \frac{\rmd f}{\rmd l} = (1-f)
	\left\{ (\mu-2) f- \frac{g}{d}\left[(d-2) f+(d-\mu)(1-f)\right] \right\}, \label{iee:f1} \\[7pt]
& \frac{\rmd g}{\rmd l} = g\left\{ 6f-4-d+g+3\mu(1-f)+ 3\, \frac{g}{d}\big[ (d-2)f+(d-\mu)(1-f) \big] \right\}.
	\label{iee:g1}
\end{align}
The ratio between the (possibly) morphologically
destabilizing parameter, $\nu$, and the stabilizing one, $\mathcal{K}$, follows
from the exact relation $1-f=\nu/(\nu+\mathcal{K})$. The condition
$\Delta_\Lambda\ge 0$ is satisfied as long as these two couplings are positive or  zero;
for this reason we can restrict the DRG flow analysis to the first quadrant of the $(f,g)$ plane.
Then, a bare parameter choice corresponding to the morphologically unstable condition $\nu(l=0) < 0$
implies $f(l=0) > 1$, while the converse condition $f(l=0) < 1$ corresponds to morphological
stability in terms of the bare parameters. The line $f=1$ (at which $\nu=0$) is actually never crossed by any
trajectory, since flow lines cannot cross each other in this autonomous flow and $f=1$ is clearly invariant
under the iteration. Note, $f=1$ implies $\nu=0$, so that no explicit renormalization
of $\nu$ from negative to positive values is allowed by the flow (\ref{iee:f1})-(\ref{iee:g1}).
Thus, dynamical stabilization of the early-time morphological instability for equation (\ref{eq_drg}) and
thus for equation (\ref{eq_gral}) cannot take place through Yakhot's mechanism (namely, ``surface tension''
renormalization) for the Kuramoto-Sivashinsky equation \cite{Yakhot:1981}. This
agrees moreover with expectations based on a naive power counting analysis of both equations that
would lead {\em not} to expect RG corrections to a $k^{\mu}$ term from a $k^2$ term that is
irrelevant in comparison for $\mu < 2$.

The fixed points of the RG flow are given by the values of the pair $(f,g)$ for which equations
$\rmd f/\rmd l=0$ and $\rmd g/\rmd l=0$ hold simultaneously. Obviously, $f=1$ and $g=0$ satisfy the first and the
second equations, respectively. Moreover, the rhs of equation \eref{iee:f1} is identically zero for
\begin{equation}
f=\frac{g(d-\mu)}{(\mu-2)(d-g)},
\label{f-fp}
\end{equation}
while relation
\begin{equation}
g=\frac{3\mu(f-1)+4+d-6f}{3\left[(d-2)f+(d-\mu)(1-f)\right]/d+1},
\label{g-fp}
\end{equation}
leads to $\rmd g/\rmd l=0$.


A detailed analysis of equations \eref{iee:f1}-\eref{iee:g1}
yields the same fixed point structure as for the KPZ equation \cite{Krug:1997}
---namely, the Edwards-Wilkinson (EW) fixed points at $(f,g)=(1,0)$ and, for appropriate values
of $d$ (see next section), the KPZ fixed point at $(f,g)=(1,1/2)$---, with the
addition of two new nontrivial fixed points. One corresponds to the morphologically stable
interfaces (we call it {\em Smooth} fixed point), being located at the origin of the $(f,g)$ plane.
Its dynamic exponent is $z=\mu$, while hyper-scaling holds, namely,
$z=2\alpha+d$. Note that a zero value for $g$ means that the non-linear term $\lambda$ is vanishing and \eref{eq_drg} becomes a linear equation, the Smooth fixed point being the one employed in our discussion on the critical dimension
in Section \ref{scaling_anal}. Actually, at the linear EW and Smooth fixed points, the equation becomes variational and hyperscaling ensues, relating the dynamic and roughness exponents. Hyperscaling is associated with non-renormalization of the noise amplitude \cite{Halpin-Healy:1995} given that, under these conditions, the equation has (in the morphologically stable condition) the asymptotic height distributions ${\cal P}_{\rm EW}[h] \propto \exp[(-{\cal K}/(2\Pi_0)) \int k^2 |h_{\mathbf{k}}|^2 \rmd\mathbf{k}]$ or ${\cal P}_{\rm Smooth}[h] \propto \exp[(-\nu/(2\Pi_0)) \int k^{\mu} |h_{\mathbf{k}}|^2 \rmd\mathbf{k}]$ at the EW and Smooth fixed points, respectively.

The second new fixed point that exists for the flow \eref{iee:f1}-\eref{iee:g1} implements, as the KPZ fixed point, the scaling relation $\alpha+z=2$ usually associated with Galilean invariance, and we thus we call it {\em Galilean} fixed point. It is the one found in simulations for the morphologically unstable condition with $\nu<0$
\cite{Nicoli:2009b}.
As in the Smooth fixed point, the dynamic exponent takes the value expected from power counting, $z=\mu$.
The coordinates of the Galilean fixed point, $(f_*,g_*)$, are given by equating \eref{f-fp} to
\eref{g-fp}, namely,
\begin{equation}
f_*=\frac{g_*(d-\mu)}{(\mu-2)(d-g_*)},\qquad
g_*=d+4-3\mu.
\label{fg:galileo}
\end{equation}
Besides, this fixed point has to satisfy the conditions
$f_* \geq 0$ and $g_* \geq 0$, hence it is not defined
for some values of $\mu$ in the $(0,2]$ interval.
Indeed, the condition $g_* \geq 0$ leads to an admissible value for
this coupling for $\mu \leq (d+4)/3$. For a fixed value of $\mu$,
this is equivalent to the $d > d_c$ condition using the value of the
critical dimension derived in Section \ref{scaling_anal} for the
morphologically stable case. The second condition,
i.e.\ $f_* >0$, has to be analyzed carefully. In fact, for $d<4/3$,
the Galilean fixed point occurs for $\mu$ within the intervals
$\big(0,d\big]\bigcup \big(4/3,(d+4)/3\big]$ (for $d=1$, see
the left panel of figure \ref{fig:fg-galileo}). In the case $4/3\leq d<2$,
the admissibility intervals are $\big(0,4/3\big)\bigcup \big[d,(d+4)/3\big]$. Finally, for
$d\geq 2$, we have only one condition for the existence of
the Galilean fixed point, namely, $\mu<4/3$, as illustrated on the right panel of
figure \ref{fig:fg-galileo}.

In order to study the stability of these fixed points, we
have to compute the derivatives of \eref{iee:f1}-\eref{iee:g1}
with respect to the couplings, and evaluate the eigenvalues of the linear stability matrices.
So far, we have outlined the main steps we have taken to obtain the fixed points of
the DRG flow. In the next section we provide details on their stability for various
substrate dimensions (in particular, $d=1$, $d=2$, and $d>2$).



\section{Fixed point stability and dependence with dimension}\label{results}

The study of the trajectories of the RG flow provides useful information on the
stability of the different fixed points. Although the stable case has been already examined in the context
of the fractal KPZ (FKPZ) equation \cite{Katzav:2003} [which is obtained from \eref{eq_gral} when $\nu >0$ and $\mathcal{K}=\mathcal{N}=0$], our DRG study supplies additional information. In \cite{Katzav:2003} the FKPZ equation was studied by means of a self-consistent expansion (SCE) for spatially correlated noise in the case of arbitrary substrate dimension. The limit of Gaussian white noise is recovered when the exponent of the noise correlations introduced in \cite{Katzav:2003} is equal to zero. Although the SCE allows to identify the strong coupling KPZ fixed point with a reasonable accuracy ($\alpha=0.295$ and $z=1.705$ for $d=2$, compatible with the most reliable estimates
for this value of $d$ \cite{Marinari:2000}), the SCE method has some drawbacks. Even though all the fixed
points reported in the previous section appear within the SCE analysis, this method cannot assess
which fixed point is selected by the dynamics during the growth process. For example, in the case of $d=2$,
three different phases are found \cite{Katzav:2003}, each one with different values for the critical exponents.
For some value of the exponent of the fractional Laplacian ($\mu$ in our notation, that is equivalent to $2-\rho$ in \cite{Katzav:2003}),
all these phases are admissible, there being no analytical criterion for the selection of one of them.
The present section is precisely devoted to the analysis of the DRG predictions on the fixed point that is selected
asymptotically by the dynamics for each substrate dimension, and as a function of bare parameter values.


\subsection{One-dimensional interfaces}

\begin{figure}[h!]
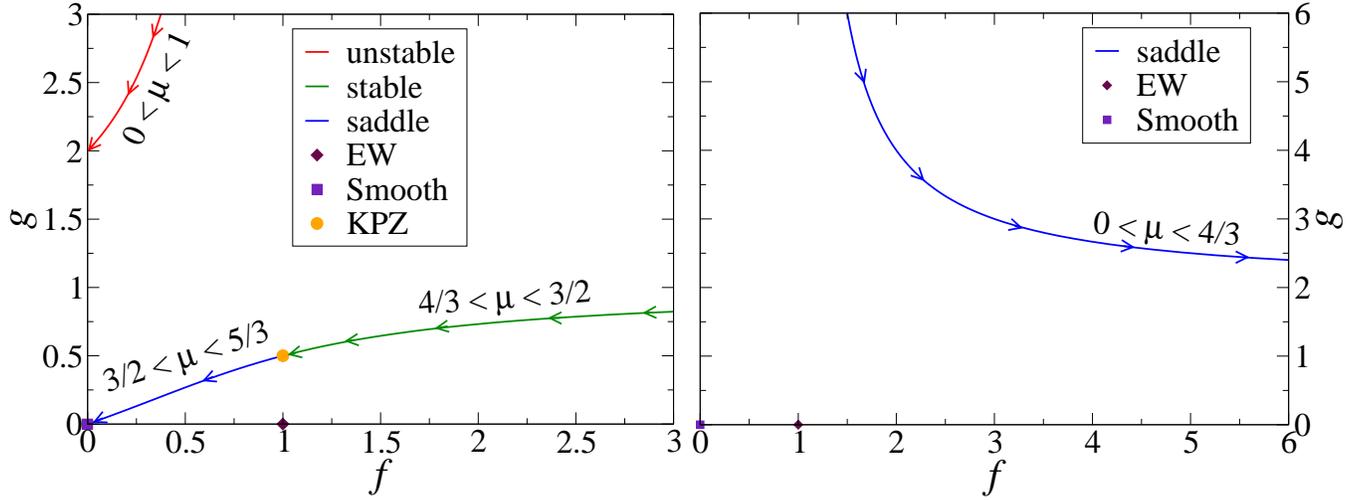

\begin{minipage}{.5\textwidth}
	\begin{center}
	\epsfig{file=./moving-galileo-d1.eps,width=\textwidth,clip=}
	\end{center}
\end{minipage}
\begin{minipage}{.485\textwidth}
	\begin{center}
	\epsfig{file=./moving-galileo-d2.eps,width=\textwidth,clip=}
	\end{center}
\end{minipage}
\caption{Position of the Galilean fixed point on the $(f,g)$ plane for $d=1$ (left panel)
and $d=2$ (right panel). Arrows and colors indicate, respectively, the displacement of this fixed
point for increasing $\mu$, and its stability. The intervals of $\mu$ specified near the solid lines
provide the values of $\mu$ at which stability changes for this fixed point.}
\label{fig:fg-galileo}
\end{figure}

\begin{figure}[!h]
\begin{center}
	\epsfig{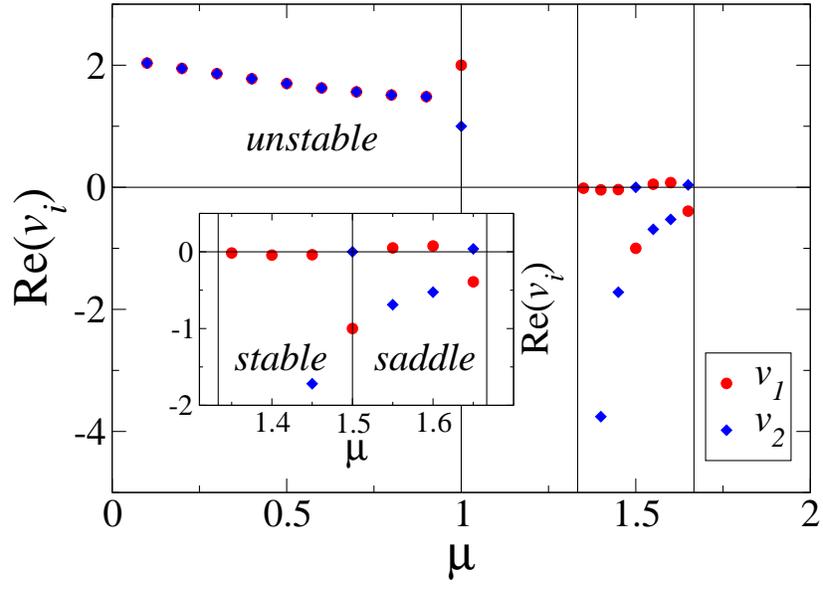}
\end{center}
\caption{\label{fig:auto-d1}Real parts of the eigenvalues of the linear stability
matrix of the Galilean fixed point for $d=1$, as functions of $\mu$. The inset is a zoom
of the region $4/3<\mu<5/3$ within which the fixed point changes stability from
stable to saddle at $\mu=3/2$.}
\end{figure}

For $d=1$ we obtain three fixed points whose coordinates do not
depend on the exponent $\mu$, i.e., the Smooth, EW, and KPZ fixed points
(see left panel of figure \ref{fig:fg-galileo}). On the contrary,
the Galilean fixed point changes location in the $(f,g)$ by changing the value of $\mu$.
In table \tref{fp-table:1} we summarize all the fixed points that exist for $d=1$, their critical exponents,
and the eigenvalues of their stability matrix. The (real parts of the) two eigenvalues of the Galilean fixed point,
$v_{1,2}^{(1)}$, have been obtained numerically; we show in figure \ref{fig:auto-d1} how they change as functions of $\mu$.

\begin{table}[h]
\begin{center}
\begin{tabular}{|c|c|c|c|c|c|c|c|c|}
\hline
\T Name &  $f$ & $g$ & $z$ & $\alpha$ & $\beta$& Scaling relation & $\lambda_1$ & $\lambda_2$ \\[3pt]
\hline
\hline
\T EW &  $1$ & $0$ & $2$ & $1/2$ &  $1/4$& $2\alpha+d=z$ & $2-\mu$ & $1$\\[5pt]
KPZ & $1$ & $1/2$ & $3/2$ & $1/2$ & $1/3$ & $\alpha+z=2$ & $3/2-\mu$ & $-1$\\[5pt]
S & $0$ & $0$ & $\mu$ & $\frac{\mu-1}{2}$ & $\frac{\mu-1}{2\mu}$&
	$2\alpha+d=z$ & $\mu-2$ & $3\mu-5$ \\[7pt]
G & $\frac{(5-3\mu)(1-\mu)}{(\mu-2)(3\mu-4)}$ &
 $5-3\mu$  & $\mu$ & $2-\mu$ & $\frac{2-\mu}{\mu}$& $\alpha+z=2$ & $v_1^{(1)}$ & $v_2^{(1)}$   \\[10pt]
\hline
\end{tabular}
\end{center}
\caption{Fixed points in $1+1$ growth dimensions ($d=1$).
In the first column, G stands for Galilean and S for Smooth, according
to the terminology introduced in the main text. In the last column, $\lambda_{1,2}$ are
the eigenvalues of the linear stability matrix at the corresponding fixed point. Their real parts
$v_{1,2}^{(1)}$, are plotted in figure \ref{fig:auto-d1} in the case of the Galilean
fixed point. The latter is not defined in the intervals $1< \mu \leq 4/3$ and $5/3< \mu\leq 2$. }
\label{fp-table:1}
\end{table}



\begin{figure}[!h]
\begin{center}
	\epsfig{file=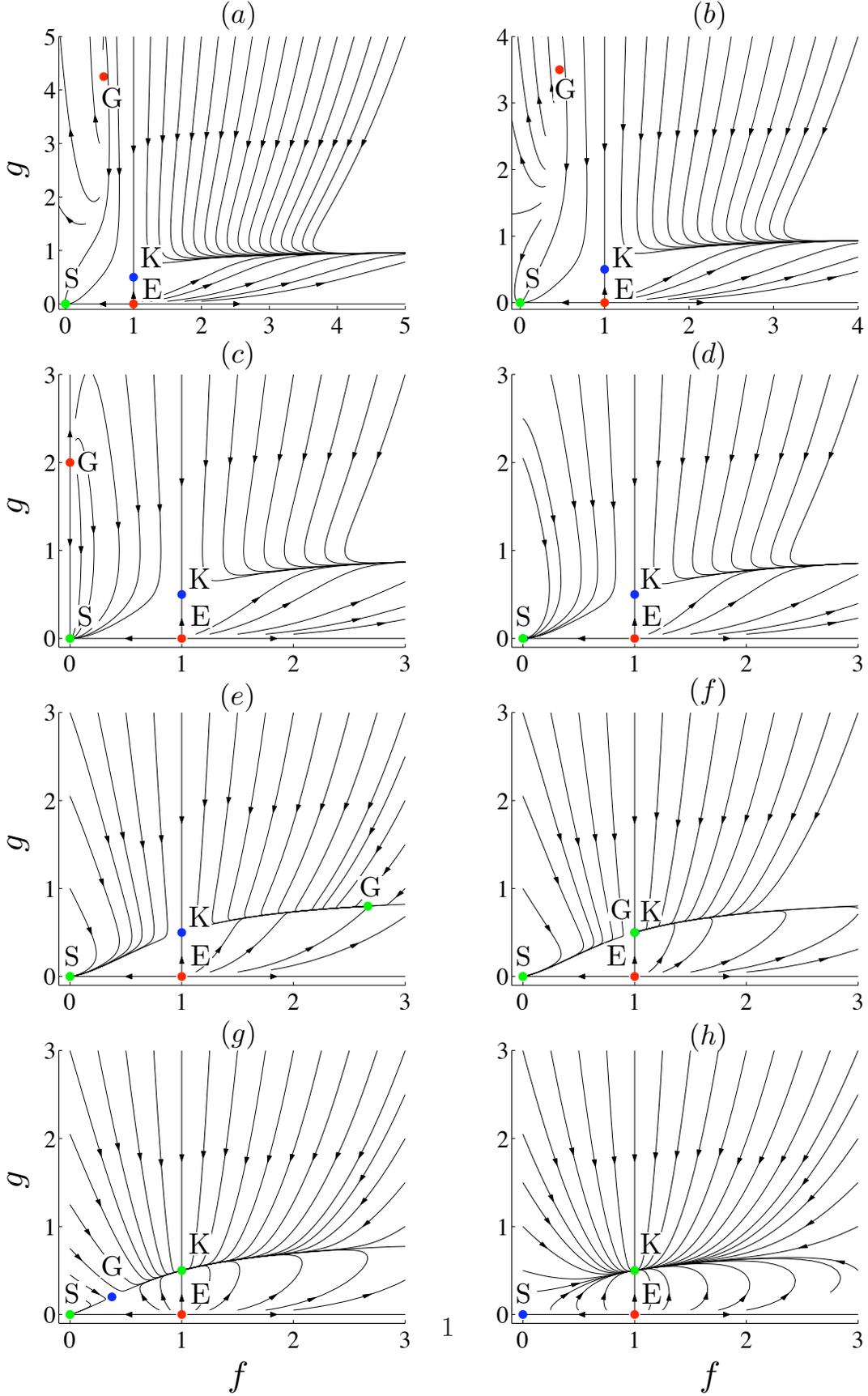,width=.8\textwidth,clip=}
\end{center}
\caption{Numerical integration of equations \eref{iee:f1}-\eref{iee:g1} for $d=1$
and different values of the exponent $\mu$: (a) $\mu=1/4$,
(b) $\mu=1/2$, (c) $\mu=1$, (d) $\mu=1.15$,
 (e) $\mu=1.4$, (f) $\mu=3/2$, (g) $\mu=1.6$, (h) $\mu=1.8$.
The values of the coupling variables at the different fixed points
(in the graphs, E stands for Edwards-Wilkinson, K for KPZ, S for Smooth, and G for
Galilean) and the associated critical exponents are reported in table \ref{fp-table:1}.
The color of each fixed point represent its stability (red stands for unstable, blue for saddle, and
green for stable). Note that the Galilean fixed point changes with $\mu$, merging with the KPZ fixed
point at $\mu=3/2$.}
\label{fig:flowd1}
\end{figure}

In figure \ref{fig:flowd1} we show eight representative snapshots of
the RG flow for $d=1$. For $0<\mu\leq 1$, the Galilean fixed point is unstable and moves
on the sector of the $(f,g)$ plane associated with morphologically stable bare conditions
$f<1$, following the red line in the left panel of figure \ref{fig:fg-galileo}.
Qualitatively, within this interval of $\mu$, the flow behaves as
in figures \ref{fig:flowd1}(a)-(c), in which the DRG flow is shown for the
cases $\mu=1/4,1/2,1$. Notice that the only finite fixed point for $f<1$
and $0<\mu<4/3$ is the Smooth fixed point, as expected from power counting. For $\mu \in (1,4/3]$,
the Galilean fixed point is not an admissible solution of the DRG flow equations \eref{iee:f1} and \eref{iee:g1},
and it disappears [for example, see figure \ref{fig:flowd1}(d)] until $\mu$ becomes larger than $4/3$,
when it reenters the phase portrait as a fixed point for morphologically unstable bare conditions $f>1$,
see e.g.\ figure \ref{fig:flowd1}(e). Meanwhile, the basin of attraction of the Smooth fixed point changes as a function of $\mu$, see figures \ref{fig:flowd1}(a)-(c).

\begin{figure}[!h]
\begin{center}
	\epsfig{file=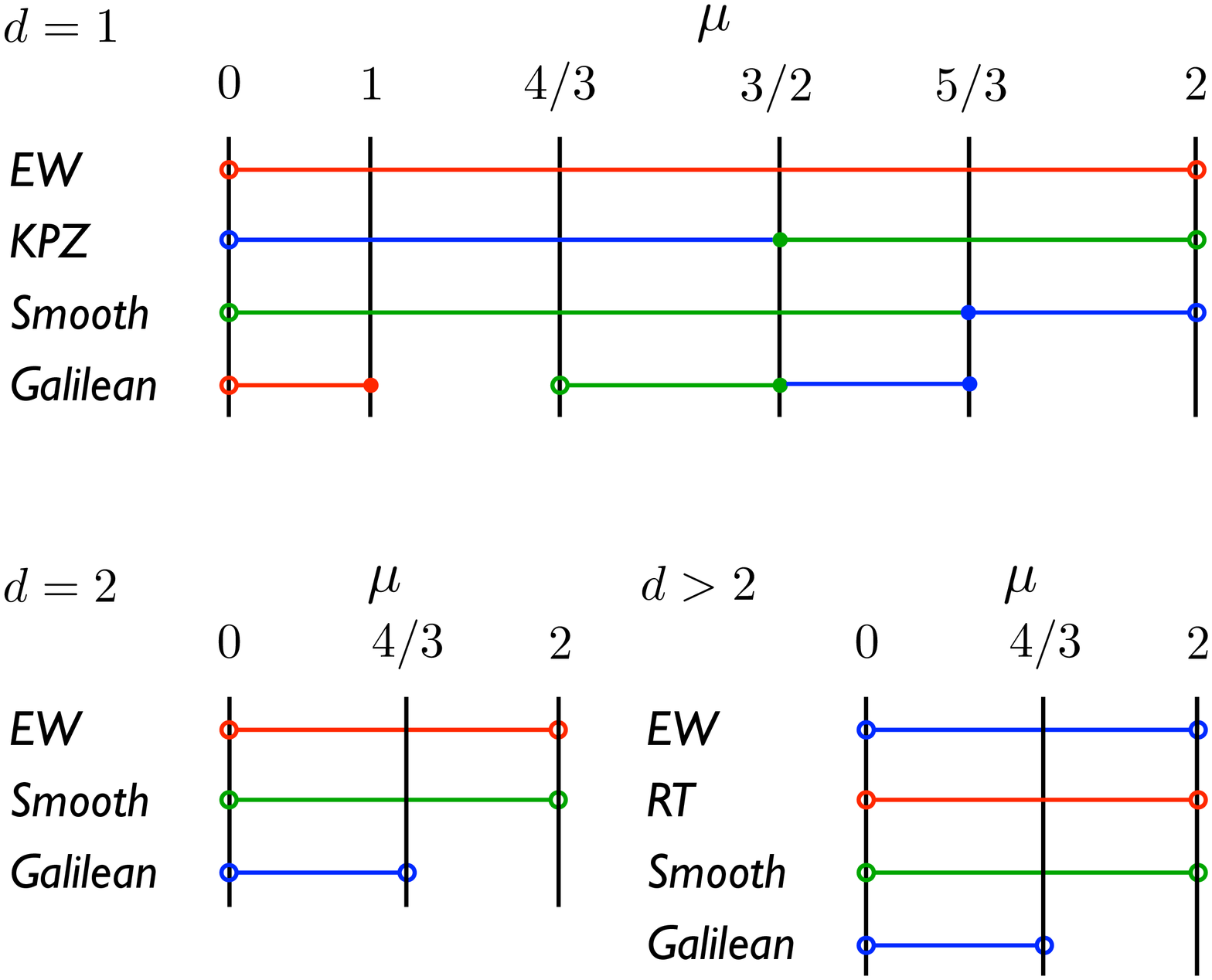,width=.6\textwidth,angle=-90,clip=}
\end{center}
\caption{\label{fig:stab}Stability of the fixed points for $d=1,2$ and $d>2$. As before the red color stands
for unstable, blue for saddle and green for stable fixed point.}
\end{figure}

For larger $\mu$ values in the interval $4/3<\mu\leq 3/2$, the Galilean fixed point
becomes stable, attracting all the trajectories with $f>1$; this condition is
shown in figure \ref{fig:flowd1}(e). The remaining trajectories are attracted
by the Smooth fixed point. Actually, the Galilean fixed point
merges with the KPZ fixed point at exactly $\mu=z_{\rm\tiny KPZ}(d=1)=3/2$ [see left panel of figure
\ref{fig:fg-galileo} and figure \ref{fig:flowd1}(f)], loosing stability in
favor of the latter for larger values, $z_{\rm\tiny KPZ}(1) \leq \mu \leq 2$,
justifying the observation of KPZ scaling in this range of $\mu$, as
obtained in numerical simulations under morphologically unstable parameter
choices \cite{Nicoli:2009b}. For $3/2<\mu\leq 5/3$, the Galilean fixed point becomes a saddle, being located
on the separatrix between the basins of attraction of the two stable
fixed points for the DRG flow, the Smooth and the KPZ fixed points, see
figure \ref{fig:flowd1}(g), back on the morphologically stable sector. For $\mu>5/3$,
the Galilean fixed point is no longer an admissible solution of the RG equations, and disappears.
The only stable fixed point for these values of $\mu$ is the KPZ fixed point [see figure \ref{fig:flowd1}(h)].
As mentioned above, indeed it is KPZ scaling that is obtained in numerical simulations \cite{Nicoli:2009b}
performed for this range of $\mu$ values for morphologically unstable conditions. Actually, preliminary
simulations performed for stable parameter conditions seem to agree with this conclusion also \cite{Nicoli:unpubl},
which in turn was expected from the power counting analysis performed in section
\ref{scaling_anal}  above.
In the first row of figure \ref{fig:stab} we summarize the DRG stability of the four fixed
point as function of $\mu$. Note, numerical simulations for the morphologically unstable condition
predict scaling as for the Galilean fixed point when $0<\mu<4/3$ \cite{Nicoli:2009b}, and thus do not agree with
the DRG flow, probably due to the assumptions made in the perturbative expansion.
Nevertheless, the DRG analysis gives significant hints about the critical behavior of this
non-local class of growth equations, in particular the change of universality class
from that associated with the Galilean fixed point for $0 < \mu < z_{\rm\tiny KPZ}(d=1)$ to
KPZ scaling for $z_{\rm\tiny KPZ}(d=1) < \mu \leq 2$, generalizing in particular known
results for the noisy Kuramoto-Sivashinsky equation \cite{Cuerno:1995b,Ueno:2005}, that is the $\mu=2$ case of equation \eref{eq_gral}.


\subsection{Two-dimensional interfaces}

\begin{figure}[h]
\begin{center}
	\epsfig{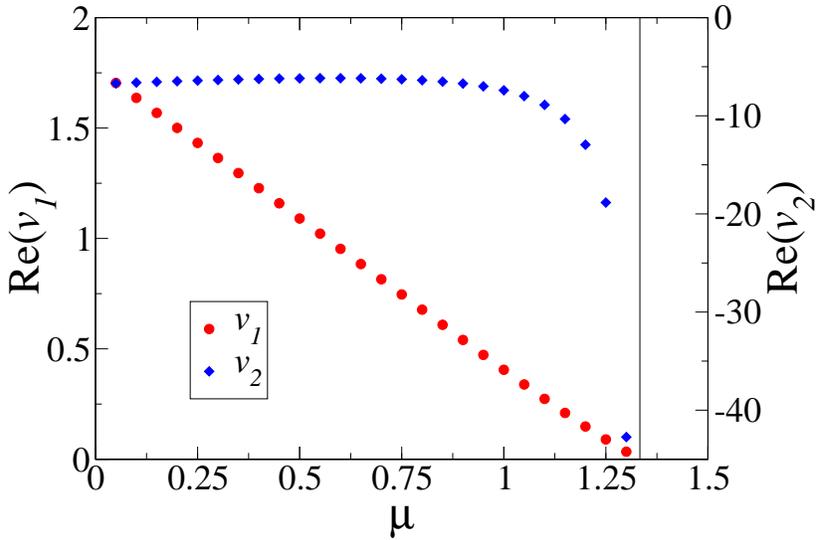}
\end{center}
\caption{Real parts of the eigenvalues of the linear stability matrix $S({\rm G})$ for the Galilean
fixed point for $d=2$, as functions of $\mu$. Note the different scales used in the left [for ${\rm Re}\left(v_1\right)$]
and the right [for ${\rm Re}\left(v_2\right)$] vertical axes of the graph.}
\label{eig-2dG}
\end{figure}

The situation changes drastically for $d=2$. As expected for perturbative
approaches such as DRG \cite{Medina:1989} and even field-theoretical schemes \cite{Frey:1994,Sun:1994},
there is no finite KPZ fixed point for the RG flow, while the Galilean fixed point remains finite
only up to $\mu=4/3 < z_{\rm\tiny KPZ}(2) \simeq 1.61$. For $\mu=4/3$, this fixed point is at
$f=\infty$ and $g=2$, see the right panel of figure \ref{fig:fg-galileo}.
As in the $d=1$ case, the position of the Galilean fixed point depends on the value of $\mu$,
as shown in the right panel of figure \ref{fig:fg-galileo}. In table \tref{fp-table:2} we summarize
the results for the fixed points in the two-dimensional case, showing in figure \ref{eig-2dG}
the real parts of the eigenvalues of the Galilean fixed point, $v_{1,2}^{(2)}$, as functions of $\mu$.
We observe that the stability of the three fixed points does not change with $\mu$ (see the second row of figure \ref{fig:stab}).
\begin{table}[!h]
\begin{center}
\begin{tabular}{|c|c|c|c|c|c|c|c|c|}
\hline
\T Name & $f$  & $g$ & $z$ & $\alpha$ & $\beta$&Scaling relation & $\lambda_1$ & $\lambda_2$  \\[3pt]
\hline
\hline
\T EW & $1$ & $0$ & $2$ & $0$ &  $0$& $2\alpha+d=z$ & $2-\mu$ & $0$\\[5pt]
S & $0$ & $0$ & $\mu$ & $\frac{\mu-2}{2}$ & $\frac{\mu-2}{2\mu}$ &
$2\alpha+d=z$ & $\mu-2$ & $3\left(\mu-2 \right)$  \\[7pt]
G & $\frac{3(\mu-2)}{3\mu-4}$ &
$3(2-\mu)$  & $\mu$ & $2-\mu$ & $\frac{2-\mu}{\mu}$ &
$\alpha+z=2$ & $v_1^{(2)}$ & $v_2^{(2)}$ \\[10pt]
\hline
\end{tabular}
\end{center}
\caption{Finite fixed points in $2+1$ growth dimensions ($d=2$).
The meaning of G, S, $\lambda_{1,2}$, and $v^{(2)}_i$ is as in table \ref{fp-table:1}.
The real parts $v_{1,2}^{(2)}$, of the eigenvalues of the Galilean fixed point are
plotted in figure \ref{eig-2dG}. This fixed point is not defined for $\mu\geq 4/3$.}
\label{fp-table:2}
\end{table}

\begin{figure}[t!]
\begin{center}
	\epsfig{file=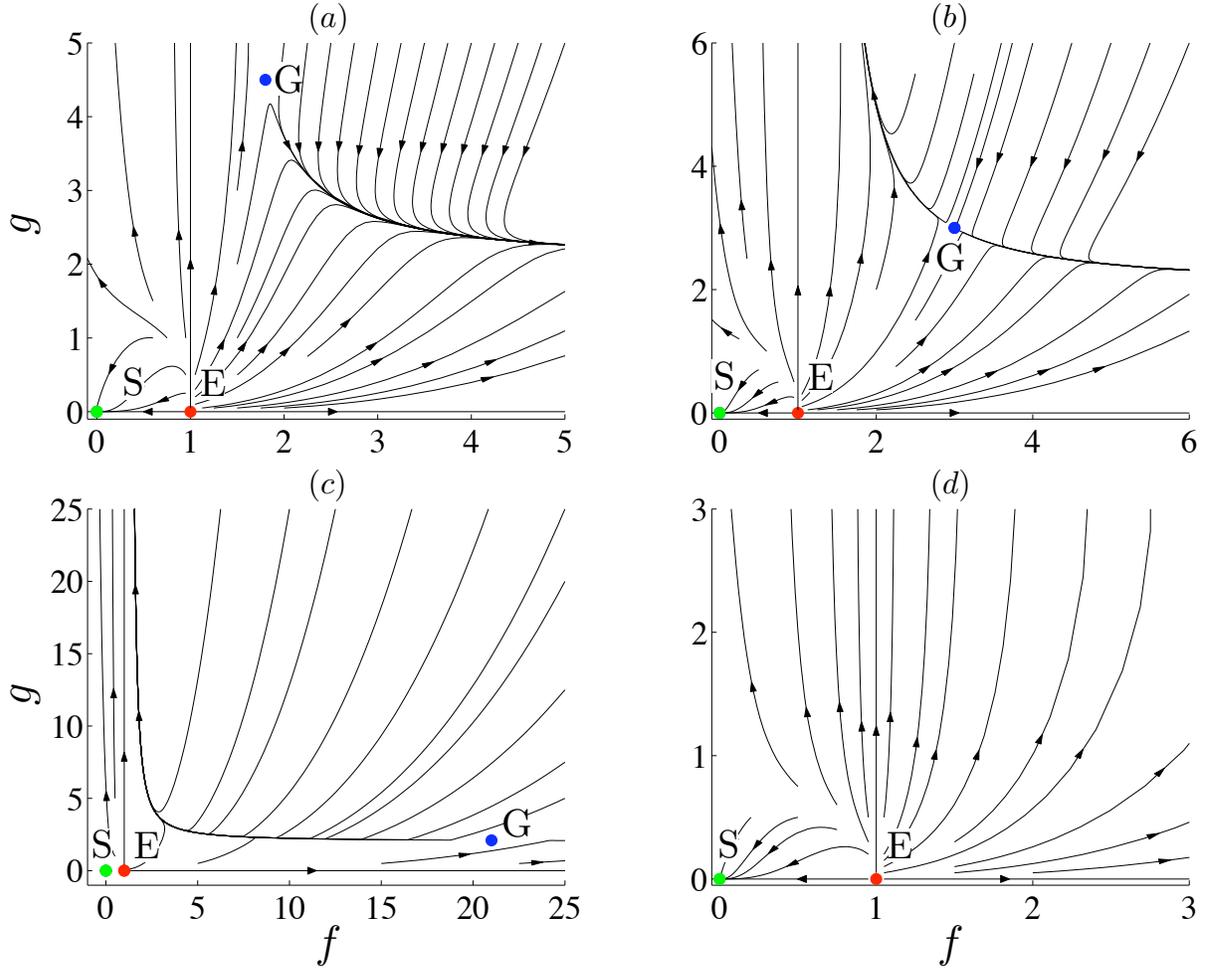,angle=90,width=.9\textwidth,clip=}
\end{center}
\caption{Numerical integration of equations \eref{iee:f1}-\eref{iee:g1} for $d=2$
and different values of the exponent $\mu$:
(a) $\mu=1/2$, (b) $\mu=1$, (c) $\mu=1.3$,
(d) $\mu=3/2$.
The values of the coupling variables at the different fixed points
(in the graphs, $E$ stands for Edwards-Wilkinson, $K$ for KPZ, $S$ for Smooth and $G$ for
Galilean) and the associated critical exponents are reported in table \ref{fp-table:2}. The color of each
fixed point represent its stability (red stand for unstable, blue for saddle and
green for stable). Note that the Galilean fixed point is not present for $\mu\ge 3/4$.}
\label{fig:flowd2}
\end{figure}

We can identify two main features in this DRG flow.
The first one is represented in Fig.\ \ref{fig:flowd2}(a)-(c). In this case the
Galilean fixed point influences all trajectories in the morphologically unstable region
$f>1$; in fact, it is located on the manifold that separates two classes of
trajectories: asymptotically, one type of trajectories flows towards $f=1$ and $g\to\infty$;
the other type of trajectories flows towards $f\to\infty$ and $g=2$, see figures \ref{fig:flowd2}(a)-(c).
Within the morphologically stable region $f<1$, the Smooth fixed point attracts some trajectories, while
those outside the basin of attraction of this fixed point flow asymptotically to $f=0$ and $g\to\infty$.


For $\mu\geq 4/3$, the Galilean fixed point is not an admissible solution
of the flow equations. In this case all trajectories with $f>1$ flow to $f\to 1$ and
$g\to\infty$, attracted by the KPZ fixed point at infinity, see figure \ref{fig:flowd2}(d).
The region with $f<1$ does not display any qualitative change as compared
to the case $\mu<4/3$. To compare with the present results of the DRG analysis,
numerical simulations of the morphologically unstable case of equation \eref{eq_gral} \cite{Nicoli:2009b}
for $d=2$ indicate Galilean scaling for $\mu< z_{\rm\tiny KPZ}(2) \simeq 1.61$ and KPZ scaling for $z_{\rm\tiny KPZ}(2) < \mu \leq 2$.


\subsection{The general case $d>2$}

For $d$ larger than two, we obtain a flow behavior that is independent on the substrate
dimension. As for the KPZ case, there is no finite fixed point associated with KPZ scaling,
while we do find an unstable fixed point located at $f=1$ (i.e., $\nu=0$) and $g=g_{\rm RT}(d)$ that is associated with the
non-equilibrium {\sl roughening transition} (RT) \cite{Barabasi:1995}, hence our terminology.
Starting from a bare parameter condition corresponding to $f(0)=1$, the parameters of equation \eref{eq_gral}
flow towards the EW fixed point in case $g(0)< g_{\rm RT}$, or towards the strong coupling regime associated
with KPZ scaling for $g(0)> g_{\rm RT}$. See table \ref{fp-table:d} for the value of $g_{\rm RT}$ as a function
of $d$. For this reason, the EW fixed point acquires a stable direction as compared with smaller values of $d$,
changing from unstable (for $d=2$) to saddle type. The two other fixed points, Smooth and Galilean, have the same stability and exponents as found for $d=2$ (take into account that a negative roughening exponent
means subleading corrections to scaling). We summarize the results for $d>2$ in table \ref{fp-table:d}.
\begin{table}[!h]
\begin{center}
\begin{tabular}{|c|c|c|c|c|c|c|c|c|}
\hline
\T Name & $f$  & $g$ & $z$ & $\alpha$ & $\beta$& Scaling relation & $\lambda_1$ & $\lambda_2$ \\[3pt]
\hline
\hline
\T EW &  $1$ & $0$ & $2$ & $(2-d)/2$ &  $(2-d)/4$& $2\alpha+d=z$
	& $2-\mu$ & $ 2-d$ \\[5pt]
 RT & $1$ & $\frac{d(d-2)}{4d-6}$ & $2+\frac{(d-2)^2}{4d-6}$ & $\frac{(d-2)^2}{6-4d}$ &
 	$\frac{(d-2)^2}{8-4d-d^2}$& $\alpha+z=2$ & $2-\mu+\frac{(d-2)^2}{4d-6}$ & $d-2$ \\[5pt]
S & $0$ & $0$ & $\mu$ & $\frac{\mu-d}{2}$ & $\frac{\mu-d}{2\mu}$ &
	$2\alpha+d=z$  & $\mu-2$ & $3\mu-d-4$\\[7pt]
G & $f_*$ & $g_*$  & $\mu$ & $2-\mu$ & $\frac{2-\mu}{\mu}$ &
	$\alpha+z=2$ & $v^{(d)}_1$ & $v^{(d)}_2$   \\[10pt]
\hline
\end{tabular}
\end{center}
\caption{Fixed points in $d+1$ growth dimensions for $d>2$.
The meaning of G, S, $\lambda_{1,2}$, and $v^{(d)}_{1,2}$ is as in tables \ref{fp-table:1}, \ref{fp-table:2}.
RT denotes the fixed point controlling the roughening transition as discussed in the
main text. The coordinates of the Galilean fixed point depend on $d$ as given by equations
\eref{fg:galileo}. This fixed point is not defined for $\mu= 3/2$.}
\label{fp-table:d}
\end{table}

\begin{figure}[h]
\begin{center}
	\epsfig{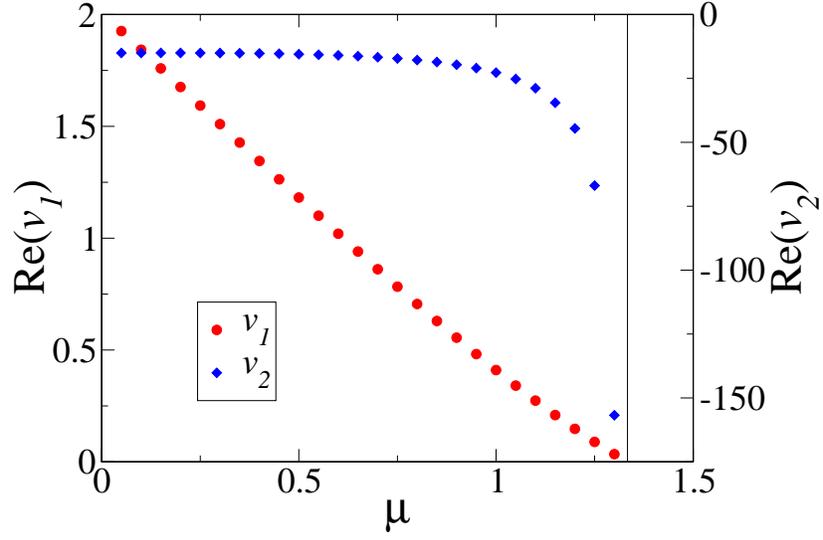}
\end{center}
\caption{Real parts of the eigenvalues of the linear stability
matrix $S({\rm G})$ for the Galilean fixed point for $d=3$, as
functions of $\mu$. Note the different scales used in the left (for $v_1$) and the right (for $v_2$) vertical
axes of the graph.}
\label{eig-3dG}
\end{figure}

As seen in the last two columns of table \ref{fp-table:d}, the signs of $\lambda_1$ and $\lambda_2$
do not depend on $d$, implying the independence of the stability of these four fixed points
on the substrate dimensionality (note that, for the RT fixed point, $\lambda_1>0$ for every $0<\mu\leq 2$).
The real parts of the eigenvalues for the Galilean fixed point are plotted in figure \ref{eig-3dG} for $d=3$.
As in the case of two-dimensional interfaces, one eigenvalue is positive whereas the other one is negative.
We have checked that this behavior does not change by increasing the value of $d$.

\begin{figure}[t!]
\begin{center}
	\epsfig{file=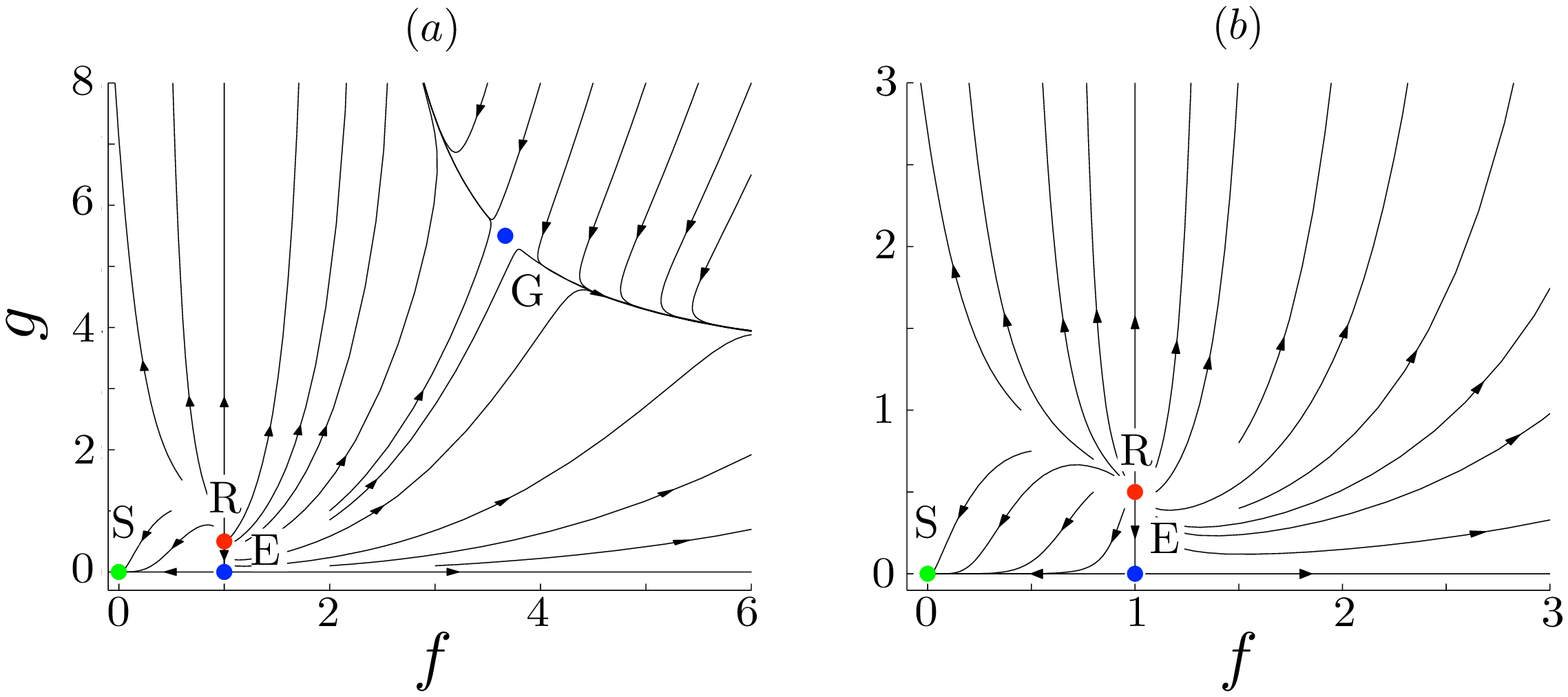,angle=-90,width=.9\textwidth,clip=}
\end{center}
\caption{Numerical integration of equations \eref{iee:f1}-\eref{iee:g1} for $d=3$
and different values of the exponent $\mu$:
(a) $\mu=1/2$, (b) $\mu=3/2$.
The values of the coupling variables at the different fixed points
(in the graphs E stands for Edwards-Wilkinson, R for the fixed point related to
the roughening transition, S for Smooth and G for Galilean) and the associated
critical exponents are reported in table \ref{fp-table:d}. The color of each
fixed point represent its stability (red stand for unstable, blue for saddle and
green for stable). Note that the Galilean fixed point is not present for $\mu= 3/2$.}
\label{fig:flowd3}
\end{figure}
\begin{figure}[t!]
\begin{center}
	\epsfig{file=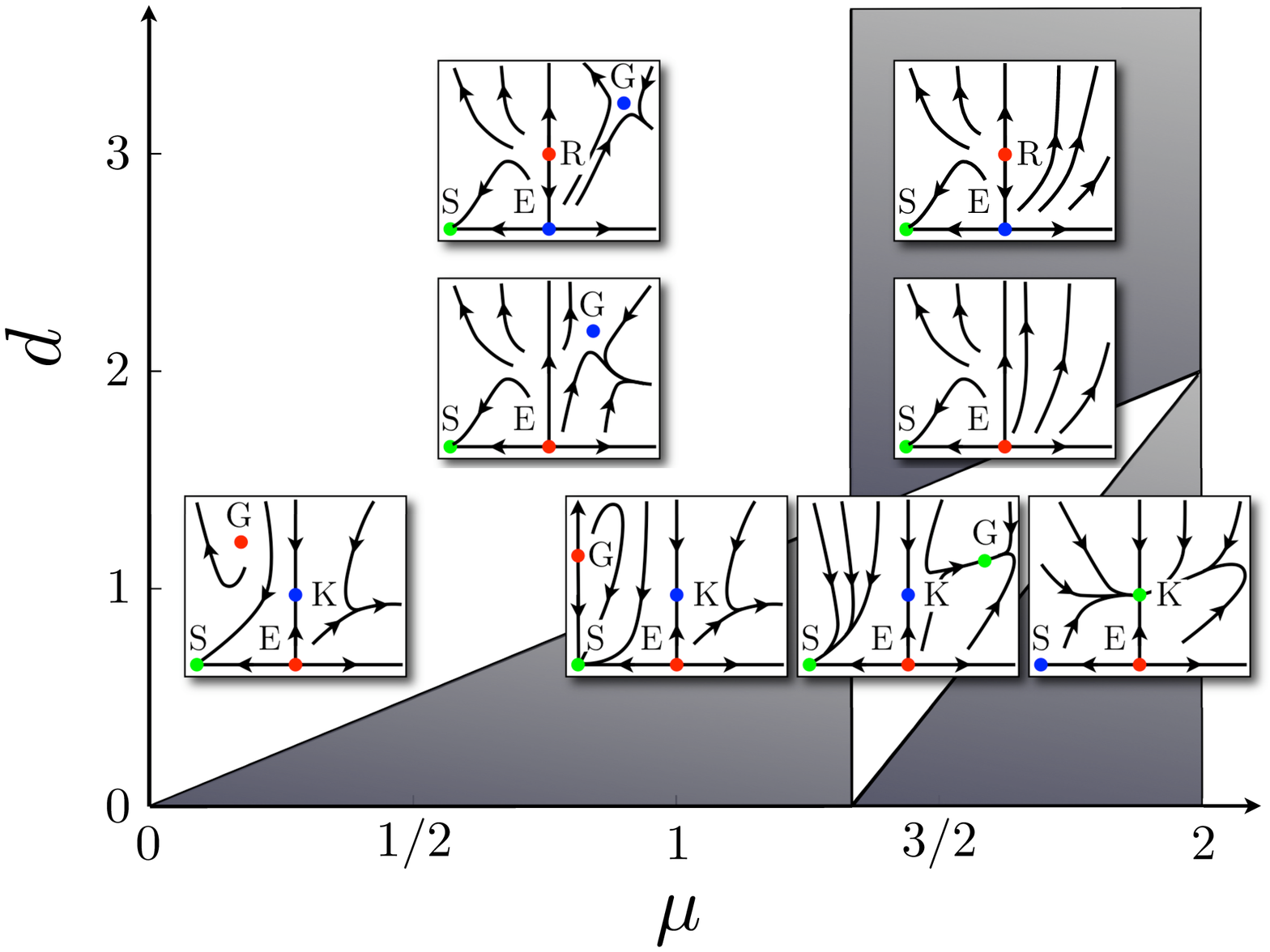,angle=-90,width=.95\textwidth,clip=}
\end{center}
\caption{Qualitative behavior of the DRG flow for different values of $\mu$
and the substrate dimension $d$. The color of each fixed point represent its stability
(red stand for unstable, blue for saddle and green for stable). The shaded region of
the plane corresponds to the values of $\mu$ (as function of $d$) within the Galilean
fixed point is not defined.}
\label{fig:resumen-flow}
\end{figure}

Beyond fixed point stability, and similarly to the KPZ equation, for any $d>2$ the structure of the RG flow is
almost identical to the two dimensional case, the only difference being that the $f=1$ line acquires the RT fixed point,
giving the possibility to observe EW scaling for $\nu=0$. In figure \ref{fig:flowd3} we report two
representative flows for $d=3$. By comparing them with those reported in figures \ref{fig:flowd2}(a) and \ref{fig:flowd2}(d), one cannot appreciate any qualitative change for flow trajectories outside the
$f=1$ line. Unfortunately, we are not aware of any numerical study of equation
\eref{eq_gral} for $d>2$, so that we cannot compare predictions of our DRG flow with numerical
simulations for these values of dimensionality. In any case, we expect that the numerical results
obtained for $d=1,2$ \cite{Nicoli:2009b} generalize for any substrate dimension. A qualitative summary of the
DRG flows as obtained for different integer dimensions is provided in figure \ref{fig:resumen-flow}.


\section{Discussion and conclusions} \label{discussion}

The DRG analysis of equation \eref{eq_gral} yields a number of non-trivial
results that can be classified with respect to their relevance to several issues on
surface kinetic roughening: {\em (i)} the fixed point structure for a class of non-local equations,
and its dependence on dimensionality, as compared to available results from numerical simulations;
{\em (ii)} general consequences for stochastic equations that feature a KPZ non-linearity, in particular
the issue of Galilean invariance; {\em (iii)} new mechanisms for the dynamical stabilization of morphologically unstable systems; {\em (iv)} the ensuing predictions on scaling behavior of realistic systems for which a continuum description can be expected to apply that fits within the framework of equation \eref{eq_gral}. We will devote this section to a brief discussion of these issues, in a way that allows us to summarize the results of the paper and put them into a wider perspective. In general, interesting new results can be anticipated from further studies of the class on non-local equations \eref{eq_gral} along the lines discussed in the three subsections that follow.

\subsection{Asymptotic properties}

For different dimensions, $d$, our DRG analysis can be roughly summarized as a generalization
of both the KPZ and the noisy Kuramoto-Sivashinsky flows, in which additional non-trivial fixed points arise that are intrinsically associated with the non-local nature of the equations that we are studying. Thus, at the Smooth and
the Galilean fixed points the dynamic exponent $z$ takes a value that coincides with the one that would be expected from a
naive power counting argument, while the roughness exponent can be evaluated from the latter through a scaling relation
that is either hyperscaling or the one usually associated with Galilean invariance.

The comparison between the analytical DRG results and those from numerical simulations leads to partial agreement.
Specifically in the $d=1$ case, the change in scaling behavior that takes place for increasing $\mu$ for the morphologically stable condition \cite{Nicoli:unpubl} is well described by the DRG results. Perhaps more remarkably, the DRG seems to also capture correctly the role of $z_{\rm KPZ}(d=1)=3/2$ as a critical value for $\mu$ in the morphologically unstable case, that separates asymptotic behavior between that controlled by the Galilean fixed point for $\mu < z_{\rm KPZ}(1)$ and KPZ scaling for $\mu > z_{\rm KPZ}(1)$, also in good agreement with previous numerical results \cite{Nicoli:2009b}. However, possibly the most salient disagreement between the DRG and the numerics is the failure to generalize that role of $z_{\rm KPZ}(d)$ for general $d>1$. This is not surprising in view of the known non-perturbative effects that are expected for the KPZ nonlinearity, particularly in higher dimensions. In general, the behavior of the Galilean fixed point changes with $\mu$ and system dimensionality in a rather non-trivial way, and some of the consequences for the DRG flow may be associated with artifacts of the perturbative scheme employed. On the other hand, the dynamic renormalization group does account for the irrelevance of higher order contributions like those with parameter ${\cal N}$ in equation \eref{eq_gral}, even in the morphologically unstable case, again as observed in numerical simulations \cite{Nicoli:2009b}. We expect that the same qualitative results apply to any equation with the shape of
\eref{gen_eq} using a general dispersion relation like \eref{gen_disp_rel} such that
$r_1 < r_2 = 2 < \dots < r_n$.

From a general perspective, the fact that equation \eref{eq_gral} ---and more specifically equation \eref{super_gen_eq}---
generalizes both the KPZ and the Lai-Das Sarma-Villain equations, has allowed us to prove vertex cancellation in the last two equations to stem merely from the shape of the nonlinearity, and to be independent of an underlying continuum symmetry of
the dynamical equation. Although limited to the one-loop perturbative expansion that has been employed in our DRG approach, this result may prove informative in the current questioning of Galilean invariance as inducing vertex non-renormalization in the KPZ equation, although further evidence is needed. In general, a more detailed analytical understanding of the asymptotic properties of equation \eref{eq_gral} will require resorting to techniques that differ in nature from the perturbative DRG.


\subsection{A mechanism for dynamical stabilization}

A very remarkable feature of the present DRG results is that they describe analytically a mechanism for the dynamical stabilization of morphologically unstable systems, that differs from Yakhot's stabilization mechanism for the (noisy) Kuramoto-Sivashinsky equation. Namely, the latter requires renormalization of the negative ``surface tension" parameter $\nu$ to a positive value,  eventually leading to asymptotic behavior in the KPZ universality class. Dynamically, as seen in numerical simulations of the
noisy Kuramoto-Sivashinsky equation both in one \cite{Karma:1993,Cuerno:1995,Ueno:2005} and in two \cite{Nicoli:2010} dimensions, the periodic cell structure that appears at short times is seen to evolve in a chaotic fashion that leads to height disorder and kinetic roughening properties at sufficiently long times. In marked contrast, for equation \eref{eq_gral} the corresponding unstable $\nu$ parameter cannot change sign for $\mu < 2$, as implied by the first equation in \eref{ecs:nu,lambda}, or equivalently by the fact that $f=1$ is an invariant of the DRG flow, see equations \eref{iee:couplings}-\eref{iee:f1}. Still, the system evolves from an initial periodic morphology into a late time topography that is also disordered and rough. For any value of $\mu<2$, this process is more similar to cell coarsening by creation and annihilation of cusps than to the familiar chaotic cell dynamics characteristic of the ($\mu=2$) nKS case. See movies, e.g., in the Supplemental Material provided with \cite{Nicoli:2009} and with \cite{Nicoli:2009b} for one and two-dimensional systems, respectively. Cusp annihilation and creation has been studied in detail in the particular case of the stochastic and deterministic Michelson-Sivashinsky equations, namely, for $\mu=1$ with $m=2$, see e.g.\ \cite{Kupervasser:unpubl} and references therein, but does not seem to have been studied explicitly for other values of $\mu$. Note that, in spite of such strong differences in the dynamical evolution between the nKS equation and equation \eref{eq_gral}, they share the same KPZ asymptotic state, provided $\nu< 0$ and (as conjectured) $z_{\rm KPZ}(d) < \mu < 2$.

\subsection{Comparison with experimental and other model systems}


In view of the previous results on scaling properties of equation \eref{eq_gral}, a natural question concerns their experimental relevance. We believe that the kinetic roughening properties of a number of experimental systems are indeed compatible with the predictions from equation \eref{eq_gral}. For instance, in the morphologically stable condition, viscous flow induced on glassy films by ion irradiation leads to the scaling exponents obtained from \eref{eq_gral} for $\mu=1$ \cite{Mayr:2001}.

Perhaps more interestingly, a number of actual thin film systems that are characterized by large values of the growth exponent $\beta=\alpha/z$ \cite{Barabasi:1995} are compatible with our continuum framework in the morphologically unstable case, possibly due to the occurrence of geometrical shadowing effects \cite{Bales:1990}.
For instance, in \cite{Zhao:1999} Si(100) surfaces were plasma etched, finding $\alpha = 0.96\pm 0.06$, $\beta=0.91\pm 0.03$, and, $z = 1.05\pm 0.09$, consistent with the predictions of equation \eref{eq_gral} for $\mu=1$ at the Galilean fixed point. Very similar exponent values have been recently found in chemical vapor deposition (CVD) growth of silica films \cite{Castro:2011}.
CVD is an experimental technique within which shadowing instabilities are quite generically expected. Thus, e.g.\ in \cite{Dalakos:2005} amorphous hydrogenated silicon (a-Si:H) was deposited on glass via a plasma enhanced CVD technique, finding $\beta = 1.13\pm0.04$ for the condition of low-silane concentration in the gas mixture. Similarly, \cite{Buijnsters:2008} reports the roughness evolution of hydrogenated amorphous carbon (a-C:H) films grown by electron cyclotron resonance CVD. A growth exponent value $\beta = 0.74$ was measured, which suggests that an underlying morphological instability is responsible for the cauliflower-like morphology observed at the late stages of the growth process, that looks quite similar to that found in \cite{Castro:2011}. Further, in \cite{Hormann:2009} a large growth exponent ($\beta = 0.9\pm 0.1$) is reported for hydrogenated diamond-like carbon (a-C:H) coatings deposited through plasma enhanced CVD. In general, it would be interesting to search for experimental parameters/techniques that allow to tune the value of $\mu$ characterizing a putative continuum description through equation \eref{eq_gral}, in order to probe the corresponding continuum line of fixed points. A potential candidate might be the sticking probability in CVD systems, that seems able to tune scaling properties from KPZ-type \cite{Ojeda:2000} (``large" $\mu$ values) to sMS-type ($\mu=1$) \cite{Castro:2011}, although values of the effective $\mu$ parameter should be still identified that differ from $\mu=1,2$.

In connection with kinetic roughening in the presence of shadowing instabilities, we note that other continuum and discrete  models are available \cite{Krug:1997}. Typically many of them involve the calculation of the so-called exposure angle, which is a non-linear and non-local function of the height profile, that is very expensive to compute. By means of a small slopes expansion, the exposure angle can be seen to lead to a term of the form $k h_{\bi{k}}$ in the height equation \cite{Karunasiri:1989}, leading to a non-local dispersion relation as in \eref{eq_gral} (with $\nu<0$, $\mu=1$, and, $m=4$). Note that for this choice of $\mu$, the Galilean fixed point leads to the exponents $\alpha = \beta = z =1$. However, other authors have included the non-locality of the shadowing process in the non-linear part of the evolution equation through a non-local redistribution process of the incident particle flux \cite{Zhao:1999}. The latter model has been studied numerically with a Monte Carlo method \cite{Drotar:2000b,Drotar:2000a}, and analytically \cite{Chattopadhyay:2002}, finding the same critical exponents, i.e., $\alpha\simeq \beta\simeq z\simeq 1$, and $\alpha=1.04$, $\beta = 1.08$, and, $z=0.96$, respectively. In spite of the different formulation of the shadowing instability, the critical exponents coincide.

Overall, we believe that equation \eqref{eq_gral} provides a very compact description of a very wide class of systems undergoing non-conserved interface growth, both under conditions of morphological stability and in the presence of
pattern forming effects. Note that, in spite of the presence of the KPZ nonlinearity in the equation of motion, KPZ scaling only occurs under the, relatively restricted, simultaneous conditions of morphological {\em instability} and (as conjectured) sufficiently large $\mu > z_{\rm KPZ}(d)$ values. This fact, together with the long transients induced by morphological instabilities, may account for the surprising experimental scarcity of KPZ scaling \cite{Cuerno:2001,cuerno:2007}. Understanding the degree to which this is truly the case will require further exploration of the present and related continuum models.


\label{sec:Galileo}

\acknowledgements
We gratefully acknowledge discussions with C.\ Escudero. This work has been partially supported through Grants No.\ FIS2009-12964-C05-01 and  No.\ FIS2009-12964-C05-03 (MICINN, Spain). M.\ N.\ acknowledges support by Fundaci\'on Carlos III (Spain) and by Fondazione Angelo Della Riccia (Italy).

\appendix
\section{Coarse-graining of propagator and noise variance for Eq.\ (\ref{eq_drg})}
\label{ax:nonlocal}

For simplicity and as justified in more detail in the next Appendix, we consider equation \eref{eq_drg}, namely,
equation \eref{eq_gral} with $\sigma_\bi{k}=-\nu k^\mu-\mathcal{K} k^2$, as the simplest representative for the class of non-local equations studied in the main text.
Here we present the detailed calculations for the coarse-graining step in the renormalization of the propagator and
the noise variance for equation \eref{eq_drg}.

In equation \eref{iee:renorm_eq} for the slow modes, the contribution from
the elimination of the fast modes is contained in the function $\Sigma$.
In order to evaluate expression \eref{sigma}, we symmetrize the
wave-vector integral by introducing the standard
substitution $(\bi{q},\Omega)\to (\bi{j}+\bi{k}/2,\Omega+\omega/2)$
\cite{Forster:1977}.
We begin the calculation by considering the various contributions separately.

The wave-vector contribution due to the vertex
is readily obtained in the new symmetrized variables as
\begin{equation}
[\bi{q}\cdot (\bi{k}-\bi{q})] [-\bi{q}\cdot\bi{k}]=
    j^3 \cos(\theta) k+\frac{1}{2} j^2 k^2+\Or(k^3),
\label{momentpart}
\end{equation}
where $\theta$ is the angle between $\bi{k}$ and $\bi{j}$.
Next, we expand the contribution from the bare propagators by
taking the long time limit (i.e., $\omega\to 0$),
\begin{align}
\lim_{\omega \to 0} \left|G_0\left(\hat{q}\right)\right|^2  G_0\left(\hat{k}-\hat{q}\right) &
=\left\{ \left[\nu j^\mu\left(1+x \cos(\theta)+\frac{x^2}{4}\right)^{\mu/2} \hskip -0.5cm +
	\mathcal{K} j^2\left(1+x \cos(\theta)+\frac{x^2}{4}\right)^{\mu/2}\right]^2+\Omega^2
	\right\}^{-1} \nonumber \\[5pt]
&  \hskip 1 cm \times \left[
	\nu j^\mu\left(1-x \cos(\theta)+\frac{x^2}{4}\right)^{\mu/2}\hskip -0.5 cm +
	\mathcal{K} j^2\left(1-x \cos(\theta)+\frac{x^2}{4}\right)^{\mu/2}+\rmi\Omega
	\right]^{-1} \nonumber\\[5pt]
& \sim
	\left[1+\left(\frac{1}{\Delta_j+\rmi z}-\frac{2\Delta_j}{\Delta_j^2+z^2}\right) \mathcal{C}_j
	\cos(\theta) \frac{k}{j}\right]
	\,
	\left[
	\left(\mathcal{K}j^2\right)^3 \left(\Delta_j^2+z^2\right)\left(\Delta_j+\rmi z\right)
	\right]^{-1},
\end{align}
where we have introduced variables $z=\Omega/\mathcal{K}j^2$ and $x=k/j$, and the function
$\mathcal{C}_j=1+\nu\mu/2\mathcal{K}|j|^{2-\mu}$ in order to
shorten the notation [$\Delta_j$ has been already defined in \eref{Delta}].
We can multiply the last result by \eref{momentpart} and
split the integral in $\bi{q}$ into an angular integral in $d-1$-dimensions and a
one dimensional radial integral, where we approximate the shell to a (hyper-)spherical shape.
Thus,
\begin{equation}
\int_>\frac{\rmd\bi{q}}{\left(2\pi\right)^d}=
\frac{S_{d-1}}{\left(2\pi\right)^d}\int_>  \rmd q\, q^{d-1}
\int_0^\pi \rmd\theta\,  \sin^{d-2}(\theta) ,
\end{equation}
where $S_d=2\pi^{d/2}/\Gamma(d/2)$
is the surface area of the $d$-dimensional
unit sphere.

We can rewrite \eref{sigma} up to second order in the external wave-vector $k$,
\begin{align}
	\hskip -0.1 cm \Sigma(\bi{k},0)=
	2\frac{\lambda^2 \Pi_0}{\mathcal{K}}  \frac{S_{d-1}}{\left(2\pi\right)^{d+1}}
	\int^\Lambda_{\Lambda /b} \rmd j\,  j^{d-3}   \int_0^\pi \rmd\theta\, \sin^{d-2}(\theta)
	\left[
	j \cos(\theta) I_{11} k+
	\left(\frac{1}{2} I_{11}+
	\mathcal{C}_j\cos^2(\theta)\left(I_{12}-2\Delta_j I_{21} \right)
	\right) k^2
	\right],
\label{sigma2}
\end{align}
where the integrals
\begin{equation}
\label{drg:ilm}
I_{lm}=\int_{-\infty}^{+\infty} \rmd z
\left(\Delta_j^2+z^2 \right)^{-l}
\left(\Delta_j+\rmi z\right)^{-m},
\end{equation}
are easily calculated through  the residue theorem as
$I_{11}=\pi/2\Delta_j^2$, $I_{12}=\pi/4\Delta_j^3$, and
$I_{21}=3\pi/8\Delta_j^4$ (taking into account
that $\Delta_j\ge 0$ for  $j\in [\Lambda/b,\Lambda]$).

Integrating \eref{sigma2} in $\theta$ using
\begin{align}
& \int_0^\pi \rmd\theta\, \sin^{d-2}(\theta)\cos(\theta) =0,  \\[5pt]
& \int_0^\pi \rmd\theta\, \sin^{d-2}(\theta)\cos^2(\theta) =
	\frac{1}{d}\int_0^\pi \rmd\theta\, \sin^{d-2}(\theta) , \\[5pt]
& K_d=\frac{S_{d-1}}{(2\pi)^d}\int_0^\pi \rmd\theta\, \sin^{d-2}(\theta) , \label{k_d}
\end{align}
we obtain
\begin{equation}
\Sigma(\bi{k},0)=\frac{\lambda^2\Pi_0 K_d}{4d\mathcal{K}^2}
\int_{\Lambda/ b}^\Lambda \rmd j\, \frac{j^{d-3}}{\Delta_j^3}
\left( d\Delta_j-2\mathcal{C}_j\right) k^2.
\end{equation}
Finally, we perform a perturbative expansion up to first order in
$1/b=\rme^{-\delta l}= 1-\delta l+\Or(\delta l^2)$, that leads to
\begin{equation}
\label{ax:sigma-renorm}
\Sigma(\bi{k},0)=
\frac{\lambda^2 \Pi_0 K_d}{4d} \frac{\Lambda^{d+2-2\mu}}
{\left(\mathcal{K}\Lambda^{2-\mu}+\nu\right)^3}
\left[ (d-2)\mathcal{K} \Lambda^{2-\mu}+(d-\mu)\nu \right]
\delta l\, k^2 .
\end{equation}

As for the coarse-graining of the noise variance,
we proceed as above using the symmetrized variables,
\begin{equation}
	\lim_{\omega \to 0} \left|G_0\left(\hat{q}\right)\right|^2 \left|G_0\left(\hat{k}-\hat{q}\right)\right|^2
	=\frac{1}{\left(\mathcal{K}j^2 \right)^4}
	\Big\{\left[\Delta_j+\mathcal{C}_j\cos(\theta)\, x+z^2\right]
 	\left[\Delta_j-\mathcal{C}_j\cos(\theta)\, x+z^2\right]
	\Big\}^{-1}
\sim
	\frac{\left(\mathcal{K}j^2\right)^{-4}}
	{\left(\Delta_j^2+z^2 \right)^2}.
\end{equation}
Putting this last result into the expansion of equation \eref{noise} up to
zero order in powers of $k$, we obtain
\begin{align}
\Phi(\bi{k},0) & =\frac{\lambda^2 \Pi^2_0}{(2\pi)^{d+1}}
	\int_{\Lambda/ b}^\Lambda \rmd j\,  j^{d-1}
	\int_{-\infty}^{+\infty} \rmd z\,
	\frac{j^4}{\left(\mathcal{K}j^2\right)^3} \frac{S_{d-1}}{(\Delta_j^2+z^2)^2}
	\int_0^\pi \rmd\theta\, \sin^{d-2}(\theta)
  =\frac{\lambda^2 \Pi_0^2 K_d}{4\mathcal{K}^3}\int_{\Lambda/ b}^\Lambda \rmd j\,
	\frac{j^{d-3}}{\Delta_j^3} \nonumber\\[7pt]
&	\sim \dfrac{\lambda^2 \Pi_0^2}{4}
	\frac{K_d\Lambda^{d+4-3\mu}}{\left(\mathcal{K}\Lambda^{2-\mu}+\nu\right)^3}\,
	\delta l.
\label{ax:noise-renorm}
\end{align}
The modified parameters are obtained from equation \eref{ax:sigma-renorm}
and equation \eref{ax:noise-renorm}
\begin{align}
&\nu^< = \nu, \qquad \lambda^< = \lambda, \nonumber\\[5pt]
&\mathcal{K}^< =\mathcal{K}
	\left\{
	1-\frac{\lambda^2\Pi_0 K_d}{4d}
	\frac{\Lambda^{d+2-2\mu}}
	{\left(\mathcal{K}\Lambda^{2-\mu}+\nu\right)^3}
	\left[ (d-2)\mathcal{K} \Lambda^{2-\mu}+(d-\mu)\nu
	\right]\delta l
	\right\},\nonumber\\[5pt]
&\Pi_0^< = \Pi_0\left\{1+
	\lambda^2 \Pi_0
	\frac{K_d\Lambda^{d+4-3\mu}}{4\left(\mathcal{K}\Lambda^{2-\mu}+\nu\right)^3}
	\delta l
	\right\}.
\end{align}
Now we can perform the rescaling according to the usual DRG transformation.
Proceeding as indicated in the main text, the parameter flow finally reads
\begin{align}
&\frac{\rmd\nu}{\rmd l}=\nu\left[z-\mu\right],\qquad
	\frac{d\lambda}{dl}=\lambda\left[\alpha+z-2\right],\nonumber \\[5pt]
&\frac{\rmd\mathcal{K}}{\rmd l}=\mathcal{K}
	\left\{
	z-2-\frac{\lambda^2\Pi_0 K_d}{4d}
	\frac{\Lambda^{d+2-2\mu}}{\left(\mathcal{K}\Lambda^{2-\mu}+\nu\right)^3}
	\left[ (d-2)\mathcal{K} \Lambda^{2-\mu}+(d-\mu)\nu
	\right]
	\right\}, \nonumber\\[5pt]
&\frac{\rmd\Pi_0}{\rmd l}=\Pi_0
	\left\{z-2\alpha-d+
	\lambda^2 \Pi_0
	\frac{K_d\Lambda^{d+4-3\mu}}{4\left(\mathcal{K}\Lambda^{2-\mu}+\nu\right)^3}
	\right\}. \label{flow:noloc}
\end{align}
The set of equations \eref{ecs:nu,lambda}-\eref{ec:D} in the main text
follows from \eref{flow:noloc} by imposing an unit wave-vector cut-off, i.e. $\Lambda=1$.

\section{Irrelevant stabilizing terms}
\label{ax:complex}

In this appendix we present the DRG calculations for a case of equation \eref{eq_gral}
in which the linear dispersion relation includes additional stabilizing terms (proportional to
$k^3$ and $k^4$) for 1+1 growth dimensions. We will see that the linear terms parametrized by
$\mathcal{N}\neq 0$ and exponent $n$ larger than $2$ are irrelevant with respect to a term
like $-k^2 h_{\bi{k}}$, and do not change the hydrodynamic scaling behavior of the equation.
This result can be generalized straightforwardly to the full family of nonlocal equations
considered in the main text, although for the sake of concreteness we provide the
detailed DRG calculations for the case $\mu=1$ only. This result is not surprising, since
these terms are linear, and simple dimensional analysis anticipates such results.
However, we have included these details for completeness, and to explicitly show how the non-trivial
fixed points that control the dynamics of equation \eref{eq_gral} already emerge when considering only
the ``surface tension'' term proportional to $k^2$, as done in sections \ref{DRGflow} and \ref{results},
and appendix \ref{ax:nonlocal}.

To begin with, we consider a linear
dispersion relation of the form
\begin{equation}
\sigma_k=-C_1 |k| -C_2 k^2 - C_3 |k|^3 -C_4 k^4,
\end{equation}
so that in the coarse-grained propagator we have to retain
powers of wave-vector up to fourth order (note that here $k$ is the
one-dimensional wave-vector and {\em not} its modulus).
Later on we will consider the choices $C_3=0$ and $C_4=0$ as particular cases.
We proceed as in appendix \ref{ax:nonlocal}, using the standard symmetrization and
considering one by one the various contributions to the integrand in
equation \eref{sigma}. The first contribution is readily obtained as
\begin{equation}
k(q-k) q^2=k\left( j+\frac{k}{2}\right)\left( j^2-\frac{k^2}{4}\right)
=j^3k+\frac{1}{2}j^2 k^2-\frac{1}{4} j k^3-\frac{1}{8}k^4.
\label{B:vertex}
\end{equation}
The product of the first two bare propagator factors in
equation \eref{sigma} is evaluated considering that $|k|\ll|j|$:
\begin{eqnarray}
 & &|j+k/2| = |j|+{\rm sgn}(j)\,  k/2, \\[7pt]
& &|j+k/2|^3 = |j|^3+3 j^2 {\rm sgn}(j)\, k/2+3 |j| k^2/4+{\rm sgn}(j)\, k^3/8.
\end{eqnarray}
Using these identities, we are ready to expand the first product of two bare propagators
appearing in \eref{sigma} up to third order
\begin{align}
\lim_{\omega \to 0} \left|G_0\left(\hat{q}\right)\right|^2 & =
	\frac{1}{\left(C_4 j^4\right)^2}
	\left[\left(\Delta_j+a_j x+b_j x^2+c_j x^3\right)^2+z^2
	\right]^{-1} \nonumber \\[5pt]
& \sim
    \frac{1}{\left(C_4 j^4\right)^2}
    \frac{1}{\Delta_j^2+z^2}\Bigg\{1-\left(\frac{2a_j\Delta_j}{\Delta_j^2+z^2}\right) x+
    \frac{1}{\Delta_j^2+z^2}
    \left[ \frac{(2 a_j \Delta_j)^2}{\Delta_j^2+z^2}-2 b_j \Delta_j-a_j^2
		\right] x^2 \nonumber \\[7pt]
& \hskip 1 cm +\frac{1}{\Delta_j^2+z^2}
    \left[\frac{4 a_j \Delta_j}{\Delta_j^2+z^2}\left(2 b_j\Delta_j+a_j^2\right)-
    2(c_j \Delta_j+a_j b_j)-
    \frac{\left(2 a_j \Delta_j\right)^3}{\left(\Delta_j^2+z^2\right)^2}
    \right] x^3\Bigg\},
\label{bare:1}
\end{align}
where
\begin{align}
& \Delta_j=  {\displaystyle 1+\frac{C_3}{C_4 |j|}+\frac{C_2}{C_4 j^2}+\frac{C_1}{C_4 |j|^3},} \\[5pt]
& a_j=  {\displaystyle 2+\frac{3}{2}\frac{C_3}{C_4 |j|}+\frac{C_2}{C_4 j^2}+
    \frac{C_1}{2C_4 |j|^3},} \\[5pt]
& b_j=  {\displaystyle \frac{3}{2}+\frac{3}{4}\frac{C_3}{C_4 |j|}+\frac{C_2}{4C_4 j^2},}\\[5pt]
& c_j=  {\displaystyle \frac{1}{2}+\frac{C_3}{8 C_4 |j|},}
\label{funzc}
\end{align}
and we have introduced two new variables: the non-dimensional expansion parameter
$x=k/j$, and the rescaled frequency variable $z=\Omega/(C_4 j^4)$. The
same procedure as applied to the remaining factor in the integrand of equation \eref{sigma}
leads to
\begin{align}
\lim_{\omega \to 0} G_0\left(\hat{k}-\hat{q}\right) & =
	\frac{1}{C_4 j^4} \left(\Delta_j-a_j x+  b_j x^2-c_j x^3+\rmi z\right)^{-1}
     \nonumber \\[5pt]
&  \sim \frac{1}{C_4j^4}\frac{1}{\Delta_j+\rmi z} \Bigg\{1+\left(\frac{a_j}{\Delta_j+\rmi z}\right) x+
	\frac{1}{\Delta_j+\rmi z}\left( \frac{a_j^2}{\Delta_j+\rmi z}-b_j\right) x^2
&\nonumber \\[5pt] & \hspace{2cm}
+\frac{1}{\Delta_j+\rmi z}\left[c_j- \frac{2a_j b_j}{\Delta_j+\rmi z}+
	\frac{a_j^3}{\left(\Delta_j+\rmi z\right)^2}\right] x^3\Bigg\}.
\label{bare:1'}
\end{align}
After multiplication of \eref{bare:1} and \eref{bare:1'}, we can integrate out the
rescaled frequency after the change of variable $\Omega\to z=\Omega/(C_4 j^4)$.
Note that the frequency integral in the new variable is
\begin{equation}
\int_{-\infty}^{+\infty}\, \rmd\Omega\to
C_4 j^4\,\int_{-\infty}^{+\infty}\, \rmd z.
\end{equation}
We can use the compact notation introduced before for the integrals $I_{lm}$ involved in the
evaluation of the coarse-grained propagator [see equation \eref{drg:ilm}]. After some algebra,
we obtain the contribution to equation \eref{sigma} as
\begin{align}
\int_{-\infty}^{+\infty} \rmd\Omega\,  \Bigg|G_0
    \left(\hat{j}+\frac{\hat{k}}{2}\right)
    \Bigg|^2   \, &
    G_0\left(\frac{\hat{k}}{2}-\hat{j}\right)\sim
	\frac{1}{\left(C_4 j^4\right)^2} \,
    \Bigg\{ I_{11}+a_j \left(I_{12}-2\Delta_j I_{21}\right) x \nonumber\\[5pt]
& \hskip -2cm + \Bigg[
    a_j^2 I_{13}-b_j I_{12}-2a_j^2\Delta_j I_{22}+
    \left(2a_j\Delta_j\right)^2 I_{31}-
    2b_j\Delta_j I_{21} -a_j^2  I_{21}
    \Bigg] x^2  \nonumber\\[5pt]
& \hskip -2cm  + \Bigg[
    c_j I_{12}-2a_j b_j I_{13}+a_j^3 I_{14}-2a_j\Delta_j
    \left(a_j^2 I_{23}-b_j I_{22} \right)
 +\left[
	\left(2a_j\Delta_j\right)^2 I_{32}-2b_j\Delta_j I_{22}-a_j^2 I_{22}	
    \right] a_j \nonumber \\[5pt]
& \hskip -1cm  +    4a_j\Delta_j \left( 2b_j\Delta_j+a_j^2\right) I_{31}-
     2\left( c_j \Delta_j+a_j b_j\right) I_{21}-
     \left(2a_j\Delta_j\right)^3 I_{41}
     \Bigg] x^3	
     \Bigg\}.
  \label{B:prop}
\end{align}
The $I_{lm}$ integrals are easily calculated using the residue theorem, giving
the contributions listed below:
\begin{equation*}
\left.
\begin{array}{lllll}
I_{11}=\pi/ 2\Delta_j^2, & I_{12}=\pi /  4\Delta_j^3 ,
& I_{13}=\pi/ 8\Delta_j^4 , &  I_{14}=\pi/ 16\Delta_j^5 , &
I_{21}=3\pi /  8\Delta_j^4 , \\[10pt]
I_{22}= \pi /  4\Delta_j^5 , & I_{23}=5\pi / 32\Delta_j^6 , &
I_{31}=5\pi / 16\Delta_j^6 , & I_{32}=15\pi /  64\Delta_j^7 , &
I_{41}=35\pi /  128\Delta_j^8 .
\end{array}
\right.
\end{equation*}
Using these relation and multiplying contributions \eref{B:vertex},
\eref{B:prop} together, we can write $\Sigma$ as sum of four terms,
\begin{equation}
\Sigma(k,0)=  \frac{\lambda^2\Pi_0}{4\pi}
\int_> \rmd j\  \left[f_1(j) k+f_2(j) k^2+f_3(j) k^3+f_4(j)k^4 \right],
\label{sigma1:2}
\end{equation}
where
\begin{align}
\label{funzf1}
& f_1(j)=  \frac{1}{\left(C_4\Delta_j\right)^2 j^5},
	\qquad f_2(j)=  \frac{1}{\left(C_4\Delta_j j^3\right)^2}
    \left( \frac{1}{2}-\frac{a_j}{\Delta_j}\right), \\[5pt]
& f_3(j)=   \frac{1}{\left(C_4\Delta_j\right)^2 j^7}
    \left[
    \frac{1}{\Delta_j}\left( \frac{a_j^2}{\Delta_j}-2b_j\right)-
    \frac{a_j}{2\Delta_j}-\frac{1}{4}
    \right],  \\[5pt]
& f_4(j)=  \frac{1}{\left(C_4\Delta_j j^4\right)^2}
    \left[
    \frac{1}{2\Delta_j}\left( \frac{a_j^2}{\Delta_j}-2b_j\right)+
    \frac{a_j}{4\Delta_j}-\frac{1}{8}-\frac{1}{\Delta_j}
    \left( c_j-\frac{3}{\Delta_j}a_j b_j+\frac{a_j^3}{\Delta_j^2}\right)
    \right].
\label{funzf4}
\end{align}
Considering the parity of the various functions of $j$ that appear,
we can simplify expression \eref{sigma1:2}. In fact,  $a_j$, $b_j$, $c_j$,
and $\Delta_j$ in \eref{funzf1}-\eref{funzf4} are even in $j$: for this reason the
functions $f_1$, $f_3$ are odd, whereas $f_2$ and $f_4$ are also even.
The momentum shell is a symmetric interval around $j$ equal to zero,
hence the integrals of $f_1(j)$ and $f_3(j)$ vanish.

Another important point to address is the contribution of the shell to the renormalization
of the propagator. This calculation is only required when $k^4$ (or higher order) terms are
present, because the lowest contribution from the shell occurs, precisely, at that order.
Writing explicitly the integration limits in the positive part
of the momentum integral, and expanding for small $k$
\cite{Alonso-Sanchez:2000,Berera:2010},
\begin{align}
\int_>^+  \rmd j\,  f(j) & =  \frac{1}{2}
	\left[\,
	\int\limits^{\Lambda-k/2}_{\Lambda/ b+k/2}\hskip -0.25cm \rmd j \, f(j)  +
	\int\limits^{\Lambda+k/2}_{\Lambda/ b-k/2}\hskip -0.25cm \rmd j\,  f(j)
	\right] \nonumber \\[5pt]
& \sim
	\int_{\Lambda/ b}^\Lambda \rmd j \, f(j)
 + \frac{k}{4}
 	\left[
	-f\left(\Lambda-\frac{k}{2}\right)-f\left(\frac{\Lambda}{b}+\frac{k}{2}\right)
	+f\left(\Lambda+\frac{k}{2}\right)+f\left(\frac{\Lambda}{b}-\frac{k}{2}\right)
	\right]\Bigg|_{k=0}  \label{bare:2}\nonumber  \\[5pt]
& \hskip 3 cm+ \frac{k^2}{16}
	\left[
	f'\left(\Lambda-\frac{k}{2}\right)-f'\left(\frac{\Lambda}{b}+\frac{k}{2}\right)
	+f'\left(\Lambda+\frac{k}{2}\right)-f'\left(\frac{\Lambda}{b}-\frac{k}{2}\right)
	\right]\Bigg|_{k=0}\hskip -0.5cm+\dots,
\end{align}
where $f$ is any function of $j$ (the symbol $+$ in the integral means
that the integration interval is restricted to positive $j$). With this simple result,
we see that if a function $f_i$ in equation \eref{sigma1:2} contributes at order $n$
in powers of $k$, then through the expansion of the shell it also contributes at every order
$n+2m$, with $m$ a positive integer. Thus, in our calculation we only need to consider
the contribution of $f_2$ to the $k^4$ term as induced by the shell,
\begin{eqnarray}
{\displaystyle \int_>^+ \rmd j\, f_2(j)\sim
F_2(j)\Big|_{\Lambda /b}^\Lambda+ } &
{\displaystyle \frac{k^2}{8}
    \left[ f_2'(\Lambda)-f_2'(\Lambda/b)\right]
\sim \left[f_2(\Lambda)+
\frac{k^2}{8} f_2''(\Lambda)\right] \Lambda \delta l.}
\end{eqnarray}
Here, the function $F_2$ is the primitive of $f_2$.
We now introduce three new coupling variables that help us to write \eref{sigma1:2}
in a more compact way,
\begin{equation}
A=\frac{C_1}{C_4 |j|^3},
    \qquad F=\frac{C_2}{C_4 j^2}, \qquad B=\frac{C_3}{C_4 |j|},
\end{equation}
so that we can express $f_2$, its second derivative, and $f_4$ using these
variables,
\begin{align}
&  f_2(j) =  -\frac{1}{2\left(C_4 j^3\right)^2}
  	\frac{3+2 B+F}{\left(1+A+B+F\right)^3}, \\[5pt]
& f_4(j)=  \frac{1}{8\left(C_4 j^4\right)^4}
 	 \big[
  	 11+40A+27B+29F+2A^2+19B^2+11F^2+47AB+21AF +35BF \nonumber \\[5pt]
& \hskip 3 cm  +\left(2B+F\right)
     \left(
        A^2+2B^2+F^2+7AB+3AF+4BF
     \right) \left(1+A+B+F\right)^{-5}, \\[5pt]
& f''_2(j)=  -\frac{1}{\left(C_4 j^4\right)^4}
	\big[
	63-90A+119B+27F+9A^2+82B^2+15F^2-95AB-27AF+55BF+3F^3+ 2A^2B \nonumber \\[5pt]
 &\hskip 3 cm	-3AF^2+20B^3+14BF^2
	-26AB^2-17ABF+25B^2F
  	\big]
	\left(1+A+B+F\right)^{-5}.
\end{align}
Considering that the shell is composed of  positive and  negative wave-vectors
we have to count twice the previous contribution,
\begin{equation}
\Sigma(k,0)=  {\displaystyle \frac{\lambda^2\Pi_0}{4\pi}
\left[
2 f_2(\Lambda) k^2+
\left( 2f_4(\Lambda)+\frac{1}{4} f''_2(\Lambda) \right) k^4
\right] \Lambda \delta l=
\left(
\Sigma_2 k^2 +\Sigma_4 k^4
   \right) \delta l,}
\label{sigma3}
\end{equation}
where $\Sigma_2$ and $\Sigma_4$ are
\begin{align}
&\Sigma_2 = -\frac{\lambda^2\Pi_0}{4\pi}
	\frac{(3+2B+F)}{C_4^2\Lambda^5\left(1+A+B+F\right)^3}, \label{cf:sigma2} \\[5pt]
& \Sigma_4 = -\frac{\lambda^2\Pi_0}{16\pi C_4^2\Lambda^7}
	\Big[
	    52-130A+92B-2F+7A^2+63B^2+4F^2+ 16B^3+2F^3 -142AB -48AF \nonumber \\[5pt]
&\hskip  3 cm +20BF-30ABF-40AB^2-6AF^2
	+8BF^2-A^2F+15B^2F
	    \Big]
  \left(1+A+B+F\right)^{-5}.
\label{cf:sigma4}
\end{align}

In our DRG program, the next step is calculating the contributions arising from the coarse-graining
of the noise variance. We start by considering equation
 \eref{noise} for $d=1$. On the other hand,
\begin{equation}
\left[q (k-q)\right]^2=\left(\frac{k^2}{4}-j^2 \right)^2=
j^4-\frac{1}{2}j^2 k^2+\frac{1}{16} k^4,
\label{momentnoise}
\end{equation}
while the first factor due to  the bare propagator has been  already obtained,
see equation \eref{bare:1}. The other contribution is exactly the square
of the absolute value of equation \eref{bare:2},
\begin{align}
\lim_{\omega\to 0}
	\left|G_0\left(\frac{\hat{k}}{2}-\hat{j}\right)\right|^2 & \sim
	\frac{\left(\Delta_j^2+z^2\right)^{-1}}{\left(C_4 j^4\right)^2}
	 \Bigg\{1+\frac{2a_j\Delta_j}{\Delta_j^2+z^2} x
	+
 	\left[ \frac{(2 a_j \Delta_j)^2}{\left(\Delta_j^2+z^2\right)^2}-
	\frac{2 b_j \Delta_j-a_j^2}{\Delta_j^2+z^2}
	 \right] x^2  \nonumber\\[5pt]
&\hskip 3 cm
+
	\left[  2\frac{c_j\Delta_j+a_j b_j}{\Delta_j^2+z^2}
	 +\frac{\left(2 a_j \Delta_j\right)^3}{\left(\Delta_j^2+z^2\right)^3}
    	 -\frac{4 a_j \Delta_j}{\left(\Delta_j^2+z^2\right)^2}\left(2 b_j\Delta_j+a_j^2\right)
	\right] x^3
\Bigg\}.
\end{align}
The coarse-graining of the non-conserved noise involves only
the zeroth order term of this expansion, hence
\begin{equation}
\Phi(k,0)=\lambda^2 \Pi_0^2 \int_>  \frac{\rmd j}{2\pi}
\int\frac{\rmd z}{2\pi}\, j^4 \left(C_4 j^4\right)^{-3}\,
\left(\Delta_j^2+z^2\right)^{-2}.
\label{phi1}
\end{equation}
After integration in $z$
and, due to the parity of the functions in \eref{phi1}, we can finally write the coarse-grained noise variance
in the following form,
\begin{equation}
\Phi(k,0)=\frac{\lambda^2 \Pi_0^2}{4\pi C_4^3}\int^\Lambda_{\Lambda/ b}  \,
\frac{\rmd j}{j^8 \Delta_j^3}=\frac{\lambda^2 \Pi_0^2}{4\pi C_4^3}
\frac{\delta l}{\Lambda^7 \Delta_j^3}.
\end{equation}

As seen in the main text, there is no renormalization for the
KPZ vertex for any dispersion relation that can be written as a polynomial in $k$,
at least within our one-loop expansion. Now we are in a position to apply the usual
rescaling, and to calculate the renormalized parameters
\begin{eqnarray}
C_1^< &=& C_1, \qquad C_3^< = C_3, \label{renorm:mg} \\[5pt]
C_2^< &=& C_2
    \left[
    1+ \lambda^2\Pi_0 \frac{P_2(A,B,F)}{4\pi C_2 C_4^2\Lambda^5} \delta l
    \right], \label{renormnu} \\[5pt]
C_4^< &=& C_4
 \left[
 1+ \lambda^2\Pi_0\frac{P_4(A,B,F)}{16\pi C_4^3\Lambda^7} \delta l
 \right], \\[5pt]
\Pi_0^< &=& \Pi_0 \left[
 1+ \frac{\lambda^2 \Pi_0}{4\pi C_4^3}
 (1+A+B+F)^{-3}
 \frac{\delta l}{\Lambda^7}
 \right],
\label{renormD}
\end{eqnarray}
where the functions $P_2$ and $P_4$ are
\begin{align}
& P_2 = \frac{3+2B+F}{\left(1+A+B+F\right)^3}, \\[5pt]
& P_4 = \Big[
  	  52-130A+92B-2F+7A^2+63B^2+4F^2+ 16B^3+2F^3-142AB \nonumber \\[5pt]
 & \hskip 1 cm -48AF+20BF
 	-30ABF-40AB^2-6AF^2+8BF^2-A^2F+15B^2F
	    \Big]    \left(1+A+B+F\right)^{-5}.
\end{align}
In the next two sections we show that for a dispersion relation as the one appearing
in equation \eref{eq_gral} with $m=2$, additional relaxation terms with $n>2$
do not change the hydrodynamical properties of the system as obtained in appendix \ref{ax:nonlocal}.

\subsection{Irrelevance of the $k^3$ term}
\label{drg:k3}

As a first case we consider the dispersion relation that is
a third order polynomial in $|k|$,
$\sigma_k=-C_1|k|-C_2 k^2-C_3|k|^3$. The ensuing evolution equation
is an interpolation of the sMS and MSKPZ equations.
Using the results already obtained, we can write the RG flow,
\begin{eqnarray}
&{\displaystyle \frac{\rmd C_1}{\rmd l}}& = C_1 \left[ z-1\right],\qquad
\frac{\rmd C_3}{\rmd l} = C_3 \left[ z-3\right], \\[5pt]
&{\displaystyle \frac{\rmd C_2}{\rmd l}}& = C_2 \left[z-2+
\frac{\lambda^2 \Pi_0\Lambda^2}{4\pi C_2}
\left(\frac{2C_3 \Lambda+C_2}
{\left(C_1+C_2\Lambda+C_3 \Lambda^2\right)^3}
\right)
\right],\label{ec_C2}\\[5pt]
&{\displaystyle \frac{\rmd \lambda}{\rmd l}}& = \lambda \left[ \alpha+z-2\right], \\[5pt]
&{\displaystyle \frac{\rmd \Pi_0}{\rmd l}}& = \Pi_0     \left[
z-2\alpha-1+
\frac{\lambda^2 \Pi_0 \Lambda^2}{4\pi}
\left(C_1+C_2\Lambda+C_3\Lambda^2\right)^{-3}
\right].
\end{eqnarray}
From equation \eref{ec_C2}, note that, even if the bare ``surface tension'' coefficient $C_2(l=0)=0$,
a non-zero value for $C_2$ will be generated by the RG flow, provided $C_3(l=0)\neq 0$.
Introducing the coupling variables
\begin{equation}
a_1=\frac{C_1}{C_3\Lambda^2},
\qquad
f_1=\frac{C_2}{C_3\Lambda},
\qquad
g_1=\frac{\lambda^2 \Pi_0}{4\pi C_3^3\Lambda^4},
\end{equation}
the resulting flow reads
\begin{eqnarray}
&{\displaystyle \frac{\rmd a_1}{\rmd l}}& = 2 a_1,\label{a3:3} \\[5pt]
&{\displaystyle \frac{\rmd f_1}{\rmd l}}& = f_1
\left[1+\frac{g_1(2+f_1)}
{f_1\left(1+a_1+f_1\right)^3}
\right],\label{f3:3}\\[5pt]
&{\displaystyle \frac{\rmd g_1}{\rmd l}}& = g_1 \left[
4+ \frac{g_1}{\left(1+a_1+f_1\right)^3}
\right]. \label{g3:3}
\end{eqnarray}
In this case the condition $\Delta_\Lambda\ge 0$ that we
have used in order to calculate the integrals $I_{lm}$ requires $1+a_1+f_1>0$.
In fact, when the dispersion relation is a third order polynomial, the coefficient
$C_3$ has to be positive and
\begin{equation}
\label{sigma:c3}
\sigma_\Lambda=-C_3 \Lambda^3(1+a_1+f_1),
\end{equation}
has to be negative in order to have a well-defined short distance behavior,
hence the positive sign required for the parenthesis in \eref{sigma:c3}.
For this reason we do not consider fixed points of equations \eref{a3:3}-\eref{g3:3} that do not satisfy this restriction.
\begin{table}[h!]
\begin{center}
\begin{tabular}{|c|c|c|c|c|c|c|c|c|c|}
\hline
\T Name & $a_1$  & $f_1$ & $g_1$ & $z$ & $\alpha$ & $\beta$ &
 	$\lambda_1$ & $\lambda_2$ & $\lambda_3$ \\[3pt]
\hline\hline
\T Fp$_{3,0}$ & $0$ & $0$ & $0$ & $3$ & $1$ & $1/3$ &
 	$2$ & $1$ & $4$ \\[3pt]
\hline
\end{tabular}
\end{center}
\caption{Fixed point of \eref{a3:3}-\eref{g3:3} in the $(a_1,f_1,g_1)$ plane.}
\label{fp-table:3}
\end{table}
The only admissible fixed point in the $(a_1,f_1,g_1)$ parameter space, Fp$_{3,0}$,
is reported in table \ref{fp-table:3}.
This fixed point is associated with the linear limit in
which $C_1=C_2=\lambda=0$. All the eigenvalues of the stability matrix associated to
Fp$_{3,0}$ are positive, hence this fixed point is linearly unstable in every direction.

A second set of couplings that can be considered is
\begin{equation}
a_2=\frac{C_1}{C_2\Lambda},
\qquad
f_2=\frac{C_3\Lambda}{C_2},
\qquad
g_2=\frac{\lambda^2 \Pi_0}{4\pi C_2^3\Lambda},
\end{equation}
their flow reading
\begin{eqnarray}
&{\displaystyle \frac{\rmd a_2}{\rmd l}}& = a_2
\left[
1- \frac{g_2(2f_2+1)}{\left(1+a_2+f_2\right)^3}
\right], \label{C3:a2}
\\[7pt]
&{\displaystyle \frac{\rmd f_2}{\rmd l}}& = -f_2
\left[
1+ \frac{ g_2(2f_2+1)}{\left(1+a_2+f_2\right)^3}
\right], \label{C3:f2}\\[7pt]
&{\displaystyle \frac{\rmd g_2}{\rmd l}}& = g_2
\left[
1-2\frac{g_2(3f_2+1)}{\left(1+a_2+f_2\right)^3}
\right]. \label{C3:g2}
\end{eqnarray}
The corresponding fixed points and the eigenvalues of their stability matrices are listed
in table \ref{fp-table:4},
whereby they are seen to correspond to the EW ($C_1=C_3=\lambda=0$)
and KPZ ($C_1=C_3=0$) fixed points. Note they both have stable and unstable directions.
\begin{table}[h!]
\begin{center}
\begin{tabular}{|c|c|c|c|c|c|c|c|c|c|}
\hline
\T Name &  $a_2$  & $f_2$ & $g_2$ & $z$ & $\alpha$ & $\beta$
	& $\lambda_1$ & $\lambda_2$ & $\lambda_3$ \\[3pt]
\hline\hline
\T EW & $0$ & $0$ & $0$ & $2$ & $1/2$ & $1/4$
	& $1$ & $-1$ & $1$\\[3pt]
KPZ & $0$ & $0$ & $1/2$ & $3/2$ & $1/2$ & $1/3$
	& $1/2$ & $-3/2$ & $-1$ \\[3pt]
\hline
\end{tabular}
\end{center}
\caption{Fixed point of \eref{C3:a2}-\eref{C3:g2} in the $(a_2,f_2,g_2)$ plane.}
\label{fp-table:4}
\end{table}

A third convenient set of couplings is
\begin{equation}
a_3=\frac{C_3\Lambda^2}{C_1},
\qquad
f_3=\frac{C_2\Lambda}{C_1},
\qquad
g_3=\frac{\lambda^2 \Pi_0\Lambda^2}{4\pi C_1^3},
\end{equation}
their flow equations reading
\begin{eqnarray}
&{\displaystyle \frac{\rmd a_3}{\rmd l}}& = -2 a_3, \label{C3:a1}\\[7pt]
&{\displaystyle \frac{\rmd f_3}{\rmd l}}& =
\frac{g_3(2a_3+f_3)}{\left(1+a_3+f_3\right)^3}-f_3,
\label{C3:f1}\\[7pt]
&{\displaystyle \frac{\rmd g_3}{\rmd l}}& = g_3
\left[
\frac{g_3}{\left(1+a_3+f_3\right)^3}-2
\right].\label{C3:g1}
\end{eqnarray}
In this parameter space, the Smooth and the Galilean
fixed points that have been discussed in the main text
arise naturally, with features as in table
\ref{fp-table:5}.
Note, the Smooth fixed point is completely stable in every
direction.
\begin{table}[h!]
\begin{center}
\begin{tabular}{|c|c|c|c|c|c|c|c|c|c|}
\hline
\T Name &  $a_3$  & $f_3$ & $g_3$ & $z$ & $\alpha$ & $\beta$
	& $\lambda_1$ & $\lambda_2$ & $\lambda_3$\\[3pt]
\hline\hline
\T Smooth & $0$ & $0$ & $0$ & $1$ & $0$ & $0$ &
	$-2$ & $-1$ & $-2$ \\[3pt]
Galilean & $0$ & $0$ & $2$ & $1$ & $1$ & $1$ &
	$-2$ & $1$ & $2$\\[3pt]
\hline
\end{tabular}
\end{center}
\caption{Fixed point of \eref{C3:a1}-\eref{C3:g1} in the $(a_3,f_3,g_3)$ plane.}
\label{fp-table:5}
\end{table}


Using the values for exponents $\alpha$ and $z$ as calculated for
each fixed point, we can show the irrelevance of $C_3$ as compared to $C_2$.
After rescaling, equation \eref{eq_gral} reads
\begin{eqnarray}
\label{k3:rescaled}
\partial_t h_k =  b^{z-1}\left[ \left( -C_1 |k|-\frac{C_2}{b} k^2-
	\frac{C_3}{b^2} |k|^3\right) h_k
+\frac{\lambda}{2} b^{\alpha-1}
\mathcal{F}\left[ (\nabla h)^2\right]
+b^{(1-z-2\alpha)/2}\eta_k\right] ,
\end{eqnarray}
and we consider now the case of $C_2,C_3>0$. Clearly, from linear analysis the
ratio between the stabilizing terms is $C_2 b/C_3$, so that the surface tension term becomes
more relevant than the $|k|^3$ term. Moreover, the sign of the eigenvalues of the linear stability matrix
at the EW fixed point
shows that this fixed point is stable along the direction
$f_2=C_3\Lambda/C_2$.
The same situation occurs for the KPZ fixed point ($C_1=0$), in fact
it has two negative eigenvalues associated to the
two stable directions $f_2$ and $g_2$. For the Galilean fixed point, the only stable direction
is $a_1=C_3\Lambda^2 / C_1$, while for the Smooth fixed point every direction is stable.
The linear stability of these fixed points shows that Fp$_{3,0}$ can be reached only if
$C_1=C_2=\lambda=0$ and is unstable in every direction,
hence we can conclude that the $|k|^3$ term is irrelevant
as compared to $k^2$ (for $C_2,C_3>0$). We would need this term only in case of
a negative surface tension, i.e., $C_2<0$, in order to have
a dispersion relation that is well posed in the $k\to\infty$ limit.

\subsection{Irrelevance of the $k^4$ term}
\label{drg:k4}

In the last part of this appendix we show that the
parameter $C_4$ is irrelevant when the dispersion relation
contains a relaxation term with an exponent smaller than
four. Let us consider a dispersion relation of the form
\mbox{$\sigma_k=-C_1 |k|-C_2 k^2-C_4 k^4$}: in this case,
expressions \eref{cf:sigma2} and \eref{cf:sigma4} reduce to
\begin{align}
&\Sigma_2 = -\frac{\lambda^2\Pi_0}{4\pi C_4^2 \Lambda^5 }
	\left[\frac{3+F}{\left(1+A+F\right)^{3}}\right], \\[5pt]
&\Sigma_4 = -\frac{\lambda^2\Pi_0}{16\pi C_4^2\Lambda^7}
    	\Big[52  -130A+4F^2-2F+2F^3-48 AF+7 A^2 -6 AF^2-FA^2\Big]   \left(1+A+F\right)^{-5}.
\end{align}
Hence, after rescaling,
\begin{align}
& \frac{\rmd C_1}{\rmd l} = C_1\left[ z-1\right], \qquad  \frac{\rmd \lambda}{\rmd l}
	= \lambda\left[ \alpha+z-2\right] \\[5pt]
& \frac{\rmd C_2}{\rmd l} = C_2\left[ z-2+\frac{\lambda^2\Pi_0\Lambda^2}{4\pi C_2}
	\frac{3C_4\Lambda^2+C_2}{\big(C_1+C_2\Lambda+C_4\Lambda^3\big)^3}\right],
\label{ec_C2'} \\[5pt]
& \frac{\rmd C_4}{\rmd l} = C_4  \bigg[ z-4+\frac{\lambda^2\Pi_0}{16\pi C_4}
	\Big(  52 C_4^3 \Lambda^8  - 130 C_1 C_4^2\Lambda^5 -2 C_2 C_4^2\Lambda^6 +
	7C_1^2 C_4\Lambda^2+4 C_2^2 C_4 \Lambda^4
	 \nonumber \\[5pt]
	& \hskip 2 cm
	 + 2 C_2^3 \Lambda^2 - 48 C_1 C_2 C_4 \Lambda^3 - 6 C_1 C_2^2\Lambda
	-C_1^2 C_2\Big)   \left(C_1+C_2 \Lambda+C_4\Lambda^3\right)^{-5} \bigg], \\[5pt]
& \frac{\rmd \Pi_0}{\rmd l} = \Pi_0\left[ z-2\alpha-1+ \frac{\lambda^2\Pi_0\Lambda^2}{4\pi}
\left( C_1+C_2\Lambda+C_4\Lambda^3\right)^{-3}\right].
\end{align}
Analogously to the result obtained in the previous section, equation \eref{ec_C2'} implies that, even if the bare ``surface tension'' coefficient $C_2(l=0)=0$, a non-zero value will be generated for $C_2$ by the RG flow, provided $C_4(l=0)\neq 0$.
Also in parallel to our previous computation, we consider a first set of couplings
\begin{equation}
a_1=\frac{C_1}{C_4\Lambda^3},
\qquad
f_1=\frac{C_2}{C_4 \Lambda^2},
\qquad
g_1=\frac{\lambda^2 \Pi_0}{4\pi C_4^3\Lambda^7},
\end{equation}
for which the renormalization flow is
\begin{align}
&\frac{\rmd a_1}{\rmd l} =
a_1 \Bigg[ 3-\frac{g_1}{4}
    \Big(
    52  -130a_1+4f_1^2-2f_1+2f_1^3-48 a_1 f_1+7 a_1^2
    -6 a_1 f_1^2-f_1 a_1^2\Big)
\left(1+a_1+f_1\right)^{-5}
    \Bigg],  \label{k4:a4} \\[5pt]
&
\frac{\rmd f_1}{\rmd l} = 2f_1+
	\frac{g_1}{4}
	\Big(
	12-24 f_1+22 f_1^2-2f_1^4+24 a_1+12 a_1^2+56a_1 f_1^2+162 a_1 f_1
    -3 f_1 a_1^2+a_1^2 f_1^2+6 a_1 f_1^3\Big) \nonumber \\[5pt]
	&\hskip 3 cm \times   \left(1+a_1+f_1\right)^{-5}, \\[5pt]
&\frac{\rmd g_1}{\rmd l} = g_1
    \Bigg\{
    7+  \frac{g_1}{\left(1+a_1+f_1\right)^3}
	\Bigg[ 1-\frac{3}{4}
		\Big( 52-130a_1+4f_1^2-2f_1+2f_1^3-48 a_1 f_1+7 a_1^2-6 a_1 f_1^2-f_1 a_1^2\Big)
		\nonumber \\[5pt]
   &\hskip  5 cm  \times \left(1+a_1+f_1\right)^{-2}
	\Bigg]
    \Bigg\}. \label{k4:g4}
\end{align}
\begin{table}[h!]
\begin{center}
\begin{tabular}{|c|c|c|c|c|c|c|}
\hline
\T Name & $a_1$  & $f_1$ & $g_1$ & $z$ & $\alpha$ & $\beta$  \\[3pt]
\hline\hline
\T Fp$_{4,0}$ & $0$ & $0$ & $0$ & $4$ & $3/2$ & $3/8$  \\[3pt]
KPZ  & $0$ & $13.1868$ & $1064.4$ & $1.54$ & $0.46$ & $0.299$ \\[3pt]
Fp$_{4,1}$ & $7.17$ & $-6$ & $20.4$ & $1$ & $1$ & $1$ \\[3pt]
\hline
\end{tabular}
\end{center}
\caption{Fixed points of \eref{k4:a4}-\eref{k4:g4} in the $(a_1,f_1,g_1)$ plane.}
\label{fp-table:6}
\end{table}
The different fixed points in the $(a_1,f_1,g_1)$ parameter space
are given in table \ref{fp-table:6}, where the coordinates of the KPZ fixed point have been
determined using the only admissible real solution, $x_0$, of equation $x^4-8x^3-63x^2-68x-42=0$.
Hence, the KPZ fixed point is located at the point $\left(0,x_0,P(x_0)\right)$, where
$P(x)=(686+969 x+846x^2-25x^3)/97$. Likewise, the Fp$_{4,1}$ fixed point has been obtained
solving $7y^2+2y-374=0$ and using the only admissible solution $y_0$ for the components of
vector $(y_0,-6,Q(y_0))$, where $Q(y)=(13014 y-92286)/49$. The Fp$_{4,0}$ fixed point is associated
with the linear Molecular Beam Epitaxy (MBE) equation ($C_1=C_2=\lambda=0$) \cite{Barabasi:1995},
while Fp$_{4,1}$ has the same exponents as (but does not coincide with) the Galilean fixed point.
The stability matrices of these points are dense, and we report their eigenvalues in table \ref{fp-table:7}.
\begin{table}[h!]
\begin{center}
\begin{tabular}{|c|c|c|c|c|}
\hline
\T  Name  & $\lambda_1$ & $\lambda_2$ & $\lambda_3$ & Stability  \\[3pt]
\hline\hline
\T Fp$_{4,0}$ & $3$ & $2$ & $7$ & Unstable \\[3pt]
KPZ & $0.54$ & $-2.90$ & $-1.06$ & Saddle \\[3pt]
Fp$_{4,1}$ & $88.4$ & $7.33$ & $-0.24$ & Saddle \\[3pt]
\hline
\end{tabular}
\end{center}
\caption{Eigenvalues of the fixed points of table \ref{fp-table:6}.}
\label{fp-table:7}
\end{table}
All these points are unstable, or at least hyperbolic, in the three dimensional
space, but if we consider the $C_1=0$ plane, the KPZ fixed point is stable.

A second set of couplings is provided by
\begin{equation}
a_2=\frac{C_1}{C_2\Lambda},
\qquad
f_2=\frac{C_4\Lambda^2}{C_2},
\qquad
g_2=\frac{\lambda^2 \Pi_0}{4\pi C_2^3\Lambda},
\end{equation}
whose flow is
\begin{align}
&\frac{\rmd a_2}{\rmd l} = a_2\left[1- \frac{g_2(1+3f_2)}{(1+a_2+f_2)^3}\right],
	\label{k4:a2} \\[5pt]
& \frac{\rmd f_2}{\rmd l} = f_2\Bigg[\frac{g_2}{4 f_2}  \Big( 2 - 12 f_2^4+ 24 f_2^3 -162 a_2 f_2^2 -22 f_2^2 -56 a_2 f_2
- 6 a_2 - a_2^2 -12 a_2^2 f_2^2
-24 a_2 f_2^3 \Big)  \nonumber\\[5pt]
&\hskip 3 cm  \times \left(1+a_2+f_2 \right)^{-5} -2\Bigg],
	\label{k4:f2} \\[5pt]
& \frac{\rmd g_2}{\rmd l} = g_2\left[1- \frac{g_2(2+9 f_2)}{(1+a_2+f_2)^3}\right]. \label{k4:g2}
\end{align}
We provide the ensuing fixed points in table \ref{fp-table:10} (note they include the EW and KPZ fixed points),
while their stability properties are given in table \ref{fp-table:11}.
\begin{table}[h!]
\begin{center}
\begin{tabular}{|c|c|c|c|c|c|c|}
\hline
\T Name &  $a_2$  & $f_2$ & $g_2$ & $z$ & $\alpha$ & $\beta$ \\[3pt]
\hline\hline
\T EW  & $0$ & $0$ & $0$ & $2$ & $1/2$ & $1/4$ \\[3pt]
KPZ & $0$ & $0.07583$ & $0.4642$ & $1.54$ & $0.46$ & $0.299$ \\[3pt]
Fp$_{4,1}$ & $-1.1946$ & $-1/6$ & $-0.09436$ & $1$ & $1$ & $1$ \\[3pt]
\hline
\end{tabular}
\end{center}
\caption{Fixed points of \eref{k4:a2}-\eref{k4:g2} in the $(a_2,f_2,g_2)$ plane.}
\label{fp-table:10}
\end{table}
\begin{table}[h!]
\begin{center}
\begin{tabular}{|c|c|c|c|c|}
\hline
\T  Name  & $\lambda_1$ & $\lambda_2$ & $\lambda_3$ & Stability  \\[3pt]
\hline\hline
\T EW & $1$ & $-2$ & $1$ & Saddle \\[3pt]
KPZ & $0.54$ & $-2.90$ & $-1.05$ & Saddle \\[3pt]
Fp$_{4,1}$ & $88.4$ & $7.32$ & $-0.24$ & Saddle \\[3pt]
\hline
\end{tabular}
\end{center}
\caption{Eigenvalues of the fixed points of table \ref{fp-table:10}.}
\label{fp-table:11}
\end{table}

Finally, a further informative set of coupling parameters is
\begin{equation}
a_3=\frac{C_4\Lambda^3}{C_1},
\qquad
f_3=\frac{C_2\Lambda}{C_1},
\qquad
g_3=\frac{\lambda^2 \Pi_0\Lambda^2}{4\pi C_1^3},
\end{equation}
whose flow is
\begin{align}
& \frac{\rmd a_3}{\rmd l} =
    -3a_3 +\frac{g_3}{4}
    \Big[52a_3^3 -2f_3 a_3^2+4f_3^2a_3+2f_3^3-130a_3^2-48a_3 f_3
+7a_3 -6 f_3^2-f_3
        \Big]    \left(1+a_3+f_3\right)^{-5},\label{k4:a1} \\[5pt]
& \frac{\rmd f_3}{\rmd l} = -f_3+
    \frac{g_3\left(3a_3+f_3\right)}{\left(1+a_3+f_3\right)^3}, \\[5pt]
& \frac{\rmd g_3}{\rmd l} = g_3
	\left[
	    \frac{g_3}{\left(1+a_3+f_3\right)^3}-2
	\right]. \label{k4:g1}
\end{align}
In this parameter space the fixed points and their exponents are in table \ref{fp-table:8},
\begin{table}[h!]
\begin{center}
\begin{tabular}{|c|c|c|c|c|c|c|}
\hline
\T Name &  $a_1$  & $f_1$ & $g_1$ & $z$ & $\alpha$ & $\beta$ \\[3pt]
\hline\hline
\T Smooth  & $0$ & $0$ & $0$ & $1$ & $0$ & $0$  \\[3pt]
Galilean  & $0$ & $0$ & $2$ & $1$ & $1$ & $1$ \\[3pt]
Fp$_{4,1}$ & $0.1395$ & $-0.8371$ & $0.05534$ & $1$ & $1$ & $1$ \\[3pt]
\hline
\end{tabular}
\end{center}
\caption{Fixed points of \eref{k4:a1}-\eref{k4:g1} in the $(a_3,f_3,g_3)$ plane.}
\label{fp-table:8}
\end{table}
and the associated eigenvalues in table \ref{fp-table:9}.
The Fp$_{4,1}$ fixed point is obtained by solving equation
$374 x^2-2x-7=0$ and using the only admissible solution
$x_1=(1+\sqrt{291})/374$ for the components of vector
$(x_1,-6 x_1, L(x_1))$, where $L(x)=(167238-1184895 x)/34969$.
\begin{table}[h!]
\begin{center}
\begin{tabular}{|c|c|c|c|c|}
\hline
\T  Name  & $\lambda_1$ & $\lambda_2$ & $\lambda_3$ & Stability  \\[3pt]
\hline\hline
\T Smooth & $-1$ & $-3$ & $-2$ & Stable \\[3pt]
Galilean & $2$ & $\frac{3}{4}+i\frac{\sqrt{47}}{4}$ &
    $\frac{3}{4}+i\frac{\sqrt{47}}{4}$ & Unstable \\[3pt]
Fp$_{4,1}$ & $1.09$ & $2.28$ & $-1.98$ & Saddle \\[3pt]
\hline
\end{tabular}
\end{center}
\caption{Eigenvalues of the fixed points of table \ref{fp-table:8}.}
\label{fp-table:9}
\end{table}

As done for the $|k|^3$ term, we can show the irrelevance of the
$k^4$ ``surface diffusion'' term, compared to the $k^2$ ``surface tension'' one for $C_2,C_4>0$. After
rescaling of equation \eref{eq_gral}, the ratio between these two terms
is equal to $C_2 b^2/C_4$. In the presence of surface diffusion, we obtain
several fixed points which are not present in its absence. However, the only
fixed point with exponents different from those found in section
\ref{DRGflow} is Fp$_{4,0}$, i.e.\ the linear MBE fixed point, while Fp$_{4,1}$ has
the same exponents of the Galilean fixed point, $z=1$ and $\alpha=1$.
The stability matrices show the latter fixed point has one stable direction,
while the Galilean and Fp$_{4,0}$ fixed points are both unstable.
The KPZ fixed point has two stable directions and becomes the only stable
fixed point of the DRG flow if we restrict our analysis to the $C_1=0$ case, i.e.\ $a_4=0$.
Finally, as for the $|k|^3$ term, surface diffusion does not introduce
any unexpected behavior in case $C_1\neq 0$ and $C_2>0$, hence we conclude that
the $C_4$ term is required only when $C_2<0$, in order to to have
a dispersion relation that is well posed in the $k\to\infty$ limit.

In summary, within the limitations of the DRG technique, we have shown that
the main conclusions obtained for a linear dispersion relation of the $k-k^2$
type also apply in the presence of higher order polynomial contributions, as long
as they have the proper stabilizing signs. Only when the $k^2$ surface tension term
is destabilizing, such higher order terms are required for physical consistency.
A similar conclusion is expected for $\mu\neq 1$ values. This irrelevance
(and also the self-consistency of the DRG calculations) qualifies equation
\eref{eq_drg} as a faithful representative of the scaling behavior of \eref{eq_gral},
as employed in the main text.

\section*{References}

\bibliography{biblio}

\begin{thebibliography}{79}
\expandafter\ifx\csname natexlab\endcsname\relax\def\natexlab#1{#1}\fi
\expandafter\ifx\csname bibnamefont\endcsname\relax
  \def\bibnamefont#1{#1}\fi
\expandafter\ifx\csname bibfnamefont\endcsname\relax
  \def\bibfnamefont#1{#1}\fi
\expandafter\ifx\csname citenamefont\endcsname\relax
  \def\citenamefont#1{#1}\fi
\expandafter\ifx\csname url\endcsname\relax
  \def\url#1{\texttt{#1}}\fi
\expandafter\ifx\csname urlprefix\endcsname\relax\def\urlprefix{URL }\fi
\providecommand{\bibinfo}[2]{#2}
\providecommand{\eprint}[2][]{\url{#2}}

\bibitem[{\citenamefont{Barab\'asi and Stanley}(1995)}]{Barabasi:1995}
\bibinfo{author}{\bibfnamefont{A.-L.} \bibnamefont{Barab\'asi}}
  \bibnamefont{and} \bibinfo{author}{\bibfnamefont{H.~E.}
  \bibnamefont{Stanley}}, \emph{\bibinfo{title}{Fractal concepts in surface
  growth}} (\bibinfo{publisher}{Cambridge University Press},
  \bibinfo{address}{Cambridge}, \bibinfo{year}{1995}).

\bibitem[{\citenamefont{Krug}(1997)}]{Krug:1997}
\bibinfo{author}{\bibfnamefont{J.}~\bibnamefont{Krug}}, \bibinfo{journal}{Adv.
  Phys.} \textbf{\bibinfo{volume}{46}}, \bibinfo{pages}{139}
  (\bibinfo{year}{1997}).

\bibitem[{\citenamefont{Cuerno et~al.}(2007)\citenamefont{Cuerno, Castro,
  Mu{\~n}oz-Garc\'{\i}a, Gago, and V\'azquez}}]{cuerno:2007}
\bibinfo{author}{\bibfnamefont{R.}~\bibnamefont{Cuerno}},
  \bibinfo{author}{\bibfnamefont{M.}~\bibnamefont{Castro}},
  \bibinfo{author}{\bibfnamefont{J.}~\bibnamefont{Mu{\~n}oz-Garc\'{\i}a}},
  \bibinfo{author}{\bibfnamefont{R.}~\bibnamefont{Gago}}, \bibnamefont{and}
  \bibinfo{author}{\bibfnamefont{L.}~\bibnamefont{V\'azquez}},
  \bibinfo{journal}{Eur. Phys. J. Special Topics}
  \textbf{\bibinfo{volume}{146}}, \bibinfo{pages}{427} (\bibinfo{year}{2007}).

\bibitem[{\citenamefont{Misbah et~al.}(2010)\citenamefont{Misbah, Pierre-Louis,
  and Saito}}]{Misbah:2010}
\bibinfo{author}{\bibfnamefont{C.}~\bibnamefont{Misbah}},
  \bibinfo{author}{\bibfnamefont{O.}~\bibnamefont{Pierre-Louis}},
  \bibnamefont{and} \bibinfo{author}{\bibfnamefont{Y.}~\bibnamefont{Saito}},
  \bibinfo{journal}{Rev. Mod. Phys.} \textbf{\bibinfo{volume}{82}},
  \bibinfo{pages}{981} (\bibinfo{year}{2010}).

\bibitem[{\citenamefont{Cuerno and Castro}(2001)}]{Cuerno:2001}
\bibinfo{author}{\bibfnamefont{R.}~\bibnamefont{Cuerno}} \bibnamefont{and}
  \bibinfo{author}{\bibfnamefont{M.}~\bibnamefont{Castro}},
  \bibinfo{journal}{Phys. Rev. Lett.} \textbf{\bibinfo{volume}{87}},
  \bibinfo{pages}{236103} (\bibinfo{year}{2001}).

\bibitem[{\citenamefont{Nicoli et~al.}(2008)\citenamefont{Nicoli, Castro, and
  Cuerno}}]{Nicoli:2008}
\bibinfo{author}{\bibfnamefont{M.}~\bibnamefont{Nicoli}},
  \bibinfo{author}{\bibfnamefont{M.}~\bibnamefont{Castro}}, \bibnamefont{and}
  \bibinfo{author}{\bibfnamefont{R.}~\bibnamefont{Cuerno}},
  \bibinfo{journal}{Phys. Rev. E} \textbf{\bibinfo{volume}{78}},
  \bibinfo{eid}{021601} (\bibinfo{year}{2008}).

\bibitem[{\citenamefont{Krug and Meakin}(1991)}]{Krug:1991}
\bibinfo{author}{\bibfnamefont{J.}~\bibnamefont{Krug}} \bibnamefont{and}
  \bibinfo{author}{\bibfnamefont{P.}~\bibnamefont{Meakin}},
  \bibinfo{journal}{Phys. Rev. Lett.} \textbf{\bibinfo{volume}{66}},
  \bibinfo{pages}{703} (\bibinfo{year}{1991}).

\bibitem[{\citenamefont{Biler et~al.}(1999)\citenamefont{Biler, Karch, and
  Woyczynski}}]{Biler:1999}
\bibinfo{author}{\bibfnamefont{P.}~\bibnamefont{Biler}},
  \bibinfo{author}{\bibfnamefont{G.}~\bibnamefont{Karch}}, \bibnamefont{and}
  \bibinfo{author}{\bibfnamefont{W.~A.} \bibnamefont{Woyczynski}},
  \bibinfo{journal}{Studia Math.} \textbf{\bibinfo{volume}{135}},
  \bibinfo{pages}{231} (\bibinfo{year}{1999}).

\bibitem[{\citenamefont{Cross and Greenside}(2009)}]{Cross:2009}
\bibinfo{author}{\bibfnamefont{M.}~\bibnamefont{Cross}} \bibnamefont{and}
  \bibinfo{author}{\bibfnamefont{H.}~\bibnamefont{Greenside}},
  \emph{\bibinfo{title}{Pattern Formation and Dynamics in Nonequilibrium
  Systems}} (\bibinfo{publisher}{Cambridge University Press},
  \bibinfo{address}{Cambridge}, \bibinfo{year}{2009}).

\bibitem[{\citenamefont{Pelc\'e}(2004)}]{Pelce:2004}
\bibinfo{author}{\bibfnamefont{P.}~\bibnamefont{Pelc\'e}},
  \emph{\bibinfo{title}{New Visions on Form and Growth}}
  (\bibinfo{publisher}{Oxford University Press}, \bibinfo{address}{New York},
  \bibinfo{year}{2004}).

\bibitem[{\citenamefont{Saffman and Taylor}(1958)}]{Saffman:1958}
\bibinfo{author}{\bibfnamefont{P.~G.} \bibnamefont{Saffman}} \bibnamefont{and}
  \bibinfo{author}{\bibfnamefont{G.}~\bibnamefont{Taylor}},
  \bibinfo{journal}{Proc. R. Soc. London A} \textbf{\bibinfo{volume}{245}},
  \bibinfo{pages}{312} (\bibinfo{year}{1958}).

\bibitem[{\citenamefont{Bensimon et~al.}(1986)\citenamefont{Bensimon, Kadanoff,
  Liang, Shraiman, and Tang}}]{Bensimon:1986}
\bibinfo{author}{\bibfnamefont{D.}~\bibnamefont{Bensimon}},
  \bibinfo{author}{\bibfnamefont{L.~P.} \bibnamefont{Kadanoff}},
  \bibinfo{author}{\bibfnamefont{S.}~\bibnamefont{Liang}},
  \bibinfo{author}{\bibfnamefont{B.~I.} \bibnamefont{Shraiman}},
  \bibnamefont{and} \bibinfo{author}{\bibfnamefont{C.}~\bibnamefont{Tang}},
  \bibinfo{journal}{Rev. Mod. Phys.} \textbf{\bibinfo{volume}{58}},
  \bibinfo{pages}{977} (\bibinfo{year}{1986}).

\bibitem[{\citenamefont{Darrieus}()}]{Darrieus:1938}
\bibinfo{author}{\bibfnamefont{G.}~\bibnamefont{Darrieus}}, \bibinfo{note}{{\em
  unpublished} (1938)}.

\bibitem[{\citenamefont{Landau}(1944)}]{Landau:1944}
\bibinfo{author}{\bibfnamefont{L.~D.} \bibnamefont{Landau}},
  \bibinfo{journal}{Acta Physicochim. USSR} \textbf{\bibinfo{volume}{19}},
  \bibinfo{pages}{77} (\bibinfo{year}{1944}).

\bibitem[{\citenamefont{Clanet and Searby}(1998)}]{Clanet:1998}
\bibinfo{author}{\bibfnamefont{C.}~\bibnamefont{Clanet}} \bibnamefont{and}
  \bibinfo{author}{\bibfnamefont{G.}~\bibnamefont{Searby}},
  \bibinfo{journal}{Phys. Rev. Lett.} \textbf{\bibinfo{volume}{80}},
  \bibinfo{pages}{3867} (\bibinfo{year}{1998}).

\bibitem[{\citenamefont{Bychkov and Liberman}(2000)}]{Bychkov:2000}
\bibinfo{author}{\bibfnamefont{V.~V.} \bibnamefont{Bychkov}} \bibnamefont{and}
  \bibinfo{author}{\bibfnamefont{M.~A.} \bibnamefont{Liberman}},
  \bibinfo{journal}{Phys. Rep.} \textbf{\bibinfo{volume}{325}},
  \bibinfo{pages}{115} (\bibinfo{year}{2000}).

\bibitem[{\citenamefont{Kardar et~al.}(1986)\citenamefont{Kardar, Parisi, and
  Zhang}}]{Kardar:1986}
\bibinfo{author}{\bibfnamefont{M.}~\bibnamefont{Kardar}},
  \bibinfo{author}{\bibfnamefont{G.}~\bibnamefont{Parisi}}, \bibnamefont{and}
  \bibinfo{author}{\bibfnamefont{Y.-C.} \bibnamefont{Zhang}},
  \bibinfo{journal}{Phys. Rev. Lett.} \textbf{\bibinfo{volume}{56}},
  \bibinfo{pages}{889} (\bibinfo{year}{1986}).

\bibitem[{\citenamefont{Castro et~al.}(2007)\citenamefont{Castro,
  Mu{\~n}oz-Garc{\'\i}a, Cuerno, Garc{\'\i}a~Hern{\'a}ndez, and
  V{\'a}zquez}}]{Castro:2007}
\bibinfo{author}{\bibfnamefont{M.}~\bibnamefont{Castro}},
  \bibinfo{author}{\bibfnamefont{J.}~\bibnamefont{Mu{\~n}oz-Garc{\'\i}a}},
  \bibinfo{author}{\bibfnamefont{R.}~\bibnamefont{Cuerno}},
  \bibinfo{author}{\bibfnamefont{M.~M.}
  \bibnamefont{Garc{\'\i}a~Hern{\'a}ndez}}, \bibnamefont{and}
  \bibinfo{author}{\bibfnamefont{L.}~\bibnamefont{V{\'a}zquez}},
  \bibinfo{journal}{New J. Phys.} \textbf{\bibinfo{volume}{9}},
  \bibinfo{pages}{102} (\bibinfo{year}{2007}).

\bibitem[{\citenamefont{Cuerno et~al.}(1995)\citenamefont{Cuerno, Makse,
  Tomassone, Harrington, and Stanley}}]{Cuerno:1995}
\bibinfo{author}{\bibfnamefont{R.}~\bibnamefont{Cuerno}},
  \bibinfo{author}{\bibfnamefont{H.~A.} \bibnamefont{Makse}},
  \bibinfo{author}{\bibfnamefont{S.}~\bibnamefont{Tomassone}},
  \bibinfo{author}{\bibfnamefont{S.~T.} \bibnamefont{Harrington}},
  \bibnamefont{and} \bibinfo{author}{\bibfnamefont{H.~E.}
  \bibnamefont{Stanley}}, \bibinfo{journal}{Phys. Rev. Lett.}
  \textbf{\bibinfo{volume}{75}}, \bibinfo{pages}{4464} (\bibinfo{year}{1995}).

\bibitem[{\citenamefont{Cuerno and Lauritsen}(1995)}]{Cuerno:1995b}
\bibinfo{author}{\bibfnamefont{R.}~\bibnamefont{Cuerno}} \bibnamefont{and}
  \bibinfo{author}{\bibfnamefont{K.~B.} \bibnamefont{Lauritsen}},
  \bibinfo{journal}{Phys. Rev. E} \textbf{\bibinfo{volume}{52}},
  \bibinfo{pages}{4853} (\bibinfo{year}{1995}).

\bibitem[{\citenamefont{Ueno et~al.}(2005)\citenamefont{Ueno, Sakaguchi, and
  Okamura}}]{Ueno:2005}
\bibinfo{author}{\bibfnamefont{K.}~\bibnamefont{Ueno}},
  \bibinfo{author}{\bibfnamefont{H.}~\bibnamefont{Sakaguchi}},
  \bibnamefont{and} \bibinfo{author}{\bibfnamefont{M.}~\bibnamefont{Okamura}},
  \bibinfo{journal}{Phys. Rev. E} \textbf{\bibinfo{volume}{71}},
  \bibinfo{eid}{046138} (\bibinfo{year}{2005}).

\bibitem[{\citenamefont{Nicoli et~al.}(2010)\citenamefont{Nicoli, Vivo, and
  Cuerno}}]{Nicoli:2010}
\bibinfo{author}{\bibfnamefont{M.}~\bibnamefont{Nicoli}},
  \bibinfo{author}{\bibfnamefont{E.}~\bibnamefont{Vivo}}, \bibnamefont{and}
  \bibinfo{author}{\bibfnamefont{R.}~\bibnamefont{Cuerno}},
  \bibinfo{journal}{Phys. Rev. E} \textbf{\bibinfo{volume}{82}},
  \bibinfo{pages}{045202} (\bibinfo{year}{2010}).

\bibitem[{\citenamefont{Pradas et~al.}(2011)\citenamefont{Pradas, Tseluiko,
  Kalliadasis, Papageorgiou, and Pavliotis}}]{Pradas:2011}
\bibinfo{author}{\bibfnamefont{M.}~\bibnamefont{Pradas}},
  \bibinfo{author}{\bibfnamefont{D.}~\bibnamefont{Tseluiko}},
  \bibinfo{author}{\bibfnamefont{S.}~\bibnamefont{Kalliadasis}},
  \bibinfo{author}{\bibfnamefont{D.~T.} \bibnamefont{Papageorgiou}},
  \bibnamefont{and} \bibinfo{author}{\bibfnamefont{G.~A.}
  \bibnamefont{Pavliotis}}, \bibinfo{journal}{Phys. Rev. Lett.}
  \textbf{\bibinfo{volume}{106}}, \bibinfo{pages}{060602}
  (\bibinfo{year}{2011}).

\bibitem[{\citenamefont{Yakhot}(1981)}]{Yakhot:1981}
\bibinfo{author}{\bibfnamefont{V.}~\bibnamefont{Yakhot}},
  \bibinfo{journal}{Phys. Rev. A} \textbf{\bibinfo{volume}{24}},
  \bibinfo{pages}{642} (\bibinfo{year}{1981}).

\bibitem[{\citenamefont{McComb}(1991)}]{McComb:1991}
\bibinfo{author}{\bibfnamefont{W.~D.} \bibnamefont{McComb}},
  \emph{\bibinfo{title}{The Physics of Fluid Turbulence}}
  (\bibinfo{publisher}{Oxford University Press}, \bibinfo{address}{New York},
  \bibinfo{year}{1991}).

\bibitem[{\citenamefont{Halpin-Healy and Zhang}(1995)}]{Halpin-Healy:1995}
\bibinfo{author}{\bibfnamefont{T.}~\bibnamefont{Halpin-Healy}}
  \bibnamefont{and} \bibinfo{author}{\bibfnamefont{Y.-C.} \bibnamefont{Zhang}},
  \bibinfo{journal}{Phys. Rep.} \textbf{\bibinfo{volume}{254}},
  \bibinfo{pages}{215} (\bibinfo{year}{1995}).

\bibitem[{\citenamefont{Medina et~al.}(1989)\citenamefont{Medina, Hwa, Kardar,
  and Zhang}}]{Medina:1989}
\bibinfo{author}{\bibfnamefont{E.}~\bibnamefont{Medina}},
  \bibinfo{author}{\bibfnamefont{T.}~\bibnamefont{Hwa}},
  \bibinfo{author}{\bibfnamefont{M.}~\bibnamefont{Kardar}}, \bibnamefont{and}
  \bibinfo{author}{\bibfnamefont{Y.-C.} \bibnamefont{Zhang}},
  \bibinfo{journal}{Phys. Rev. A} \textbf{\bibinfo{volume}{39}},
  \bibinfo{pages}{3053} (\bibinfo{year}{1989}).

\bibitem[{\citenamefont{Cuerno and V\'azquez}(2004)}]{Cuerno:2004}
\bibinfo{author}{\bibfnamefont{R.}~\bibnamefont{Cuerno}} \bibnamefont{and}
  \bibinfo{author}{\bibfnamefont{L.}~\bibnamefont{V\'azquez}}, in
  \emph{\bibinfo{booktitle}{Advanced in Condensed Matter and Statistical
  Physics}}, edited by
  \bibinfo{editor}{\bibfnamefont{E.}~\bibnamefont{Korutcheva}}
  \bibnamefont{and} \bibinfo{editor}{\bibfnamefont{R.}~\bibnamefont{Cuerno}}
  (\bibinfo{publisher}{Nova Science Publishers}, \bibinfo{year}{2004}), p.
  \bibinfo{pages}{237}.

\bibitem[{\citenamefont{Tang and Ma}(2001)}]{Tang:2001}
\bibinfo{author}{\bibfnamefont{G.}~\bibnamefont{Tang}} \bibnamefont{and}
  \bibinfo{author}{\bibfnamefont{B.}~\bibnamefont{Ma}},
  \bibinfo{journal}{Physica A} \textbf{\bibinfo{volume}{298}},
  \bibinfo{pages}{257} (\bibinfo{year}{2001}).

\bibitem[{\citenamefont{Mann and Woyczynski}(2001)}]{Mann:2001}
\bibinfo{author}{\bibfnamefont{J.~A.} \bibnamefont{Mann}} \bibnamefont{and}
  \bibinfo{author}{\bibfnamefont{W.~A.} \bibnamefont{Woyczynski}},
  \bibinfo{journal}{Physica A} \textbf{\bibinfo{volume}{291}},
  \bibinfo{pages}{159} (\bibinfo{year}{2001}).

\bibitem[{\citenamefont{Katzav}(2003)}]{Katzav:2003}
\bibinfo{author}{\bibfnamefont{E.}~\bibnamefont{Katzav}},
  \bibinfo{journal}{Phys. Rev. E} \textbf{\bibinfo{volume}{68}},
  \bibinfo{pages}{031607} (\bibinfo{year}{2003}).

\bibitem[{\citenamefont{Cuerno and Castro}(2002)}]{Cuerno:2002}
\bibinfo{author}{\bibfnamefont{R.}~\bibnamefont{Cuerno}} \bibnamefont{and}
  \bibinfo{author}{\bibfnamefont{M.}~\bibnamefont{Castro}},
  \bibinfo{journal}{Physica A} \textbf{\bibinfo{volume}{314}},
  \bibinfo{pages}{192} (\bibinfo{year}{2002}).

\bibitem[{\citenamefont{Nicoli et~al.}(2009{\natexlab{a}})\citenamefont{Nicoli,
  Cuerno, and Castro}}]{Nicoli:2009b}
\bibinfo{author}{\bibfnamefont{M.}~\bibnamefont{Nicoli}},
  \bibinfo{author}{\bibfnamefont{R.}~\bibnamefont{Cuerno}}, \bibnamefont{and}
  \bibinfo{author}{\bibfnamefont{M.}~\bibnamefont{Castro}},
  \bibinfo{journal}{Phys. Rev. Lett.} \textbf{\bibinfo{volume}{102}},
  \bibinfo{eid}{256102} (\bibinfo{year}{2009}{\natexlab{a}}).

\bibitem[{\citenamefont{Samko et~al.}(2002)\citenamefont{Samko, Kilbas, and
  Marichev}}]{Samko:2002}
\bibinfo{author}{\bibfnamefont{S.~G.} \bibnamefont{Samko}},
  \bibinfo{author}{\bibfnamefont{A.~A.} \bibnamefont{Kilbas}},
  \bibnamefont{and} \bibinfo{author}{\bibfnamefont{O.~I.}
  \bibnamefont{Marichev}}, \emph{\bibinfo{title}{Fractional Integrals and
  Derivatives}} (\bibinfo{publisher}{Taylor and Francis},
  \bibinfo{address}{London}, \bibinfo{year}{2002}).

\bibitem[{\citenamefont{Silvestre}(2007)}]{Silvestre:2007}
\bibinfo{author}{\bibfnamefont{L.}~\bibnamefont{Silvestre}},
  \bibinfo{journal}{Comm. Pure Appl. Math.} \textbf{\bibinfo{volume}{60}},
  \bibinfo{pages}{0067} (\bibinfo{year}{2007}).

\bibitem[{\citenamefont{Mukamel}(2008)}]{Mukamel:2008}
\bibinfo{author}{\bibfnamefont{D.}~\bibnamefont{Mukamel}}, in
  \emph{\bibinfo{booktitle}{Proceedings of the Les Houches Summer School on
  Long-Range Interacting Systems}} (\bibinfo{publisher}{Elsevier},
  \bibinfo{address}{New York}, \bibinfo{year}{2008}).

\bibitem[{\citenamefont{Kim and Kim}(2010)}]{Kim:2010}
\bibinfo{author}{\bibfnamefont{D.~H.} \bibnamefont{Kim}} \bibnamefont{and}
  \bibinfo{author}{\bibfnamefont{J.~M.} \bibnamefont{Kim}},
  \bibinfo{journal}{J. Stat. Mech.} p. \bibinfo{pages}{P08008}
  (\bibinfo{year}{2010}).

\bibitem[{\citenamefont{Sivashinsky}(1977)}]{Sivashinsky:1977}
\bibinfo{author}{\bibfnamefont{G.~I.} \bibnamefont{Sivashinsky}},
  \bibinfo{journal}{Acta Astronaut.} \textbf{\bibinfo{volume}{4}},
  \bibinfo{pages}{1177} (\bibinfo{year}{1977}).

\bibitem[{\citenamefont{Michelson and Sivashinsky}(1977)}]{Michelson:1977}
\bibinfo{author}{\bibfnamefont{D.~M.} \bibnamefont{Michelson}}
  \bibnamefont{and} \bibinfo{author}{\bibfnamefont{G.~I.}
  \bibnamefont{Sivashinsky}}, \bibinfo{journal}{Acta Astronaut.}
  \textbf{\bibinfo{volume}{4}}, \bibinfo{pages}{1207} (\bibinfo{year}{1977}).

\bibitem[{\citenamefont{Bales et~al.}(1990)\citenamefont{Bales, Bruinsma,
  Eklund, Karunasiri, Rudnick, and Zangwill}}]{Bales:1990}
\bibinfo{author}{\bibfnamefont{G.~S.} \bibnamefont{Bales}},
  \bibinfo{author}{\bibfnamefont{R.}~\bibnamefont{Bruinsma}},
  \bibinfo{author}{\bibfnamefont{E.}~\bibnamefont{Eklund}},
  \bibinfo{author}{\bibfnamefont{R.}~\bibnamefont{Karunasiri}},
  \bibinfo{author}{\bibfnamefont{J.}~\bibnamefont{Rudnick}}, \bibnamefont{and}
  \bibinfo{author}{\bibfnamefont{A.}~\bibnamefont{Zangwill}},
  \bibinfo{journal}{Science} \textbf{\bibinfo{volume}{249}},
  \bibinfo{pages}{264} (\bibinfo{year}{1990}).

\bibitem[{\citenamefont{Kechagia et~al.}(2001)\citenamefont{Kechagia, Yortsos,
  and Lichtner}}]{Kechagia:2001}
\bibinfo{author}{\bibfnamefont{P.}~\bibnamefont{Kechagia}},
  \bibinfo{author}{\bibfnamefont{Y.~C.} \bibnamefont{Yortsos}},
  \bibnamefont{and} \bibinfo{author}{\bibfnamefont{P.}~\bibnamefont{Lichtner}},
  \bibinfo{journal}{Phys. Rev. E} \textbf{\bibinfo{volume}{64}},
  \bibinfo{pages}{016315} (\bibinfo{year}{2001}).

\bibitem[{\citenamefont{Castro et~al.}(2011)\citenamefont{Castro, Buijnsters,
  Nicoli, Cuerno, and V\'azquez}}]{Castro:2011}
\bibinfo{author}{\bibfnamefont{M.}~\bibnamefont{Castro}},
  \bibinfo{author}{\bibfnamefont{I.}~\bibnamefont{Buijnsters}},
  \bibinfo{author}{\bibfnamefont{M.}~\bibnamefont{Nicoli}},
  \bibinfo{author}{\bibfnamefont{R.}~\bibnamefont{Cuerno}}, \bibnamefont{and}
  \bibinfo{author}{\bibfnamefont{L.}~\bibnamefont{V\'azquez}},
  \bibinfo{journal}{unpublished}  (\bibinfo{year}{2011}).

\bibitem[{\citenamefont{Nicoli et~al.}(2009{\natexlab{b}})\citenamefont{Nicoli,
  Castro, and Cuerno}}]{Nicoli:2009}
\bibinfo{author}{\bibfnamefont{M.}~\bibnamefont{Nicoli}},
  \bibinfo{author}{\bibfnamefont{M.}~\bibnamefont{Castro}}, \bibnamefont{and}
  \bibinfo{author}{\bibfnamefont{R.}~\bibnamefont{Cuerno}},
  \bibinfo{journal}{J. Stat. Mech.} p. \bibinfo{pages}{P02036}
  (\bibinfo{year}{2009}{\natexlab{b}}).

\bibitem[{\citenamefont{Haselwandter and
  Vvedensky}(2007)}]{Haselwandter:2007-prl}
\bibinfo{author}{\bibfnamefont{C.~A.} \bibnamefont{Haselwandter}}
  \bibnamefont{and} \bibinfo{author}{\bibfnamefont{D.~D.}
  \bibnamefont{Vvedensky}}, \bibinfo{journal}{Phys. Rev. Lett.}
  \textbf{\bibinfo{volume}{98}}, \bibinfo{eid}{046102} (\bibinfo{year}{2007}).

\bibitem[{\citenamefont{Haselwandter and Vvedensky}(2010)}]{Haselwandter:2010}
\bibinfo{author}{\bibfnamefont{C.~A.} \bibnamefont{Haselwandter}}
  \bibnamefont{and} \bibinfo{author}{\bibfnamefont{D.~D.}
  \bibnamefont{Vvedensky}}, \bibinfo{journal}{Phys. Rev. E}
  \textbf{\bibinfo{volume}{81}}, \bibinfo{pages}{021606}
  (\bibinfo{year}{2010}).

\bibitem[{\citenamefont{Keller et~al.}(2011)\citenamefont{Keller, Nicoli,
  Facsko, and Cuerno}}]{Keller:2011}
\bibinfo{author}{\bibfnamefont{A.}~\bibnamefont{Keller}},
  \bibinfo{author}{\bibfnamefont{M.}~\bibnamefont{Nicoli}},
  \bibinfo{author}{\bibfnamefont{S.}~\bibnamefont{Facsko}}, \bibnamefont{and}
  \bibinfo{author}{\bibfnamefont{R.}~\bibnamefont{Cuerno}},
  \bibinfo{journal}{Phys. Rev. E} \textbf{\bibinfo{volume}{84}},
  \bibinfo{pages}{015202} (\bibinfo{year}{2011}).

\bibitem[{\citenamefont{Lai and Das~Sarma}(1991)}]{Lai:1991}
\bibinfo{author}{\bibfnamefont{Z.-W.} \bibnamefont{Lai}} \bibnamefont{and}
  \bibinfo{author}{\bibfnamefont{S.}~\bibnamefont{Das~Sarma}},
  \bibinfo{journal}{Phys. Rev. Lett.} \textbf{\bibinfo{volume}{66}},
  \bibinfo{pages}{2348} (\bibinfo{year}{1991}).

\bibitem[{\citenamefont{Villain}(1991)}]{Villain:1991}
\bibinfo{author}{\bibfnamefont{J.}~\bibnamefont{Villain}}, \bibinfo{journal}{J.
  Phys. I France} \textbf{\bibinfo{volume}{1}}, \bibinfo{pages}{19}
  (\bibinfo{year}{1991}).

\bibitem[{\citenamefont{Hentschel and Family}(1991)}]{Hentschel:1991}
\bibinfo{author}{\bibfnamefont{H.~G.~E.} \bibnamefont{Hentschel}}
  \bibnamefont{and} \bibinfo{author}{\bibfnamefont{F.}~\bibnamefont{Family}},
  \bibinfo{journal}{Phys. Rev. Lett.} \textbf{\bibinfo{volume}{66}},
  \bibinfo{pages}{1982} (\bibinfo{year}{1991}).

\bibitem[{\citenamefont{Kardar}(2007)}]{Kardar:2007}
\bibinfo{author}{\bibfnamefont{M.}~\bibnamefont{Kardar}},
  \emph{\bibinfo{title}{Statistical physics of fields}}
  (\bibinfo{publisher}{Cambridge University Press},
  \bibinfo{address}{Cambridge}, \bibinfo{year}{2007}).

\bibitem[{\citenamefont{Fisher et~al.}(1972)\citenamefont{Fisher, Ma, and
  Nickel}}]{Fisher:1972}
\bibinfo{author}{\bibfnamefont{M.~E.} \bibnamefont{Fisher}},
  \bibinfo{author}{\bibfnamefont{S.-K.} \bibnamefont{Ma}}, \bibnamefont{and}
  \bibinfo{author}{\bibfnamefont{B.~G.} \bibnamefont{Nickel}},
  \bibinfo{journal}{Phys. Rev. Lett.} \textbf{\bibinfo{volume}{29}},
  \bibinfo{pages}{917} (\bibinfo{year}{1972}).

\bibitem[{\citenamefont{Giada et~al.}(2002)\citenamefont{Giada, Giacometti, and
  Rossi}}]{Giada:2002}
\bibinfo{author}{\bibfnamefont{L.}~\bibnamefont{Giada}},
  \bibinfo{author}{\bibfnamefont{A.}~\bibnamefont{Giacometti}},
  \bibnamefont{and} \bibinfo{author}{\bibfnamefont{M.}~\bibnamefont{Rossi}},
  \bibinfo{journal}{Phys. Rev. E} \textbf{\bibinfo{volume}{65}},
  \bibinfo{pages}{036134} (\bibinfo{year}{2002}).

\bibitem[{\citenamefont{Gallego et~al.}(2007)\citenamefont{Gallego, Castro, and
  L\'{o}pez}}]{Gallego:2007}
\bibinfo{author}{\bibfnamefont{R.}~\bibnamefont{Gallego}},
  \bibinfo{author}{\bibfnamefont{M.}~\bibnamefont{Castro}}, \bibnamefont{and}
  \bibinfo{author}{\bibfnamefont{J.~M.} \bibnamefont{L\'{o}pez}},
  \bibinfo{journal}{Phys. Rev. E} \textbf{\bibinfo{volume}{76}},
  \bibinfo{eid}{051121} (\bibinfo{year}{2007}).

\bibitem[{\citenamefont{Forster et~al.}(1977)\citenamefont{Forster, Nelson, and
  Stephen}}]{Forster:1977}
\bibinfo{author}{\bibfnamefont{D.}~\bibnamefont{Forster}},
  \bibinfo{author}{\bibfnamefont{D.~R.} \bibnamefont{Nelson}},
  \bibnamefont{and} \bibinfo{author}{\bibfnamefont{M.~J.}
  \bibnamefont{Stephen}}, \bibinfo{journal}{Phys. Rev. A}
  \textbf{\bibinfo{volume}{16}}, \bibinfo{pages}{732} (\bibinfo{year}{1977}).

\bibitem[{\citenamefont{Lauritsen}(1995)}]{Lauritsen:1995}
\bibinfo{author}{\bibfnamefont{K.~B.} \bibnamefont{Lauritsen}},
  \bibinfo{journal}{Phys. Rev. E} \textbf{\bibinfo{volume}{52}},
  \bibinfo{pages}{R1261} (\bibinfo{year}{1995}).

\bibitem[{\citenamefont{Janssen}(1997)}]{Janssen:1997}
\bibinfo{author}{\bibfnamefont{H.~K.} \bibnamefont{Janssen}},
  \bibinfo{journal}{Phys. Rev. Lett.} \textbf{\bibinfo{volume}{78}},
  \bibinfo{pages}{1082} (\bibinfo{year}{1997}).

\bibitem[{\citenamefont{Frey and T\"auber}(1994)}]{Frey:1994}
\bibinfo{author}{\bibfnamefont{E.}~\bibnamefont{Frey}} \bibnamefont{and}
  \bibinfo{author}{\bibfnamefont{U.~C.} \bibnamefont{T\"auber}},
  \bibinfo{journal}{Phys. Rev. E} \textbf{\bibinfo{volume}{50}},
  \bibinfo{pages}{1024} (\bibinfo{year}{1994}).

\bibitem[{\citenamefont{Sun and Plischke}(1994)}]{Sun:1994}
\bibinfo{author}{\bibfnamefont{T.}~\bibnamefont{Sun}} \bibnamefont{and}
  \bibinfo{author}{\bibfnamefont{M.}~\bibnamefont{Plischke}},
  \bibinfo{journal}{Phys. Rev. E} \textbf{\bibinfo{volume}{49}},
  \bibinfo{pages}{5046} (\bibinfo{year}{1994}).

\bibitem[{\citenamefont{McComb}(2005)}]{McComb:2005}
\bibinfo{author}{\bibfnamefont{W.~D.} \bibnamefont{McComb}},
  \bibinfo{journal}{Phys. Rev. E} \textbf{\bibinfo{volume}{71}},
  \bibinfo{pages}{037301} (\bibinfo{year}{2005}).

\bibitem[{\citenamefont{Berera and Hochberg}(2007)}]{Berera:2007}
\bibinfo{author}{\bibfnamefont{A.}~\bibnamefont{Berera}} \bibnamefont{and}
  \bibinfo{author}{\bibfnamefont{D.}~\bibnamefont{Hochberg}},
  \bibinfo{journal}{Phys. Rev. Lett.} \textbf{\bibinfo{volume}{99}},
  \bibinfo{pages}{254501} (\bibinfo{year}{2007}).

\bibitem[{\citenamefont{Berera and Hochberg}(2009)}]{Berera:2009}
\bibinfo{author}{\bibfnamefont{A.}~\bibnamefont{Berera}} \bibnamefont{and}
  \bibinfo{author}{\bibfnamefont{D.}~\bibnamefont{Hochberg}},
  \bibinfo{journal}{Nucl. Phys. B} \textbf{\bibinfo{volume}{814}},
  \bibinfo{pages}{522} (\bibinfo{year}{2009}).

\bibitem[{\citenamefont{Wio et~al.}(2010)\citenamefont{Wio, Revelli, Deza,
  Escudero, and de~la Lama}}]{Wio:2010-a}
\bibinfo{author}{\bibfnamefont{H.~S.} \bibnamefont{Wio}},
  \bibinfo{author}{\bibfnamefont{J.~A.} \bibnamefont{Revelli}},
  \bibinfo{author}{\bibfnamefont{R.~R.} \bibnamefont{Deza}},
  \bibinfo{author}{\bibfnamefont{C.}~\bibnamefont{Escudero}}, \bibnamefont{and}
  \bibinfo{author}{\bibfnamefont{M.~S.} \bibnamefont{de~la Lama}},
  \bibinfo{journal}{Europhys. Lett.} \textbf{\bibinfo{volume}{89}},
  \bibinfo{pages}{40008} (\bibinfo{year}{2010}).

\bibitem[{\citenamefont{Wio et~al.}(2011)\citenamefont{Wio, Escudero, Revelli,
  Deza, and de~la Lama}}]{Wio:2011}
\bibinfo{author}{\bibfnamefont{H.~S.} \bibnamefont{Wio}},
  \bibinfo{author}{\bibfnamefont{C.}~\bibnamefont{Escudero}},
  \bibinfo{author}{\bibfnamefont{J.~A.} \bibnamefont{Revelli}},
  \bibinfo{author}{\bibfnamefont{R.~R.} \bibnamefont{Deza}}, \bibnamefont{and}
  \bibinfo{author}{\bibfnamefont{M.~S.} \bibnamefont{de~la Lama}},
  \bibinfo{journal}{Philos. Trans. R. Soc. A} \textbf{\bibinfo{volume}{369}},
  \bibinfo{pages}{396} (\bibinfo{year}{2011}).

\bibitem[{\citenamefont{Marinari et~al.}(2000)\citenamefont{Marinari, Pagnani,
  and Parisi}}]{Marinari:2000}
\bibinfo{author}{\bibfnamefont{E.}~\bibnamefont{Marinari}},
  \bibinfo{author}{\bibfnamefont{A.}~\bibnamefont{Pagnani}}, \bibnamefont{and}
  \bibinfo{author}{\bibfnamefont{G.}~\bibnamefont{Parisi}},
  \bibinfo{journal}{J. Phys. A: Math. Gen.} \textbf{\bibinfo{volume}{33}},
  \bibinfo{pages}{8181} (\bibinfo{year}{2000}).

\bibitem[{\citenamefont{Nicoli et~al.}(2011)\citenamefont{Nicoli, Cuerno, and
  Castro}}]{Nicoli:unpubl}
\bibinfo{author}{\bibfnamefont{M.}~\bibnamefont{Nicoli}},
  \bibinfo{author}{\bibfnamefont{R.}~\bibnamefont{Cuerno}}, \bibnamefont{and}
  \bibinfo{author}{\bibfnamefont{M.}~\bibnamefont{Castro}}
  (\bibinfo{year}{2011}), \bibinfo{note}{unpublished}.

\bibitem[{\citenamefont{Karma and Misbah}(1993)}]{Karma:1993}
\bibinfo{author}{\bibfnamefont{A.}~\bibnamefont{Karma}} \bibnamefont{and}
  \bibinfo{author}{\bibfnamefont{C.}~\bibnamefont{Misbah}},
  \bibinfo{journal}{Phys. Rev. Lett.} \textbf{\bibinfo{volume}{71}},
  \bibinfo{pages}{3810} (\bibinfo{year}{1993}).

\bibitem[{\citenamefont{Kupervasser and Olami}(2011)}]{Kupervasser:unpubl}
\bibinfo{author}{\bibfnamefont{O.}~\bibnamefont{Kupervasser}} \bibnamefont{and}
  \bibinfo{author}{\bibfnamefont{Z.}~\bibnamefont{Olami}}
  (\bibinfo{year}{2011}), \bibinfo{note}{arXiv:1106.0558v1}.

\bibitem[{\citenamefont{Mayr and Averback}(2001)}]{Mayr:2001}
\bibinfo{author}{\bibfnamefont{S.~G.} \bibnamefont{Mayr}} \bibnamefont{and}
  \bibinfo{author}{\bibfnamefont{R.~S.} \bibnamefont{Averback}},
  \bibinfo{journal}{Phys. Rev. Lett.} \textbf{\bibinfo{volume}{87}},
  \bibinfo{pages}{196106} (\bibinfo{year}{2001}).

\bibitem[{\citenamefont{Zhao et~al.}(1999)\citenamefont{Zhao, Drotar, Wang, and
  Lu}}]{Zhao:1999}
\bibinfo{author}{\bibfnamefont{Y.-P.} \bibnamefont{Zhao}},
  \bibinfo{author}{\bibfnamefont{J.~T.} \bibnamefont{Drotar}},
  \bibinfo{author}{\bibfnamefont{G.-C.} \bibnamefont{Wang}}, \bibnamefont{and}
  \bibinfo{author}{\bibfnamefont{T.-M.} \bibnamefont{Lu}},
  \bibinfo{journal}{Phys. Rev. Lett.} \textbf{\bibinfo{volume}{82}},
  \bibinfo{pages}{4882} (\bibinfo{year}{1999}).

\bibitem[{\citenamefont{Dalakos et~al.}(2005)\citenamefont{Dalakos, Plawsky,
  and Persans}}]{Dalakos:2005}
\bibinfo{author}{\bibfnamefont{G.~T.} \bibnamefont{Dalakos}},
  \bibinfo{author}{\bibfnamefont{J.~P.} \bibnamefont{Plawsky}},
  \bibnamefont{and} \bibinfo{author}{\bibfnamefont{P.~D.}
  \bibnamefont{Persans}}, \bibinfo{journal}{Phys. Rev. B}
  \textbf{\bibinfo{volume}{72}}, \bibinfo{pages}{205305}
  (\bibinfo{year}{2005}).

\bibitem[{\citenamefont{Buijnsters and V\'azquez}(2008)}]{Buijnsters:2008}
\bibinfo{author}{\bibfnamefont{J.~G.} \bibnamefont{Buijnsters}}
  \bibnamefont{and}
  \bibinfo{author}{\bibfnamefont{L.}~\bibnamefont{V\'azquez}},
  \bibinfo{journal}{J. Phys. D: Appl. Phys.} \textbf{\bibinfo{volume}{41}},
  \bibinfo{pages}{012006} (\bibinfo{year}{2008}).

\bibitem[{\citenamefont{Hormann et~al.}(2009)\citenamefont{Hormann, Meier, and
  Moseler}}]{Hormann:2009}
\bibinfo{author}{\bibfnamefont{C.}~\bibnamefont{Hormann}},
  \bibinfo{author}{\bibfnamefont{S.}~\bibnamefont{Meier}}, \bibnamefont{and}
  \bibinfo{author}{\bibfnamefont{M.}~\bibnamefont{Moseler}},
  \bibinfo{journal}{Eur. Phys. J. B} \textbf{\bibinfo{volume}{69}},
  \bibinfo{pages}{187} (\bibinfo{year}{2009}).

\bibitem[{\citenamefont{Ojeda et~al.}(2000)\citenamefont{Ojeda, Cuerno,
  Salvarezza, and V\'azquez}}]{Ojeda:2000}
\bibinfo{author}{\bibfnamefont{F.}~\bibnamefont{Ojeda}},
  \bibinfo{author}{\bibfnamefont{R.}~\bibnamefont{Cuerno}},
  \bibinfo{author}{\bibfnamefont{R.}~\bibnamefont{Salvarezza}},
  \bibnamefont{and}
  \bibinfo{author}{\bibfnamefont{L.}~\bibnamefont{V\'azquez}},
  \bibinfo{journal}{Phys. Rev. Lett.} \textbf{\bibinfo{volume}{84}},
  \bibinfo{pages}{3125} (\bibinfo{year}{2000}).

\bibitem[{\citenamefont{Karunasiri et~al.}(1989)\citenamefont{Karunasiri,
  Bruinsma, and Rudnick}}]{Karunasiri:1989}
\bibinfo{author}{\bibfnamefont{R.~P.~U.} \bibnamefont{Karunasiri}},
  \bibinfo{author}{\bibfnamefont{R.}~\bibnamefont{Bruinsma}}, \bibnamefont{and}
  \bibinfo{author}{\bibfnamefont{J.}~\bibnamefont{Rudnick}},
  \bibinfo{journal}{Phys. Rev. Lett.} \textbf{\bibinfo{volume}{62}},
  \bibinfo{pages}{788} (\bibinfo{year}{1989}).

\bibitem[{\citenamefont{Drotar et~al.}(2000{\natexlab{a}})\citenamefont{Drotar,
  Zhao, Lu, and Wang}}]{Drotar:2000b}
\bibinfo{author}{\bibfnamefont{J.~T.} \bibnamefont{Drotar}},
  \bibinfo{author}{\bibfnamefont{Y.-P.} \bibnamefont{Zhao}},
  \bibinfo{author}{\bibfnamefont{T.-M.} \bibnamefont{Lu}}, \bibnamefont{and}
  \bibinfo{author}{\bibfnamefont{G.-C.} \bibnamefont{Wang}},
  \bibinfo{journal}{Phys. Rev. B} \textbf{\bibinfo{volume}{61}},
  \bibinfo{pages}{3012} (\bibinfo{year}{2000}{\natexlab{a}}).

\bibitem[{\citenamefont{Drotar et~al.}(2000{\natexlab{b}})\citenamefont{Drotar,
  Zhao, Lu, and Wang}}]{Drotar:2000a}
\bibinfo{author}{\bibfnamefont{J.~T.} \bibnamefont{Drotar}},
  \bibinfo{author}{\bibfnamefont{Y.-P.} \bibnamefont{Zhao}},
  \bibinfo{author}{\bibfnamefont{T.-M.} \bibnamefont{Lu}}, \bibnamefont{and}
  \bibinfo{author}{\bibfnamefont{G.-C.} \bibnamefont{Wang}},
  \bibinfo{journal}{Phys. Rev. B} \textbf{\bibinfo{volume}{62}},
  \bibinfo{pages}{2118} (\bibinfo{year}{2000}{\natexlab{b}}).

\bibitem[{\citenamefont{Chattopadhyay}(2002)}]{Chattopadhyay:2002}
\bibinfo{author}{\bibfnamefont{A.~K.} \bibnamefont{Chattopadhyay}},
  \bibinfo{journal}{Phys. Rev. B} \textbf{\bibinfo{volume}{65}},
  \bibinfo{pages}{041405} (\bibinfo{year}{2002}).

\bibitem[{\citenamefont{Alonso-S\'anchez and
  Hochberg}(2000)}]{Alonso-Sanchez:2000}
\bibinfo{author}{\bibfnamefont{F.}~\bibnamefont{Alonso-S\'anchez}}
  \bibnamefont{and} \bibinfo{author}{\bibfnamefont{D.}~\bibnamefont{Hochberg}},
  \bibinfo{journal}{Phys. Rev. E} \textbf{\bibinfo{volume}{62}},
  \bibinfo{pages}{7008} (\bibinfo{year}{2000}).

\bibitem[{\citenamefont{Berera and Yoffe}(2010)}]{Berera:2010}
\bibinfo{author}{\bibfnamefont{A.}~\bibnamefont{Berera}} \bibnamefont{and}
  \bibinfo{author}{\bibfnamefont{S.~R.} \bibnamefont{Yoffe}},
  \bibinfo{journal}{Phys. Rev. E} \textbf{\bibinfo{volume}{82}},
  \bibinfo{pages}{066304} (\bibinfo{year}{2010}).

\end{thebibliography}


\end{document}